\RequirePackage{fix-cm}
\documentclass[onecolumn]{svjour3}          
\smartqed  
\usepackage{graphicx}
     \usepackage{color}
     \usepackage{xcolor}
\usepackage[comma]{natbib}
\usepackage{url}
\usepackage{times}
\usepackage{amssymb}
\usepackage{amsmath}
\usepackage{outlines}
\usepackage{gensymb}
\usepackage{multirow}
\usepackage{latexsym}
\usepackage{units}
\usepackage[normalem]{ulem}
\usepackage{tabu}
\usepackage{booktabs}
\usepackage{hyperref}

\newcommand\arcsec{\mbox{$^{\prime\prime}$}}%
\newcommand{\Msun}{M$_{\odot}$}
\newcommand\aap{A\&A}                
\newcommand\aapr{A\&ARv}             
\newcommand\aaps{A\&AS}              
\newcommand\actaa{Acta Astron.}      
\newcommand\aj{AJ}                   
\newcommand\apj{ApJ}                 
\newcommand\apjl{ApJ}                
\newcommand\apjs{ApJS}               
\newcommand\apss{Ap\&SS}             
\newcommand\araa{ARA\&A}             
\newcommand\icarus{Icarus}           
\newcommand\memsai{Mem. Soc. Astron. Italiana} 
\newcommand\mnras{MNRAS}             
\newcommand\nat{Nature}              
\newcommand\prl{Phys. Rev.~Lett.}    
\newcommand\pasp{PASP}               
\newcommand\ssr{Space Sci. Rev.}     
\begin{document}

\title{\textsf{Herbig Stars}}

\subtitle{\textsf{A Quarter Century of Progress}}

\author{\sf Sean D. Brittain \and Inga Kamp \and Gwendolyn Meeus \and Ren\'{e} D. Oudmaijer  \and  L.~B.~F.~M.~Waters}


\institute{S. Brittain \at
           Department of Physics and Astronomy, Clemson University, Clemson, SC, 29634-0978, USA \\
           Tel.: +1-864-633-8265 \\
           \email{sbritt@clemson.edu} \\
           \and
          I. Kamp \at
          Kapteyn Astronomical Institute, University of Groningen, Groningen, The Netherlands \\
          \and
          G. Meeus \at
          Dpto. Física Teórica, Universidad Autónoma de Madrid, 28049 Madrid, Spain \\
          Centro de Investigación Avanzada en Física Fundamental (CIAFF), Facultad de Ciencias, UAM, 28049 Madrid, Spain \\
          \and
          R. D. Oudmaijer \at
          School of Physics and Astronomy, University of Leeds, Leeds, LS2 9JT, UK \\
          \and
          L. B. F. M. Waters \at
          Institute for Mathematics, Astrophysics \& Particle Physics,    Department of Astrophysics, Radboud University, P.O. Box 9010, NL-6500 GL Nijmegen, The Netherlands \\     
          SRON Netherlands Institute for Space Research, Sorbonnelaan 2, 3584, CA Utrecht, The Netherlands \\
          }

\date{\sf Received: 17 January 2022 / Accepted: 23 January 2023}

\maketitle

\begin{abstract}
Herbig Ae/Be stars are young contracting stars on the radiative track in the HR diagram on their way to the Main Sequence. These stars provide a valuable link between high and low mass stars. Here we review the progress that has been made in our understanding of these fascinating objects and their disks since the last major review on this topic published in 1998. We begin with a general overview of these stars and their properties. We then discuss the accretion of circumstellar material onto these stars. Next we discuss the dust and gas properties of the circumstellar disk before exploring the evidence for planet formation in these disks. We conclude with a brief discussion of future prospects for deepening our understanding of these sources and propose a new working definition of Herbig Ae/Be stars.

\end{abstract}

\tableofcontents

\section{\textsf{INTRODUCTION}}
\label{sec:intro}

Understanding the formation and early evolution of stars and planetary systems is one of the key questions in astrophysics, closely linked to the origin of the solar system. Stars form in dense molecular clouds in which gravity overtakes gas pressure, resulting in the formation of a core. Conservation of angular momentum causes the formation of an accretion disk through which gas and dust is transported to the accreting star. When the molecular cloud disperses (typically after 0.5-1 Myrs) the accretion slows down and a slower phase (typically 5-15 Myrs) of pre-main-sequence evolution ensues, which ends when the star ignites hydrogen in its core \citep[see][for reviews]{Palla1993,Lada2005, McKee2007, Li2014}.

Planet formation takes place in the (remnant) accretion disk; there is growing evidence that this process begins very early when accretion onto the star is still strong \citep{HL-Tau-ALMA2015,Segura-Cox2020, Kenyon2016, Tsukamoto2017}. 
Most studies favor the core accretion growth model for the formation of gas giant exoplanets, in which dust in the disk settles to the mid-plane to form large, millimeter to centimeter sized grains, planetesimals, and a rocky core \citep[e.g.,][]{Lissauer1993, Draz2022}. When above a critical mass of 10-20 M$_{\oplus}$, the rapid accretion of a H/He envelope follows (e.g. \citealt{Pollack1996}). Planet formation may also proceed via gravitational instability in massive disks \citep{Boss1997}, explaining the presence of very massive planets in wide orbits found in some intermediate mass stars. 

This general scenario for star- and planet formation is believed to hold for solar type stars ($0.3M_{\odot} \lesssim M_{\star} \lesssim 1.5 M_{\odot}$), and there is growing evidence that it also applies to lower mass stars and brown dwarfs \citep[$M_{\star} \lesssim 0.3 M_{\odot}$;][]{luhman2012}. For higher mass stars, this is likely to break down for M$_{\star} \gtrsim 4-5M_{\odot}$ as disk lifetimes grow too short. Because the timescale for star formation decreases rapidly with increasing mass, high mass stars do not experience a visible pre-main-sequence phase. Their much higher luminosity strongly affects the physical and chemical properties of the accretion disk  \citep[e.g.,][]{Gorti2009}. Models suggest a rapid evaporation of the outer disk and no reservoir of large dust grains is able to form, inhibiting the formation of planets through core growth. However, planets may still form around massive young stars through gravitational instability. So far, observations indicate there is no evidence for close-in planets orbiting stars with mass above 4-5 M$_{\odot}$, while campaigns for planets at larger separations are underway (\citealt{janson2021A&A...646A.164J}). 

The Herbig Ae/Be stars are intermediate mass objects and as such bridge the gap between the lower mass, solar type stars  (M$_\star \lesssim 1.5$ \Msun) and the most massive stars (M$_\star \gtrsim 10's$ \Msun). 
They were first discussed as a group in the seminal paper of \citet{1960ApJS....4..337H}. In this paper, George Herbig sought to identify more massive stars (aiming at 3-20 M$_{\odot}$) by selecting   a sample of 26 A and B stars with emission lines in the spectrum (in particular H$\alpha$) that were associated with (reflection) nebulosity. These stars have since been studied in great detail and indeed many of them turned out to be intermediate mass pre-main-sequence (PMS) stars. In later literature, intermediate mass PMS stars were given his name Herbig Ae/Be stars. In the context of star formation studies, these stars form the higher mass counterparts to the solar mass T Tau stars, named after the prototype T Tau. Herbig also immediately noted the difficulty in disentangling these stars from other classes of B and A stars with circumstellar matter, such as the classical Be stars \citep{Rivinius2013} and the B$[$e$]$ stars \citep{Kraus2019}. Herbig continued to work on these objects throughout his career, as evidenced by his extensive bibliography. 

Previous reviews dedicated to Herbig Ae/Be stars can be found in the conference
proceedings \cite{The1994}, and in \cite{1997SSRv...82..407P} and
\cite{1998ARA&A..36..233W}. In addition, there have been numerous reviews on disks around young stars that are relevant to Herbig Ae/Be stars. Examples include \citet{Dullemond2010, Williams2011, Andrews2020}. For recent reviews dedicated to aspects of Herbig Ae/Be stars please see the Topical
Collection on Herbig Ae/Be stars published in 2015 in Astrophysics and
Space Science\footnote{\url{https://link.springer.com/collections/hbggjficdi}}.

In recent years a wealth of new observational and theoretical studies have shed light on the nature and evolutionary status of Herbig Ae/Be stars and their circumstellar environment. As we will summarize in this review, Herbig Ae/Be stars are at an exciting crossroads between low- and high-mass star formation. The high stellar luminosity, disk mass, and often large disks allow easier access to the relevant spatial scales to study planet formation processes when compared to lower mass objects. Indeed, the gas and dust in their circumstellar disks has been spatially resolved with unprecedented detail, revealing Keplerian disks that show convincing evidence for the presence of forming planets. As samples of known exoplanets rapidly increase, we can begin to make the link between the diversity of exoplanetary systems in intermediate mass stars and their birth sites.  Herbig Ae/Be stars also mark the upper mass limit of stars with habitable zones in which life on a planet could develop. The main sequence lifetime of 2.5\Msun\ stars is $\sim$600Myr \citep{Ekstrom2012} the lower bound for which life is thought to have evolved on Earth \citep{Lopez2005, Danchi2013}. For more massive stars, the main sequence lifetime becomes prohibitively short.


\subsection{\textsf{Definition of Herbig stars}}\label{subsec:def}
Evidence that the objects presented in \citet{1960ApJS....4..337H} were indeed intermediate mass PMS stars was provided by placing these stars on an HR diagram \citep{Strom1972} and were consistent with stellar masses spanning $\sim2-15$\Msun \citep[see also][]{Hillenbrand1992}.  Subsequent studies found emission above what is expected from the stellar photosphere at wavelengths with $\lambda \gtrsim 1\mu m$  \citep[e.g.,][]{1973MNRAS.161...97C,1980MNRAS.191..499C,Hillenbrand1992, 1998A&A...331..211M}, caused by a dusty envelope. Several papers presented criteria to observationally define the class of Herbig Ae/Be stars \citep[e.g.,][]{1994A&AS..104..315T, malfait1998, waters1998,  Vieira2003} leading to a general consensus that Herbig Ae/Be stars have a spectral type B, A, or F,  H~\textsc{i} emission lines, and an infrared excess. Most papers identify the stellar mass range represented by such objects to range from $\sim$1.5 - 10\Msun. The lower mass limit was set by the spectral type of the coolest star that was thought to reach the zero age main sequence (ZAMS) as an A9 star.
The upper mass limit is set by the maximum mass a star is expected to experience a pre-main-sequence phase while it is not enshrouded in its protostellar envelope, however, this upper mass limit is not sharply defined as many stars typically included in catalogs of Herbig Ae/Be stars have higher masses \citep[e.g.,][]{Vioque2018}.  

In this review, we will refer to the Herbig Ae/Be stars as \emph{Herbig stars}, and predominately limit ourselves to stars with the following properties:  young A or B type stars that are evolving towards the main sequence, with H$\alpha$ emission, often associated with a nebulosity, and an infrared (IR) excess due to warm ($\sim$1000~K) and/or cold ($\sim$100~K) circumstellar dust.  Herbig stars in principle do not include objects cooler than about 7000~K (i.e., stars with spectral types later than A9), but some F-type stars such as HD~142527, CQ~Tau and HD~135344B have also been discussed in the literature in the context of Herbig star samples, because of their high luminosity and associated mass. We will include these in this review. Most if not all of these stars are not yet core-hydrogen burning PMS stars; however, for the most luminous objects in the sample this is not easy to establish.  We do not consider the A and B stars with debris disks, that have lost their primordial gas and have secondary dust produced by collisions between larger bodies \citep{hughes2018}. Nor do we discuss young PMS A and B stars that show no evidence for circumstellar material any more (and would be the equivalent to the naked T~Tauri stars; e.g., \citealt{Walter1986}). 
Catalogs of (candidate) Herbig stars were published by \citet{1984A&AS...55..109F, 1988cels.book.....H, 1994A&AS..104..315T, Vieira2003}, and more recently by \citet{Vioque2018}. HArchiBe, an on-line archive of known Herbig Ae/Be stars and their properties is described in \citet{Guzman2021} \footnote{http://svo2.cab.inta-csic.es/projects/harchibe/main/}. From this catalog, we identify 31 Herbig stars (including spectral type F) within 225 pc and 87 Herbig stars within 450 pc.

\subsection{\textsf{Connection to Intermediate Mass T Tauri stars}}

\begin{figure*}
\begin{center}
\includegraphics[width=0.85\textwidth]{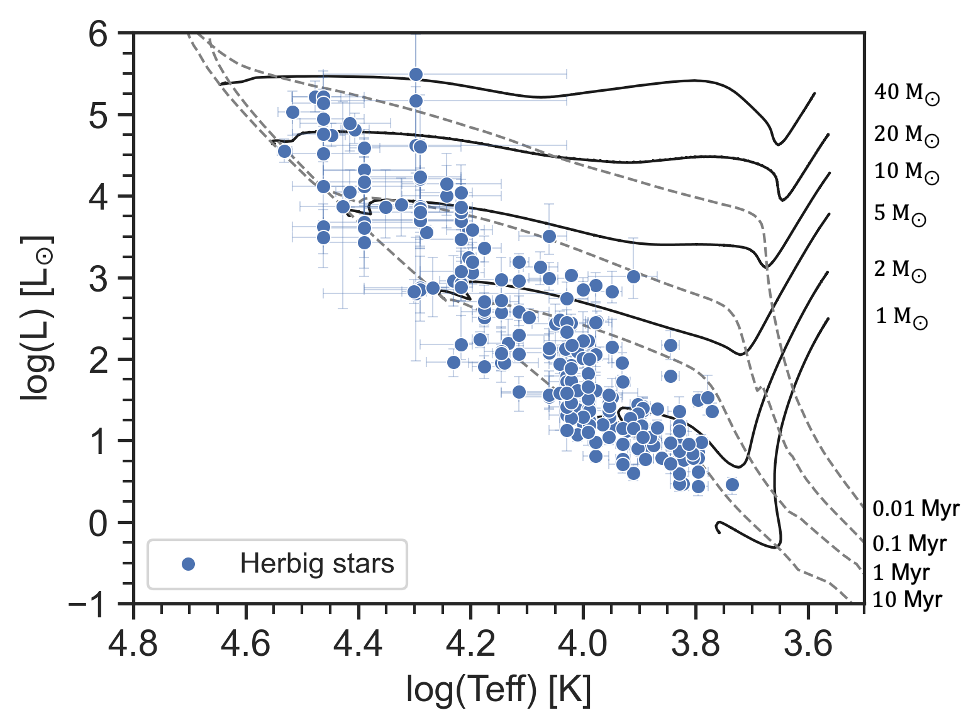}
\caption{ HR diagram containing 218  Herbig stars with high quality DR2 Gaia parallaxes (adapted from \protect\citet{Vioque2018}, their Figure 2). The solid lines represent PMS tracks \protect\citep{Bressan2012}, with the final masses indicated on the Main Sequence. The dashed lines represent isochrones taken from \protect\citet{Marigo2017}. The Zero Age Main Sequence is the region marked by the end of the PMS tracks.  Figure kindly provided by M. Vioque.}
\label{hrfirst}
\end{center}
\end{figure*}

Late F, G, and K type intermediate mass PMS stars are classified as T Tauri stars. They are the evolutionary predecessors of the Herbig stars. Since young intermediate mass stars evolve from the birth line towards the main sequence, they will have properties that classify them as T Tauri stars when still cool and shift to earlier spectral types as their temperature increases as they evolve along the radiative track on their way to the ZAMS (Figure~\ref{hrfirst}). 
In fact, T Tau itself has been shown to be of intermediate mass \citep{Duchene2006}.  \cite{1988cels.book.....H} defined a special group called ``su" stars with properties similar to that of the intermediate mass T Tauri star SU Aur. \cite{herbst1994} introduced the class of Early Type T Tau stars, that contains both T Tauri and Herbig stars. \cite{Calvet2004} used the term intermediate mass T Tauri (IMTT) stars, which has been adopted in subsequent literature to denote the low temperature progenitors of the Herbig stars. 

In a recent study, \cite{2021Valegard} compiled a list of about 50 IMTT stars from the T Tau literature (i.e. stars classified as T Tau stars) and, based on Gaia DR2 distances \citep{Gaia2016, Gaia2018} to these stars, derived basic properties such as temperature, luminosity, and IR excess. Stars with mass above 1.5 M$_{\odot}$ were selected, based on their position in the HR diagram and using PMS evolutionary tracks.  This study shows that the circumstellar environment of IMTT stars is qualitatively similar to that of the Herbig stars.   A full view on how intermediate mass stars and their environment evolve from the birth line towards the main sequence ideally includes  IMTT stars. We will return to this point in Section \ref{sec:future}. In this review, we will not consider the IMTT stars any further. 

\subsection{\textsf{Link between low mass and high mass star formation}}

Herbig stars can be found in low- and high mass star forming regions (e.g., Chamaeleon and Orion respectively). However, early type Herbig Be stars tend to be surrounded by a dense clustering of stars, while this is not the case for Herbig Ae stars \citep{Hillenbrand1995,testi1999}. There is a smooth transition in clustering between these two ranges. This points to a qualitative difference in the mode of star formation that leads to massive (M$_{\star} \gtrsim10-20$ M$_{\odot}$) and lower mass stars. Models for star formation also distinguish between two modes: isolated, low mass star formation in clouds of modest mass, and clustered, high mass star formation in giant molecular clouds \citep{motte2018}. An important difference between these modes is that in high mass star forming regions massive stars, that evolve quickly, provide a strong feedback on their environment. Their stellar winds and UV radiation fields are important already during the main accretion phase.  Stars with mass above 8-10 M$_{\odot}$ do not/are not expected to go through a visible PMS phase; although, this has not yet been revisited incorporating the latest insights relating to the modelling of the spectral energy distribution (SED) of Herbig stars. 
Radiation pressure on dust grains for instance inhibits the formation of very massive stars via the core accretion model \citep{Wolfire1987}, although non-spherical accretion can circumvent this difficulty \citep{2002ApJ...569..846Y}. Feedback is believed to become relevant for masses above 10-20 M$_{\odot}$ \citep[e.g.,][]{Hosokawa2009}.

Observations and models suggest that the core accretion model for star formation, developed for low mass stars, is also applicable to more massive stars, up to masses of 10-20 M$_{\odot}$. 
Young massive YSOs in M17 have been found to have disks with properties consistent with a remnant accretion disk (e.g. \citealt{hoffmeister2008ApJ,ochsendorf2011,Ilee2013}).   Beyond a mass of about 20 M$_{\odot}$, it is unclear which mechanism dominates: core accretion and competitive accretion \citep{Bonnell2001}
have been suggested \citep{krumholzbonnell}.

Intermediate mass stars are at the crossroads between low- and high mass star formation. As stellar mass and luminosity increase, the shortening stellar evolutionary timescales, the increased strength of the stellar UV radiation field, and the denser cluster environment that occurs for more massive stars will strongly influence the physical and chemical properties of Herbig star disks and the way they dissipate. Therefore these disks are of interest to understand how stellar mass affects the early evolution of stars and planetary systems. This is one of the main motivations to study young intermediate mass stars.

In the quarter century since the last major review of Herbig stars \citep{waters1998}, there has been enormous progress in our understanding of these systems. Here we review this progress. First we discuss the stellar properties of Herbig stars - namely their multiplicity fraction, X-ray properties, and variability (Sec. \ref{sec:properties}). We then move to the star-disk interface and discuss the stellar accretion properties of these stars (Sec. 3). Moving further out, we discuss the dust (Sec. 4) and gas (Sec. 5) properties of the disks orbiting Herbig stars. We then discuss the advances in our understanding of planet formation in these disks (Sec. 6) and conclude with summary of key future lines of investigations that arise from the topics we cover in the review (Sec. 7) and a modest proposal for a new definition of Herbig stars (Sec. 8).  

\section{\textsf{STELLAR PROPERTIES}}
\label{sec:properties}
Over the past quarter century, millimeter (mm) interferometry of gas emission from disks around young stars have enabled the measurement of dynamical masses of young stars \citep[][]{Simon2000, Schaefer2009, Guilloteau2014, Simon2017,Braun2021, Law2022}. The spatial resolution and sensitivity provided by the Atacama Large Millimeter Array (ALMA) has opened the door to measuring ever larger samples of dynamical masses that provide a valuable measure for testing stellar evolution models and thus the ages and masses inferred from them.  \citet{Braun2021} compiles the largest such sample to date and finds that stars with masses of stars ranging from 1.3-3.0\Msun\ inferred from stellar evolution models predict masses ranging from -5\% to -14\% of the mass inferred dynamically.

\begin{figure}
\begin{center}
\includegraphics[width=0.75\textwidth]{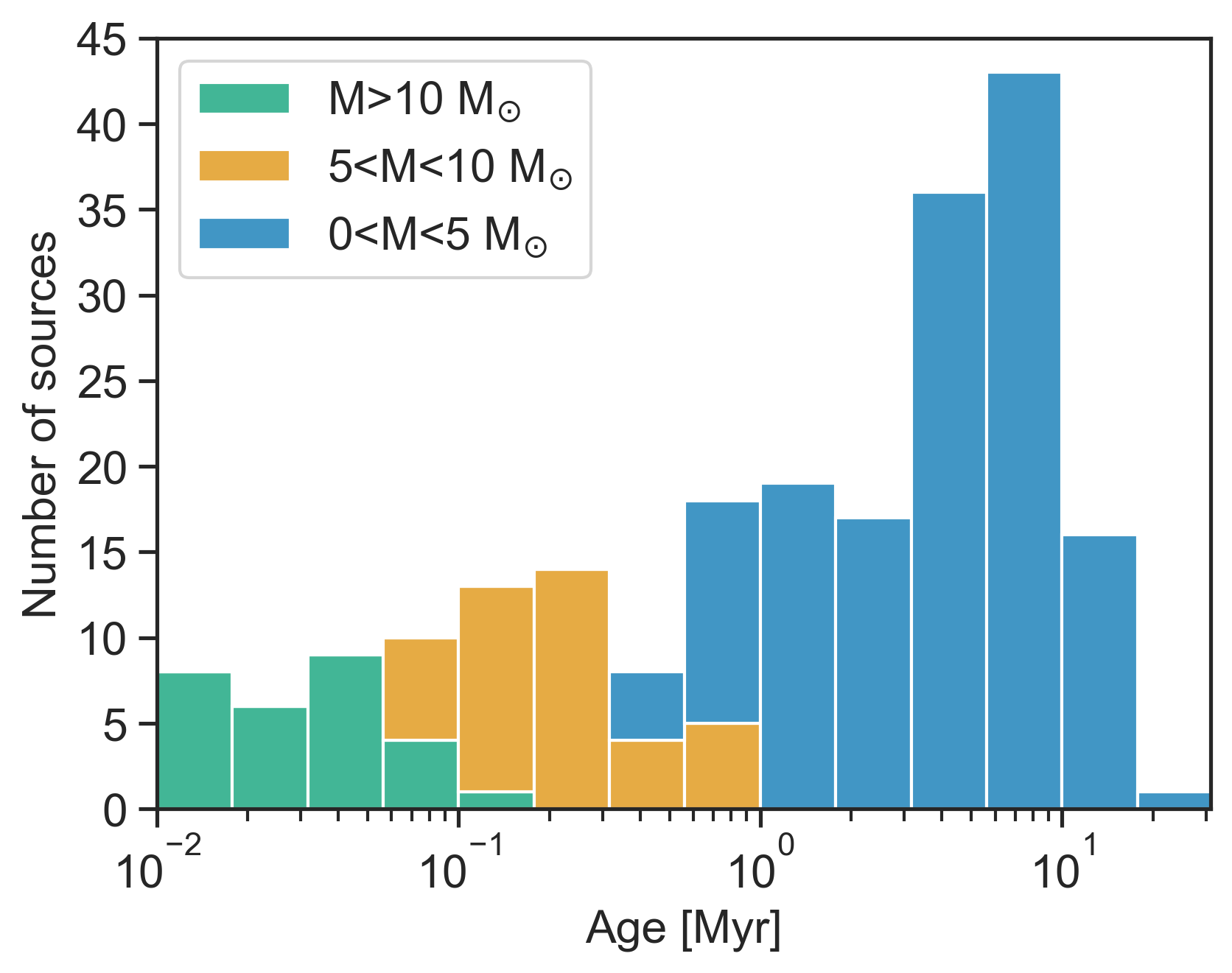}
\caption{The distribution of ages of the 218  Herbig stars as determined from the position in Figure~\ref{hrfirst} (\protect\citealt{Vioque2018}). The highest mass stars have the smallest ages. Most Herbig Ae stars, and thus their disks, have ages in excess of several Myr. Figure kindly provided by M. Vioque.}
\label{histage}
\end{center}
\end{figure}

It may be useful to see the Herbig stars in the context of their position in the HR diagram. \citet{Vioque2018} placed the 218 then known and proposed Herbig stars with high quality Gaia DR2 parallaxes \citep{Gaia2018} in the HR diagram, a version of which is reproduced in Figure~\ref{hrfirst}. To guide the eye, several evolutionary tracks and isochrones are overplotted. As expected, most Herbig objects fall at or above the ZAMS indicating that intermediate stars cease accreting prior to or just after reaching the ZAMS. The more massive Herbig Be stars are sparser than the Herbig Ae stars which can be explained by the Initial Mass Function, as well as the shorter evolutionary time scales that are associated with their evolution towards the Main Sequence.  Higher mass Herbig stars are invariably younger than their lower mass counterparts (Figure~\ref{histage}). The ages of the lower mass Herbig stars typically range from several millions of years up to 10 Myr. The latter high values have implications for the disks, their evolution and survival  which will be discussed after the overview of the stellar properties. We will proceed with the binary properties of the objects in the following sub-section.

\subsection{\textsf{Binaries among Herbig stars}}
\label{sec:binaries}

The multiplicity of Herbig stars bears on models of their formation, the origin of X-ray emission observed toward Herbig stars (cf. Sec. \ref{sec:xray}), and can affect the appearance of the disk (cf. Sec \ref{duststruct}). Herbig stars are predominately found in binary systems, and many of these are found at arcsecond scales.  However, not all separations have been equally well sampled (\citealt{Duchene2015}). The last dedicated binary studies using large samples are the spectro-astrometric surveys of \citet{Baines2006} and \citet{Wheelwright2010} who observed 31 and 45 objects respectively,
totalling 62 unique targets. The separations probed were in the $0.1 -2\arcsec$ range while a flux difference of 6 magnitudes could be
reached. Their data were tested against the (slightly less deep)
survey AO data of \citet{Leinert1997} and imaging of
\citet{Pirzkal1997}, and found to be in agreement where there was
overlap.

Both studies yielded a multiplicity of order 70\% for this parameter
range, with a hint that the Herbig Be stars are more likely to
be found in binaries. \citet{Wheelwright2010} could disentangle the
spectra of some of the otherwise unresolved binaries and determined
that the mass ratios are close to one. This value is inconsistent with a random sampling of the IMF, which would be expected if stellar capture was the main binary formation mechanism. The same team reported in \citet{Wheelwright2011} that the alignment between the binary objects and the disks surrounding the primaries was consistent with disk fragmentation. 
Similar conclusions on fragmentation being the cause of Herbig binaries were put forward by \citet{Arun2021} who reported the discovery of a wide ($6.6\arcsec$) Herbig Ae - M star binary. Given the wide separation, fragmentation at an earlier stage was favored.

The surveys with the \emph{Very Large Telescope Inter\-fero\-met\-er} (VLTI) of \citet[H-band, 51 objects]{Lazareff2017} and \citet[K band, 27 objects]{Gravity2019} probed smaller separations and similar contrasts to the above studies. These studies report few detections of binary companions. However, the surveys were designed to study disks, so the targets were selected against the presence of radial velocity binaries. Milli-arcsecond Herbig Be binaries, corresponding to sub-500 au separations, have been observed using VLTI data however \citep{Kraus2017,Koumpia2019}.  

Finally, the smallest separations can be probed using spectroscopy, where a binary 
system can be revealed through radial velocity variations or directly in a double 
lined binary. Not much work has been done in this field, with the spectroscopic 
survey by \citet{Corporon1999} of 42 Herbig stars still the largest dedicated such 
study.  The latter category includes spatially resolved objects with known 
separations of order $0.5\arcsec$, so the observed close binary fraction based on 
radial velocity variations alone is 17\%. 
 
 \bigskip
 
\noindent \textit{Summary}: The majority of Herbig stars are in binaries ($\gtrsim70\%$). The mass ratios of the binary Herbig stars indicates that the binaries form from disk fragmentation. Given that many of these are unresolved at arcsecond resolution,  this also has important implications for interpreting the X-ray properties of Herbig stars -- the topic of our next section.

\subsection{\textsf{X-ray emission}}
\label{sec:xray}

\begin{table*}[]
    \caption{Typical properties for the X-ray emission of MS OB-type, Herbig and T Tauri stars}
    \centering
    \begin{tabu} to 0.99\textwidth {XXXX}
     \toprule
         Properties & O-early B     & Herbig & T Tauri \\
    \hline     
         Plasma $T$ & $<$ 12 MK &12-60MK    & 5-30 MK\\
         $kT$& $<$ 1 keV           &1-5 keV&0.4-2.6keV\\
         log $L_{\rm{X}}$ (erg/s)&29-33&29-31.5&28-30\\
         log $(L_{\rm{X}}/L_{\rm{bol}})$ &-7&[-6,-4]&up to -3\\
         Origin & radiative wind &companion?  &$\alpha \omega$ dynamo\\
         & &shear dynamo?&\\
         Detection rate &65\%&35\%&100\%\\
    \bottomrule  
     \end{tabu}

    \label{t_xray}
\end{table*}

The more massive young objects, with spectral types O to B1, are often detected in X-rays. Their emission is soft ($kT <$ 1 keV) with fractional luminosities, 
log ($L_{\rm{X}}/L_{\rm{bol}}$) $\sim$ -7, attributed to shocks that originate in line-driven winds. 
At the other end of the mass distribution, solar-mass T Tauri stars have a magnetic dynamo that can persist due to their convective motion, giving rise to hard X-ray 
emission ($0.4\lesssim kT \lesssim 3$~keV) from the magnetically heated corona, and log ($L_{\rm{X}}/L_{\rm{bol}}$) saturating around -3 (see Table~\ref{t_xray}). 
Herbig stars, on the other hand, are thought to be fully radiative so that they cannot support a dynamo by convection, nor are they hot enough 
for a radiation driven wind; therefore, the detection of X-ray emission from Herbig stars was unexpected \citep{zinnecker1994,damiani1994}. If the stars are indeed fully radiative, this would suggest that Herbigs should not generate strong well-ordered magnetic fields as observed in their cooler, convective counterparts. However, some do have strong magnetic fields (cf., Sect.~\ref{sec:Herbig_mag}).

\citet{Hamaguchi2005} detected 30\% of the Herbig stars observed with the \textit{Advanced Satellite for Cosmology and Astrophysics} (ASCA), and 
determined an X-ray luminosity 
higher than for T Tauri stars (Table~\ref{t_xray}).
The fractional X-ray luminosity is slightly lower than that of TTS, but higher than for MS B-type stars: 
log ($L_{\rm{X}}/L_{\rm{bol}}$) = [-6,-4].  They found no evidence for a correlation between $v\sin i$ 
(an indication of rotational velocity) and X-ray luminosity, unlike what is observed in TTS where a strong 
correlation was found, supporting the $\alpha \omega$ dynamo scenario \citep{pallavicini1981}. 

The plasma temperatures of Herbig stars are between 1 and 5 keV which is too high to be produced in wind-driven 
shocks \citep{Hamaguchi2005}. Also flares were observed, which also cannot be explained by stellar winds. Since 
Herbig stars evolve from fully convective IMTT stars, some of these may possess a fossil magnetic field (see section \ref{sec:Herbig_mag}). 
Activity parameters such as 
H$\alpha$ and radio emission do not correlate with $L_{\rm{X}}$, but the amplitude of optical variability, 
$\Delta$V, does \citep{Stelzer2006}. Also, magnetospheric accretion can be ruled out as the origin of the X-ray emission because 
the free-fall velocities are too low. 
\citet{Hamaguchi2005} propose that the X-ray 
emission stems from magnetic activity, where the fossil magnetic fields of the stars reconnect with the disk (a 
star-disk magnetosphere). 

With the better spatial resolution of Chandra ($\sim$1$^{\prime\prime}$), \citet{Stelzer2006} 
observed 17 Herbig stars to 
study the companion hypothesis. They detected X-ray emission from 76\% of their sample. After correcting for the presence of known lower-mass companions, they derived an occurrence rate of 35\% which was consistent with previous work by \citet{Hamaguchi2005}. Interestingly, the
detection rate of X-rays from Herbig stars is comparable to the binary rate of Herbig stars (see section \ref{sec:binaries}). 
Further study of the binarity of X-ray emitting Herbig stars will clarify the extent to which unresolved companions can account for the X-ray emission.

In a follow-up paper, \citet{Stelzer2009} 
distinguishes between late B to A-type Herbig stars and early B-type PMS stars. They propose that the early B-type PMS
stars behave like the B-type MS stars, while the X-ray emission in the later-type Herbig stars is connected to 
magnetic fields, which have only been firmly detected in $\sim$ 20 Herbig stars (e.g. \citealt{jarvinen2019}, \ref{sec:Herbig_mag}). 

\bigskip

\noindent \textit{Summary}: The origin of the X-ray emission detected towards Herbig stars remains an unsolved mystery. An origin in a stellar wind can be excluded due to the high plasma temperatures observed. It is possible that the X-rays could come from an unseen companion or by remnant magnetic fields. A dedicated study including precise magnetic field measurements, high spatial resolution X-ray observations, and sensitive searches for companions for a large sample of Herbig stars is needed to test these hypotheses.

\subsection{\textsf{Variability}}

Herbig stars have long been known to show both photometric and spectroscopic variations. The variability can be traced back to changes in the circumstellar disks, the accretion regions, and the structure of the stars themselves. We discuss the photometric and spectroscopic behaviours separately below. 

\subsubsection{\textsf{Photometric variability}}

Some Herbig stars show non-periodic brightness variations as large as $\rm \Delta V \sim 3 \ mag$ \citep{herbst1994}. Important clues about the nature of the variations and the (inner) disk regions came from studying the brightness variations of the star UX Ori: when the star fades in brightness, its color becomes redder. However, when fading even more, this color change turns around, and the star becomes bluer, the so-called `blueing effect' \citep{bibo1990}. In tandem with the fading of the star, its radiation was found to become more polarized, hinting at starlight being scattered off circumstellar dust. These effects, referred to as the \emph{UXor phenomenon} in later literature, were further studied with simultaneous photometric, polarimetric and spectroscopic observations \citep{grinin1992,grinin1994},  supporting the hypothesis that the blueing effect is caused by an opaque dust cloud in the disk, intersecting in the line of sight towards the star. In Sect.~\ref{s_dust_innerdisc} we focus on the properties of the inner disks that were revealed by the UXor phenomenon.

In the near infrared (NIR), the variability of Herbig stars is far more modest than in the optical. \citet{Eiroa2002} observed a sample of T Tauri and Herbig Ae stars using quasi-simultaneous optical and near-infrared photometric observations, taken typically less than 1 hour and at most 2 hours apart. The peak-to-peak variations at optical wavelengths could be as high as 0.5 magnitudes during their period of observations, while the NIR photometry varied mostly at the 0.1 magnitude level. Out of the 12 Herbig Ae objects observed, only 3 were reported as non-variable, while 8 of the remaining nine showed a correlation between the optical and NIR photometric variations.  Given that the NIR emission  is most likely re-radiated by the hot circumstellar dust that absorbed the optical light, the location of the near-infrared variability can be traced to the inner circumstellar disks (see also Sect.~\ref{s_dust_innerdisc}). The remaining one object, UX Ori, did not show a correlation between the photometric variations, which was interpreted as possibly being due to rotational modulation of star spots on the stellar photosphere.  \cite{Vioque2018} later determined that roughly 25\% of Herbig stars can be classified as having such UXor variations, while Herbig Be stars are markedly less variable than the Herbig Ae stars. 

Finally, variability at milli-magnitude amplitudes has been positively detected towards a small sample of Herbig Ae stars (\citet{zwintz2014}; see also the compilation by \citealt{Steindl2021}). Such observations are challenging as any signatures at the milli-magnitude level not only require space-based photometry \citep{Zwintz2009,Casey2013} but need a careful disentangling of such data with the non-periodic variations described above. The Herbig Ae stars appear to show $\delta$ Scuti type variability \citep{Marconi2017}. This variability arises among stars whose temperature and mass lead to an ionization structure that leads to radial oscillations resulting in variability with a period on the order of a few hours. Such stars fall in a well defined region of the HR diagram known as the instability strip. Herbig stars in this region are no exception. The milli-magnitude photometric data have the prospect of probing into these pre-Main Sequence stars using asteroseismology. An interesting example is the case of HD~139614, whose age was estimated from asteroseismology to be about 11 Myrs \citep{Murphy2021}. This old age confirms the results of other studies, and shows that gas-rich disks can survive for a substantial period of time. We will return to this in Sect.~\ref{sec:gasmassevolution}.

\subsubsection{\textsf{Spectroscopic variability}}
\label{s_variability}

In addition to photometric variability, the emission lines observed towards Herbig stars have long been known to vary (for a review, see \citealt{Catala1994}). 
Spectroscopic variability studies of individual objects have revealed variations in the emission lines and their profiles which can be traced back to rotational modulation of the accretion regions \citep{Garcialopez2016,Scholler2016}. It would appear  that the variations are less extreme than those observed for T Tauri stars whose variations are connected to the magnetically controlled accretion \citep{Kurosawa2011}.

Dedicated studies of larger samples of Herbig objects have been sparse.
\citet{Costigan2014} investigated the H$\alpha$ line variability for a sample of, mostly, Herbig Ae and T Tauri stars.  
The amplitude of the variability increases from minute timescales to level off at the days-weeks timescales, even over periods as long as years. This indicates that the variability occurs in a small volume around the star and \citet{Costigan2014} attribute the variability mainly to changes in the accretion rate.
Overall, it would appear that the variability of the emission lines is connected to the accretion processes occurring in Herbig stars, the topic of the next section. 

Lastly, another type of spectral variability has been observed towards Herbig stars which was first found towards $\beta$~Pic. Multi-epoch spectroscopy revealed that metallic absorption lines would change on timescales of hours to months. In particular, lines could suddenly appear in absorption and the absorption itself would become more red-shifted in matters of hours. This behaviour was interpreted  as evidence of infalling material from cometary bodies \citep{Ferlet1987}.
\citet{Grady1996} presented spectra of 41 Herbig stars and found that 22 of them showed variability consistent with the presence of gas released from infall of evaporative bodies. We will return to this issue in Sec.~\ref{sec:feb}, where this phenomenom is discussed as an indirect tracer for the presence of planets in the disks of Herbig stars.

\bigskip

\noindent \textit{Summary}: Herbig stars display photometric and spectroscopic changes on timescales ranging from hours to months and longer. The larger amplitude photometric changes can be traced back to variable dust obscuration from an orbiting disk, while variability at smaller, milli-magnitude, amplitudes can be analysed with asteroseismology. Spectroscopically,  variable line emission is mostly associated with the accretion process, while variable absorption lines can be due to changes in the accretion process itself, but can also be caused by infalling rocky material such as exo-comets.


\begin{figure*}
\begin{center}
\includegraphics[width=0.99\textwidth,angle=0]{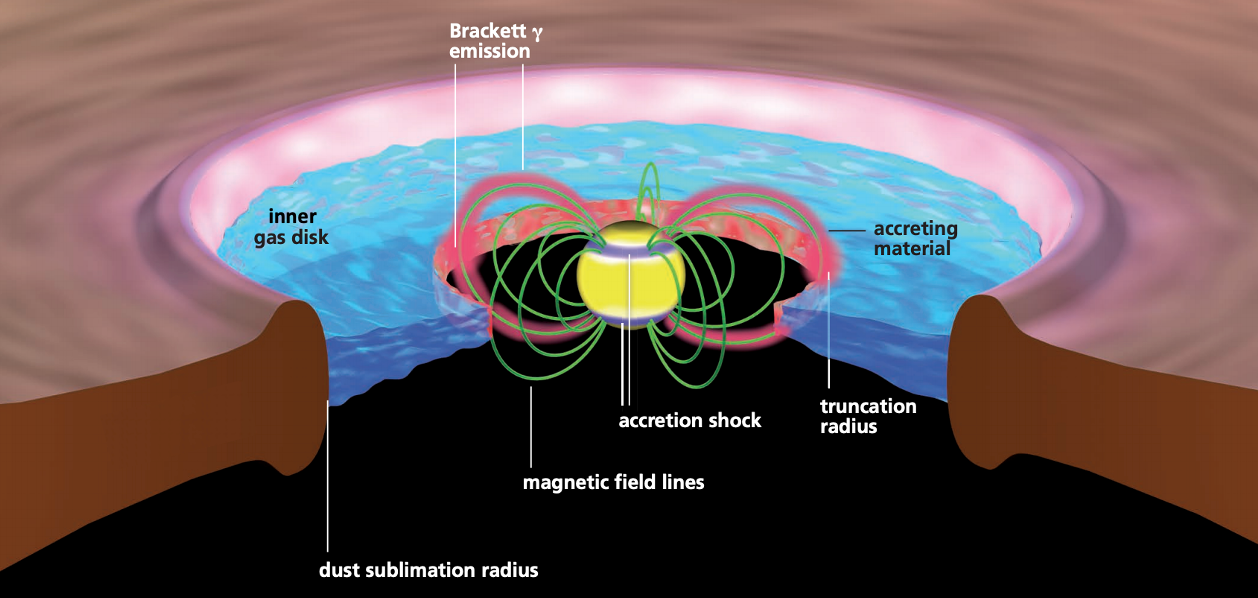}
\caption{Cartoon of the immediate circumstellar environment of a HAe star. Initially designed to illustrate the magnetospheric accretion paradigm for CTTSs, this cartoon also applies to HAe stars. The dust in the disk is sublimated when it reaches a temperature of $\sim$1500~K. Interior
to this region the disk is gaseous and heavily ionized. The disk is truncated by the magnetic field where the pressure from accreting gas is balanced by magnetic pressure. The accreting gas then falls into the star along field lines from which hydrogen recombination emission emerges and creates an accretion shock near the surface of the star. Figure based on \protect\citet{GravityTTau2020}. } 
\label{accretioncartoon}
\end{center}
\end{figure*}

\section{\textsf{ACCRETION ONTO HERBIG STARS}}
\label{sec:acc}

In this section we discuss the accretion and accretion mechanisms of
Herbig stars. Given that the intermediate mass Herbig stars
are, for all intents and purposes, the high mass counterpart of the
solar mass T Tauri stars, let us begin with a description of the paradigm for understanding accretion onto T
Tauri stars. Dedicated reviews of these objects can be found in
\citet{Bertout1989} and \citet{Bouvier2007}, recent reviews on the accretion process
in pre-Main Sequence stars in general can be found in
\citet{beltran2016}, \citet{hartmann2016} respectively, while \citet{mendigutia2020}
focuses on accretion onto Herbig stars.

T Tauri stars are well known to sustain a magnetic field due to their
internal dynamo and exhibit both evidence for accretion onto the stars
and bi-polar outflows powered by the accretion.  The
accretion itself is magnetically mediated. The stellar magnetic field 
truncates the disk where the ram pressure from accretion balances the 
opposing pressure from the stellar magnetic field. Material from the 
disk free-falls along
the magnetic field lines through accretion funnels onto the star.  With typical disk-truncation radii of 2-3
stellar radii, the free-fall velocities are very high, of order 100s
km~s$^{-1}$.  The gravitation potential energy of the material released when it
crashes onto the photosphere is therefore substantial.  The resulting shock
produces X-ray radiation that is absorbed by surrounding particles.
Then, these particles heat up and re-radiate at ultraviolet
wavelengths, producing a UV-excess that can be observed.  A cartoon
of the situation is provided in Fig.~\ref{accretioncartoon} (see e.g.
\citealt{Bouvier2007} for more details).

The relatively cool T Tauri stars radiate most of their
photospheric energy at optical and longer wavelengths, so the observed UV
emission can be used as a proxy for the accretion luminosity, which
reflects the potential energy released by the infalling matter. Assuming the material is falling from infinity, the mass accretion rate is given by, 

\begin{equation} \label{Macc}
\dot M_{\rm acc}=\frac{L_{\rm acc}R_*}{GM_*},
\end{equation}

\noindent where $L_{acc}$ is the accretion luminosity, $R_*$ is the stellar radius, $M_*$ is the stellar mass, and $G$ is the gravitational constant. 

It may be clear that a proper characterisation of the stellar
parameters is important. As will be discussed later, the Gaia
satellite, which provides parallaxes, from which distances are
determined, for more than a billion stars has had an important impact
on the field. 

The magnetospheric accretion (MA) paradigm has been applied
extensively over the past years to determine the accretion rates of
Herbig stars. In the following we give not only an overview of
the results, but also critically analyse the assumptions made in this
endeavour.  We will demonstrate that while the MA paradigm seems to
be applicable to most Herbig stars, there is compelling evidence
that it cannot act for some extreme cases, and that there is strong
evidence for a change in accretion mechanism at higher masses.

\subsection{\textsf{Magnetic properties of Herbig stars}}
\label{sec:Herbig_mag}

Let us first discuss an immediate potential critique that can be fielded against the use of the MA paradigm to Herbig stars. Conventional wisdom is that no magnetic field is expected to be generated in Herbig stars because the envelopes of A and B stars are radiative. The B-field is typically generated in a dynamo which usually occurs by convection in cooler stars. Indeed, magnetic fields have only fairly recently been detected towards a limited number of objects \citep[e.g.,][]{Alecian2013}. Based on the compilation of \citet{Hubrig2015}, \citet{mendigutia2020} presents all reported magnetic field detections of Herbig stars and their stellar parameters. The  list contains a total of 18 objects with measured B-fields, which constitutes roughly 10\% of all known and proposed Herbig stars (viz. \citealt{Vioque2018}).  Intriguingly, this fraction is similar to a 10\% detection rate in field A stars \citep{Landstreet1992}. Most of these B-fields have  strengths of order 100 G, while a few Herbig stars  have magnetic field strengths that are significantly higher (e.g., V380 Ori (2kG), and Z CMa (1.2 kG)). 
As already alluded to  in Sec.~\ref{sec:xray}, rather than being generated, the  magnetic field in these magnetic objects could be fossil in nature, either from the original collapsing parental cloud or the remnant field of an earlier IMTT star phase.

\begin{figure*}
\begin{center}
\includegraphics[width=0.85\textwidth]{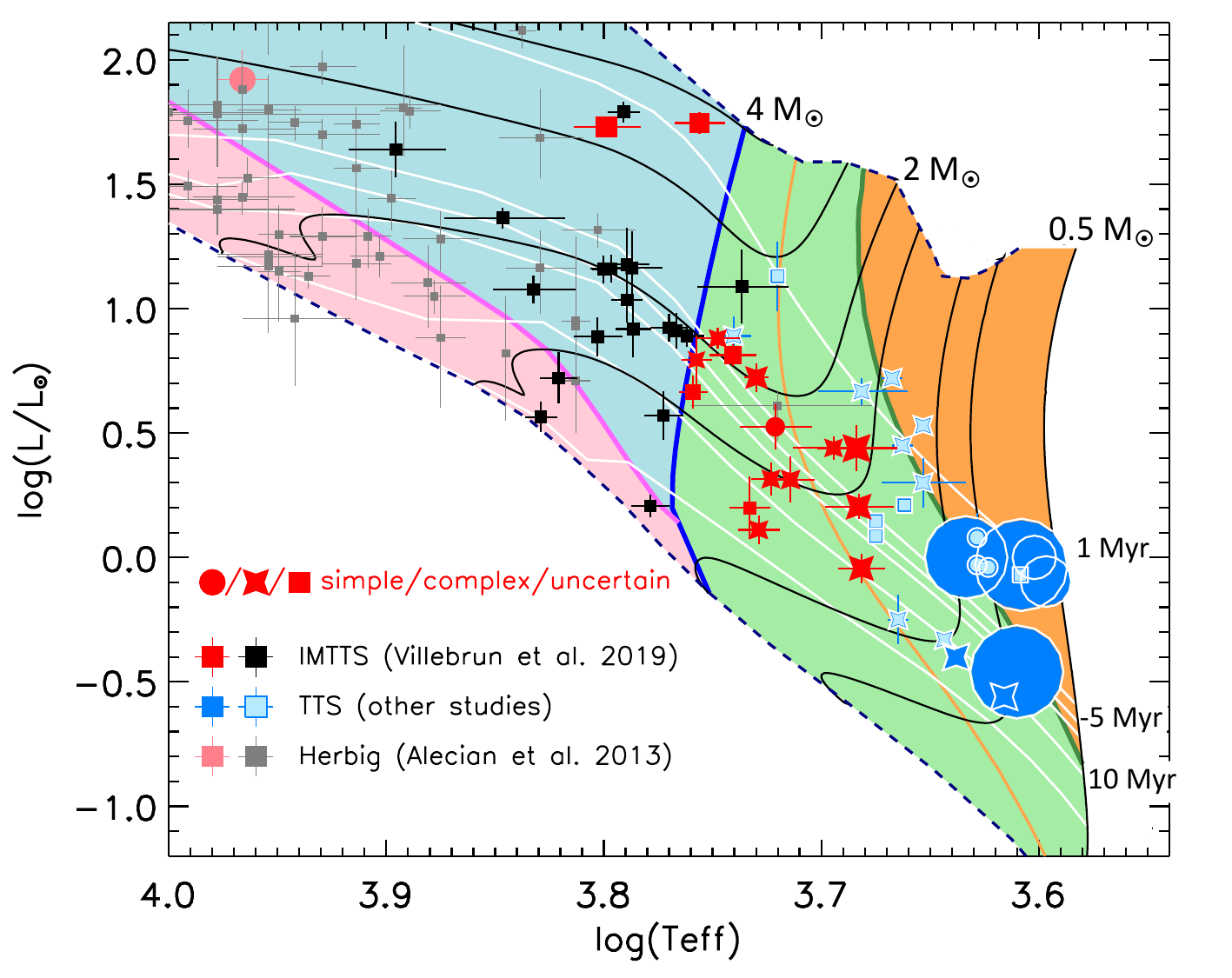}
\caption{The B-fields of T Tauri stars and Herbig Ae stars according to \protect\citet{Villebrun2019}.  
The black and gray symbols are stars with non-detections, while the red and blue symbols represent objects with detections with symbol sizes proportional to the B-field strengths. The orange zone indicates the area where stars are fully convective, the blue zone where they are fully radiative, while the other colors represent a mix; green (radiative core and  convective envelope); pink  (convective core and radiative envelope).  The various evolutionary tracks (black lines) and isochrones (white lines) are from \citet{Behrend2001}. Figure kindly provided by E. Alecian.} 

\label{bfieldhr}
\end{center}
\end{figure*}

Whatever the cause of the magnetic fields, it should be noted that
most B-field determinations concern the global integrated magnetic
field, a non-detection is therefore not necessarily evidence for the
absence of it; a complex geometry would inevitably result in a lower
observed strength. \citet{Villebrun2019} mapped the B-fields of the
IMTT stars, and find that the
fields become weaker, but also more complex and at higher
temperatures (see also Fig.~\ref{bfieldhr}). The possibility that higher temperature pre-Main Sequence stars
harbour strong magnetic fields is therefore a realistic prospect.  As
a consequence, the MA scenario which is well established for T Tauri
stars (e.g. \citealt{Bouvier2007}), may also be applicable to hotter
objects.

In the next section, we will go into more detail on similarities
between T Tauri stars and Herbig objects with emphasis on their
accretion properties. We then consider the determination of the
accretion rates of Herbig stars and provide global inferences
resulting from these.

\bigskip

\noindent \textit{Summary}: About 10\% of Herbig stars have measurable magnetic fields. While most of these are of order 100~G, a few have magnetic field strengths about an order of magnitude larger. These B-field determinations are measurements of the global integrated field. If the magnetic fields of Herbigs tend to have a complex geometry, it is possible that the  surface magnetic field strength of the typical Herbig star is much higher than these measurements suggest.

\subsection{\textsf{Accretion Geometry - similarity with T Tauri stars}}

The earliest investigation into MA onto Herbig stars was
carried out by \citet{Muzerolle2004}.  They applied their MA models to
the Herbig Ae star UX Ori, and found that the computed emission line
profiles were in qualitative agreement with the observed emission line profiles. In
parallel, they also concluded that the magnetic field strength would
be weaker and its geometry should be more complex than the usual
dipole field in T Tauri stars, just as was later shown to be the case by
\citet{Villebrun2019}.  Further supporting evidence that Herbig Ae stars may be accreting in a similar fashion as T Tauri
stars comes from various observational studies. These often stem from
comparative investigations into the accretion-related properties of T
Tauri stars on the one hand and Herbig Ae and Herbig Be stars on the
other hand. An additional result from these studies is that Herbig Be
stars often have very different properties than their Ae counterparts.

Early suggestions that the accretion processes of Herbig Ae stars and T Tauri stars are similar were reported based on results from
linear spectropolarimetry.  Although the accretion funnels are
still too small to be directly imaged, indirect linear
spectropolarimetric methods are able to probe these regions at scales
of order stellar radii \citep[e.g.][]{Vink2015,Ababakr2016,Ababakr2017}. The similarity of the linear spectropolarmetric properties of the H$\alpha$ line for Herbig Ae stars and T Tauri stars was reported by \citet{Vink2002, Vink2005}. The observations were explained with line emission from accretion hot spots
scattering off truncated disks. Indeed, the inferred sizes of order
stellar radii of the inner holes \citep{Vinkmodel2005} are consistent
with those expected for a magnetically truncated disk.  In contrast, the
very different effects observed towards the Herbig Be stars are more
consistent with disks reaching onto the central star - hinting at a
different accretion mechanism in those. \citet{Ababakr2017} presented
the largest sample of Herbig stars observed with linear
spectropolarimetry around H$\alpha$, which allowed them to identify
the transition from truncated disks to disks reaching into the stellar surface around spectral type B7/8.

Other studies related to accretion diagnostics are broadly consistent
with these findings. \citet{Cauley2014} studied the He~{\sc i} 1.083
$\mu$m line profiles of a large sample of Herbig stars. Due to its
high excitation energy, this transition is particularly sensitive to
accretion. These authors found that both T Tauri stars \citep[observed
  by][]{Edwards2006} and Herbig Ae stars had similar line profiles and comparable
fractions of red-shifted absorption lines, tracing infall, all of which could
be explained in the context of MA, while the Herbig Be stars did
not.  Furthermore, they inferred that the Herbig Ae stars could have smaller
magnetospheres than their T Tauri counterparts.

As mentioned earlier, Herbig 
stars display spectroscopic variability. \citet{Costigan2014}  found that both the timescales and amplitude of the
variability were similar for T Tauri and Herbig Ae stars, which led these authors to suggest that the
mode of accretion of Herbig Ae stars is also similar to that of T Tauri
stars.  Although less extensive in time coverage, the H$\alpha$
variability study of \citet{Mendigutia2011a} extends to the more massive
Herbig Be stars. These objects were found to have different
H$\alpha$ variability properties than the Herbig Ae stars. Of note is
that these authors not only considered the H$\alpha$ line strength variations,
but also those of the line width, which, given the high free-fall velocities 
involved may be more effective in tracing the accretion flows. Its variability was found to be
considerably smaller for the Herbig Be stars than for the Herbig Ae
stars, which,  in turn, is smaller than for T Tauri stars
\citep{Fang2013}. 

\citet{Kreplin2018} find from modelling their high
resolution infrared spectrointerferometry, that the inner radii of
the disk-winds are consistent with the Alfv\'{e}n radii computed using
the low B-field strengths observed for a few Herbig Ae stars. \citet{Guimaraes2006} observed 15 Herbig Ae stars (some with spectral type B9Ve) and found four that showed red-shifted absorption in the Balmer lines. The authors demonstrated that this absorption can be qualitatively explained with magnetospheric accretion if the disk were truncated relatively close to the star.

Detailed studies on individual targets have reached similar
conclusions as above. \citet{Grady2010} found the location of the
accretion footprint on a Herbig Ae star to be very like those on T
Tauri stars.  The in-depth spectroscopic variability study of a Herbig
Ae star showed it is undergoing magnetospheric accretion in the same
manner as classical T Tauri stars \citep{Scholler2016}. From their
modelling of spectra, \citet{Moura2020} find evidence that Herbig Ae stars
undergo MA, while they do not find this to be the case for a Herbig Be
star. On the hotter side, \citet{Patel2017} model the inner gaseous
disks of several Herbig Be stars, and find they do not need
magnetospheres to reproduce the emission line spectra.

\bigskip

\noindent \textit{Summary:} It has been empirically shown that Herbig
Ae stars exhibit many similar accretion-related characteristics as T
Tauri stars. This provides indirect evidence for magnetically
controlled accretion operating for these objects as well. In turn, it
validates the use of MA shock modelling to derive accretion rates for at least the later-type Herbig stars.

\begin{figure}
\begin{center}
\includegraphics[trim=1in 2in 0in 2in,clip, width=0.65\textwidth]{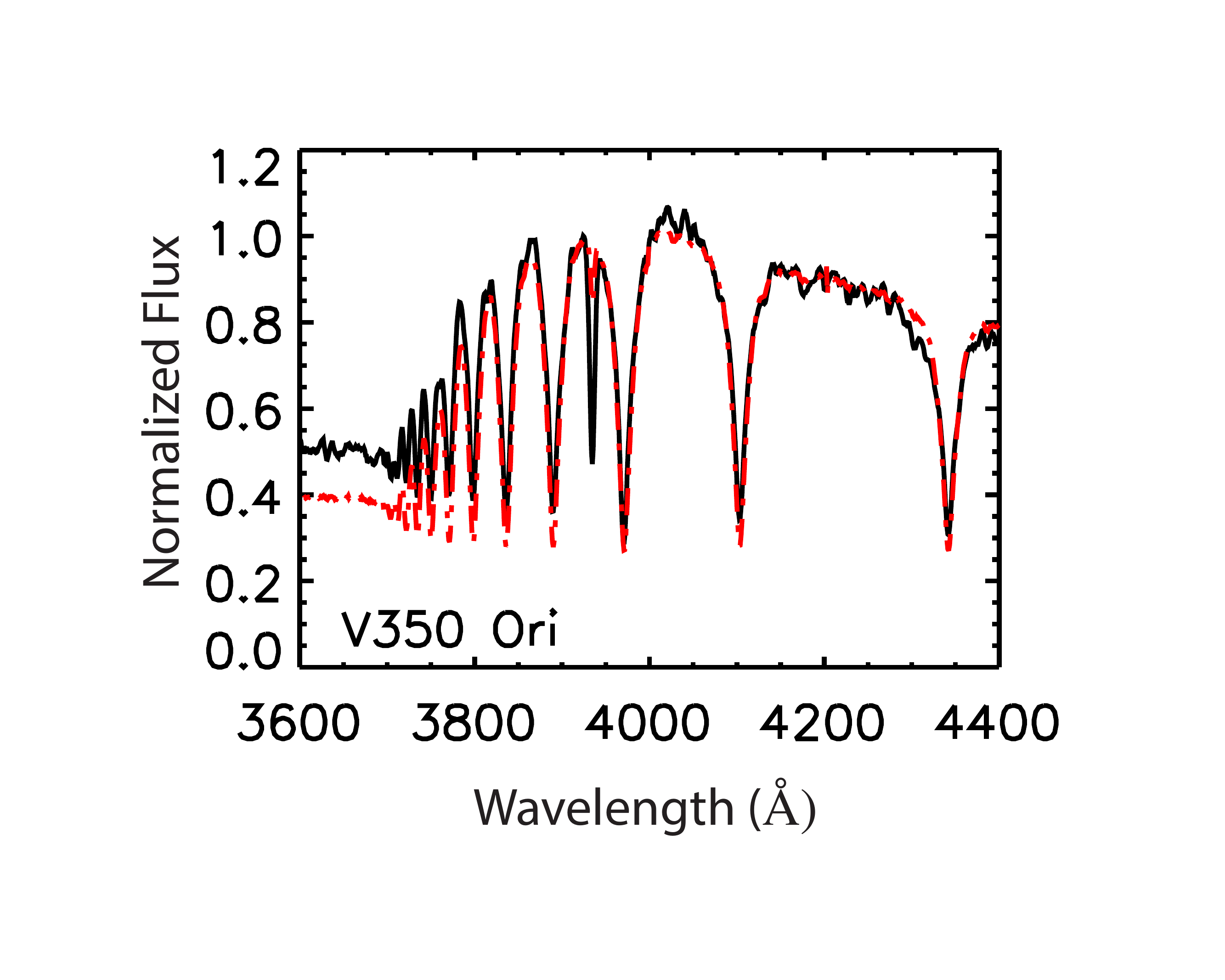}
\caption{Figure from \protect\citet{Donehew2011}. The spectrum of V350~Ori (black line) is compared to a template of a non-accreting A star (red line). While there is a good match between V350 Ori and the template long-ward of 3800\AA\, V350~Ori is brighter than the template at shorter wavelengths. This excess Balmer continuum radiation is due to the accretion shock over and above that of a normal star with the same spectral type.}
\label{donehewfig}
\end{center}
\end{figure}

\subsection{\textsf{Measurement of mass accretion rates}}

Let us now turn to the determination of the accretion rates of 
Herbig stars using the MA paradigm. As a word of caution, we stress
that it is used for all Herbig stars, regardless of their spectral
type. However, as will be discussed later, this approach serves as an additional
means to identify where a change in accretion properties may be
present.

\begin{figure*}
\begin{center}
\includegraphics[width=0.85\textwidth]{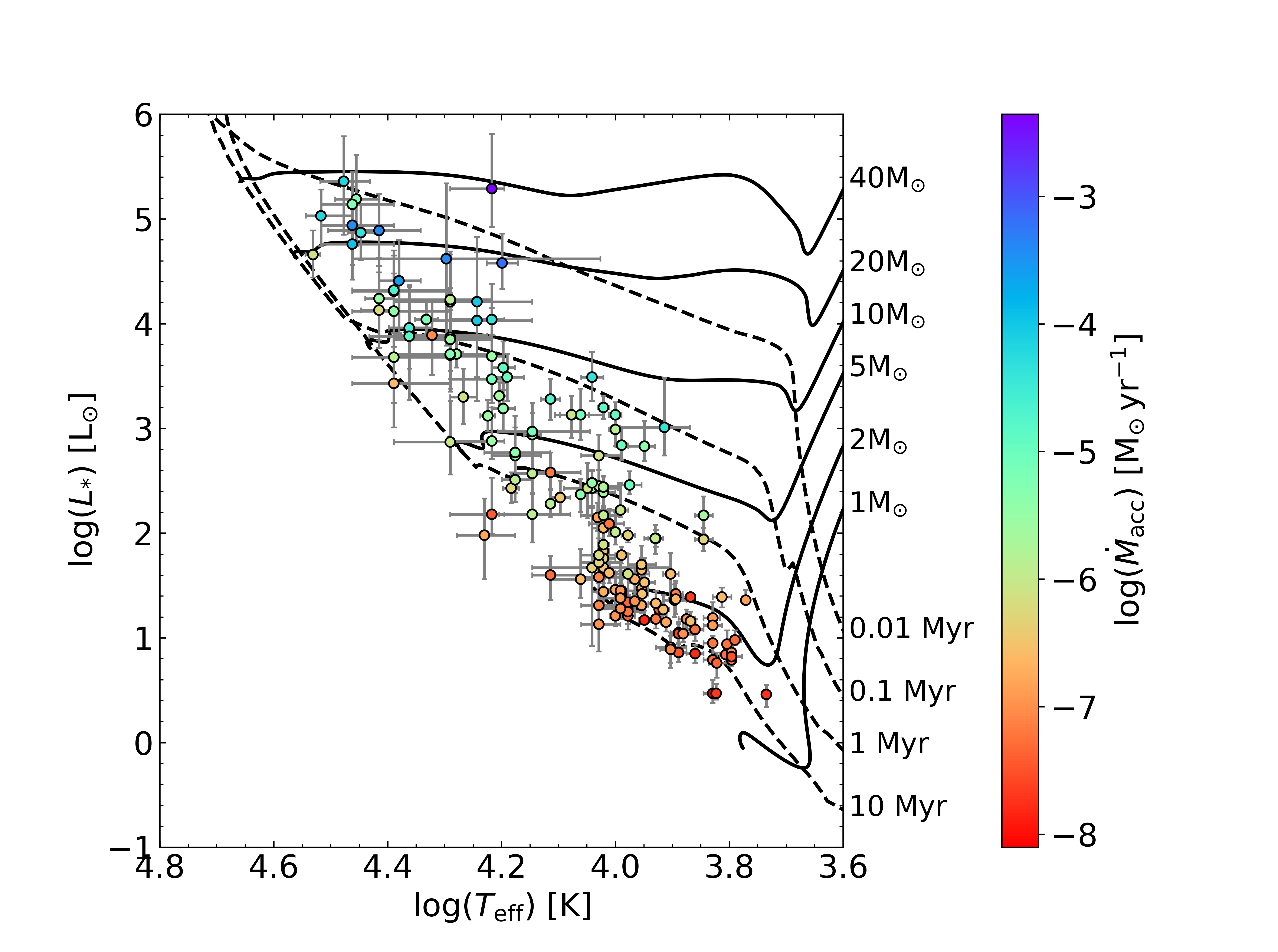}
\caption{HR diagram containing the 163 Herbig stars with good Gaia DR2 parallaxes and their accretion rates. The solid lines represent PMS tracks \protect\citep{Bressan2012,Tang2014}, with the final masses indicared on the Main Sequence. The dashed lines are isochrones of 0.01, 0.1, 1, and 10\,Myr \protect\citep{Marigo2017}.  The color map denotes the mass accretion rate. While it is evident that the younger stars are accreting at higher rates, the younger stars are also much more massive. Characterization of the evolution of the stellar accretion rate requires disentangling stellar mass and age. Figure based on \protect\citet{chumpon2020}}
\label{hrmacc}
\end{center}
\end{figure*}

\subsubsection{\textsf{Direct determination of the accretion luminosity}}

As can be seen from Eq.\ref{Macc}, once we know the stellar mass and
radius, the mass accretion rate can be derived from the accretion
luminosity.  For the cool T Tauri stars, this emission is most readily
identified by UV-excess emission over and above the stars'
photospheric radiation. In addition, it can also be seen at
optical wavelengths, where the excess emission due to accretion fills in the
underlying absorption lines. In the absence of UV observations, a value
for the excess emission can be obtained by determining the
extent of this ``veiling''.

In contrast to the T Tauri stars, the determination of the
contribution of the accretion shock to the total blue/UV emission is less
straightforward for Herbig objects. By virtue of their higher
temperatures, these stars are more luminous at UV wavelengths. Thus the contrast between the accretion shock and the photosphere at UV wavelengths is much smaller for Herbigs than for CTTSs. Because of this smaller contrast, disentangling the contributions from the stellar and accretion properties 
to the UV emission is challenging. Veiling measurements also prove problematic. The
accretion luminosity can exceed the stellar luminosity of T Tauri
stars, allowing a comparatively straightforward determination of the
veiling. However Herbig stars are brighter and the photosphere dominates the
light output at optical wavelengths hampering the identification and
measurement of the veiling. 
Despite these challenges and the
rather labor-intensive methodologies of determining the UV-excess in
these objects, the number of Herbig stars with directly
determined accretion luminosities is surprisingly large; \citet[33
  objects]{Donehew2011} and \citet[91 stars]{Fairlamb2015} obtained
blue spectra for Herbig stars and determined the accretion
luminosities for most targets by carefully comparing their Balmer
continuum (excess) emission with spectra of standard stars (see Fig.~\ref{donehewfig}).  \citet[38
  objects]{Mendigutia2011b} used broadband {\it UBV} photometry to
determine the Balmer excess emission and accretion luminosity. One advantage of using the Balmer 
continuum emission to measure the accretion luminosity is that this 
property is minimally affected by reddening \citep{Garrison1978ApJ, Muzerolle2004}.

The lowest accretion luminosities that could be measured are
of order 1 $L_{\odot}$ and the corresponding accretion rates are of order
10$^{-8} \, M_{\odot} \rm yr^{-1}$. The lower limits are due to the
sensitivity of the observations and the ability to determine the
excess emission. This sensitivity descreases for the higher
temperature objects. It is interesting to see that more than 70\% of
all the Herbig stars investigated in this manner exhibit Balmer
continuum excess emission. As it is non-trivial to find alternative
explanations for this excess, this observation constitutes
independent supporting evidence that these objects are accreting matter.

\subsubsection{\textsf{Indirect determinations of accretion luminosity}}

It turns out that it is not strictly necessary to obtain dedicated UV
or Balmer continuum observations to arrive at estimates of the
accretion rates. A correlation between the brightness of emission
lines and the accretion luminosity exists for IMTT stars
\citep{Calvet2004}. Indeed, by using (ideally) simultaneous UV and line
observations, this correlation can be used to calibrate the line
luminosity to arrive at the accretion rate
(e.g. \citealt{Ingleby2013}). \citet{Garcia2006} extrapolated this finding 
to Herbig stars and determined their mass accretion rates using the Br$\gamma$ 
line luminosity. Their approach was validated when \citet{Donehew2011} and 
\citet{Mendigutia2011b} confirmed that this correlation between accretion 
luminosity (as derived from the UV and Balmer excess) and the luminosity 
of emission lines also extends to Herbig stars.

\citet{Fairlamb2017} demonstrated that the strength of most, if not
all, emission lines in the spectra of Herbig stars correlate
with the Balmer-excess based accretion luminosity. They
provided best-fit parameters for many lines, which cover the
wavelength range from the UV to the NIR, allowing accretion
luminosities to be worked out when observations of only limited parts
of the spectrum are available. 

The relationship between accretion luminosity and line luminosity goes as (cf. e.g. \citealt{Mendigutia2011b}):

\begin{equation} \label{Lacc}
\log\left( \frac{L_{\rm 
acc}}{\rm L_{\odot}} \right)=A+B \times \log\left( \frac{L_{\rm line}}{\rm L_{\odot}} \right),
\end{equation}

\noindent where $A$ and $B$ correspond to the intercept and the gradient of the
(straight-line) relation between $\log(L_{\rm acc}/\rm L_{\odot})$ and
$\log(L_{\rm line}/\rm L_{\odot})$ respectively.  As example,
\citet{Fairlamb2017} determined the constants to $A=2.09\pm0.06$ and
$B=1.00\pm0.05$ in the case of H$\alpha$.

\bigskip

\noindent \textit{Summary}: Because Herbig stars are much more luminous than their cooler counterparts, measurement of excess UV emission is challenging. For stellar accretion rates exceeding $10^{-8}$\Msun yr$^{-1}$, the accretion luminosity can be inferred from measurement of the excess Balmer continuum. There luminosity of emission lines has been shown to scale with this excess enabling a simpler measurement of the accretion luminosity.

\subsection{\textsf{Mass accretion rates of Herbig stars}
\label{sec:AccretionRates}}

As outlined above, in order to compute the mass accretion rate of a
Herbig star, it is important to have well-determined stellar
parameters (mass, luminosity, or radius and temperature, as well as
extinction). In addition, we need a value of the accretion luminosity
such as derived from the UV excess, which is the only direct observational measure
of accretion, while an emission line luminosity provides an indirect
value via Eq.~\ref{Lacc}. As noted in Section \ref{sec:properties}, significant progress was made in the determination of Herbig stellar parameters through the {\em Gaia} astrometric study by \cite{Vioque2018}. 

\citet{Arun2019} used these data and H$\alpha$ Equivalent Widths reported in the literature 
to determine accretion rates for many of the objects. \citet{chumpon2020} used a slightly
larger sample to expand the spectroscopically determined temperatures of \citet{Fairlamb2015} 
with an additional 30 objects. They rederived the line-of-sight extinction to the objects, 
as well as measured the H$\alpha$ emission line strength from the same spectra, yielding a 
large sample to be analysed in a homogeneous way. These were supplemented by objects from 
\citet{Vioque2018}, resulting in a sample of 163 Herbig Ae/Be stars with well-determined
stellar parameters and as homogeneously derived accretion rates as possible. 

\begin{figure*}
\begin{center}
\includegraphics[width=0.85\textwidth]{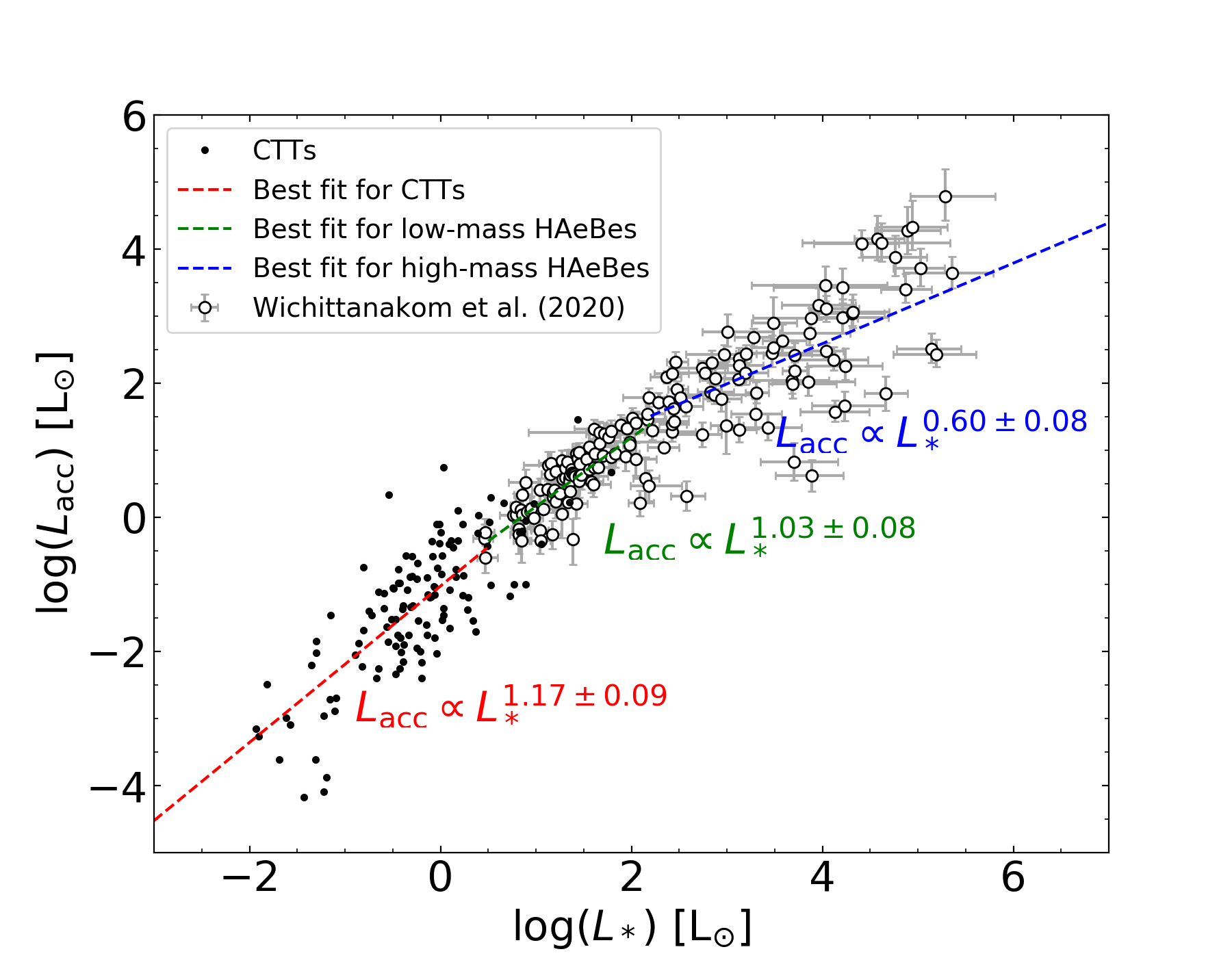}
\caption{Accretion luminosity versus stellar luminosity for 163 Herbig stars from \protect\citet{chumpon2020} and classical T Tauri stars from the literature.  Note the similarity between T Tauri and Herbig Ae stars, while the Herbig Be stars behave differently. Figure based on \protect\citet{chumpon2020}. }
\label{lacclstar}
\end{center}
\end{figure*}

An illustration of the results is provided in
Fig.~\ref{hrmacc}. It can be seen that the higher mass objects have
larger accretion rates.  Fig.~\ref{lacclstar} shows the accretion luminosities 
of the Herbig stars and several samples of T Tauri stars taken from the
literature \citep{Hartmann1998, White2003,Calvet2004, Natta2006}, as a
function of stellar luminosity. The T Tauri data are not based on Gaia
parallaxes, so although not directly comparable to the Herbig star
measurements, these objects are mostly selected to be members of
clusters and at similar distances. The dependencies of the accretion 
luminosity on the total luminosity should therefore, at least to
first order, be similar.

As was already known, the accretion luminosity increases with stellar
luminosity. However, a striking feature in this graph is that the
Herbig Ae stars follow the T Tauri stars closely; the best-fitting
slopes of the respective $L_{acc} - L_*$ relations for both groups of
objects are similar to within the error bars, while both slope and
scatter are significantly different for the higher mass Herbig Be
stars. 
As was described earlier, the Herbig Ae stars share many accretion-related properties with T Tauri
stars, and the dependence of the accretion luminosity on the stellar
luminosity proves no exception. The MA accretion paradigm should thus
be applicable to the Herbig Ae star regime.

The Herbig Be stars again differ in properties from their lower mass
counterparts. To this we add that \citet{Fairlamb2015} noted that a
number of Herbig Be stars required very large surface covering factors
of the accretion shocks in their MA models, with values approaching
100\% and higher. As such, these solutions are unphysical and their
UV/blue excess cannot be explained within the MA framework. It would
appear that the accretion rates as derived in the MA paradigm further
support the notion that Herbig Be stars accrete their material in a
different manner than the T Tauri and Herbig Ae stars. The break as 
seen in Fig.~\ref{lacclstar}, is also present in $\dot{M}_{acc} - M_*$ 
space, and occurs around 4~M$_{\odot}$ \citep{chumpon2020}, in 
agreement with the independent study
of \citet{Ababakr2017}, who found a marked change in linear
spectropolarimetric properties around spectral type B7/8.

This break at  4~M$_{\odot}$ is also seen in the high spectral resolution study of Br$\gamma$ by \citet{grant2022}. They found that the accretion rate increases with stellar mass following a trend similar to that of T Tauri stars for stellar masses $\rm \leq 4~M_{\odot}$. 
They also found that the distribution of line shapes of Br$\gamma$ was consistent across stellar mass and spectral type, though they caution that inclination affects may obscure any underlying systematic differences.

\bigskip

\noindent \textit{Summary}: The relationship between the stellar luminosity and the accretion luminosity is consistent between T Tauri stars and Herbig Ae stars. There is a break in this relationship around 4\Msun. A similar break is seen in the mass dependence of the stellar accretion rate, and is consistent with the marked change in the linear spectropolarimetric properties aound spectral type B7/B8. The mean accretion rate of Herbig Ae stars is 10$^{-7}$\Msun yr$^{-1}$. 

\subsection{\textsf{Mass accretion rate as a function of stellar age, disk dispersal times}}

We continue this discussion with a word of caution. In order to understand the
evolution of pre-Main Sequence stars and the disk dispersion
timescales, it has been common to study the mass accretion rate as a
function of age of the stars \citep[e.g.][]{Hartmann1998}. Due to the
comparatively small mass range and large samples of objects from
clusters with well-defined ages, it is feasible to determine this
relationship empirically for T Tauri stars. Several attempts to do so
have been made for Herbig stars, and, indeed, it has been found
that the mass accretion rate decreases with the stellar age
\citep[e.g.][]{Fairlamb2015,Arun2019}. However, one should be extremely
careful in deriving conclusions from this result.

As can be seen from the isochrones in Fig.~\ref{hrmacc}, the PMS
evolution across the HR diagram is much faster for high mass stars
than for low mass stars. As a result, a higher mass Herbig star 
inevitably has a younger age than a lower mass object. This in 
turn means that when investigating
relationships of various parameters with ``stellar age'', the age
essentially acts as proxy for the stellar mass. Given that, as
mentioned earlier, the higher mass Herbig stars have larger mass
accretion rates, we can expect this effect to dominate any age-related
correlations.  In order to disentangle these effects, and study the
mass accretion rate as a function of age, one needs to select a very
narrow mass range to remove or minimize this effect.

\citet{chumpon2020} struggled finding a statistical sample of stars
within a mass range narrow enough to isolate the time dependence of the mass
accretion rate.  The best subsample (having a mass range
2-2.5M$_{\odot}$) hinted at a decrease, but with generous error bars, $\dot
M_{\rm acc}\propto Age^{-1.95\pm0.49}$. Improved mass determinations,
which are still dominated by uncertainties in the parallax and extinction, and larger samples of Herbig
stars are needed to make progress in this area. 

A second caveat to keep in mind is that Herbig stars are selected on the basis of 
possessing emission lines. Once these stars cease to accrete, it is very difficult to 
identify PMS stars in this mass range. Thus it is likely that as these stars age, the
sample of PMS intermediate mass stars becomes less complete such that the Herbig stars 
represent the tip of a distribution of sources with long-lived accretion disks. If 
this is true, measurement of the time dependence of mass accretion from this biased 
sample is an upper limit. Next to the identification of non-accreting 
PMS intermediate mass stars, consideration of accretion rates of the direct progenitors of the Herbig stars, the IMTT stars, will help clarify our understanding on the time dependence of 
the stellar accretion rate. 

\bigskip

\noindent \textit{Summary}: Inferring the time dependence of the stellar accretion rate for Herbig stars is challenging because of the strong mass dependence of this relationship. For a narrow mass range (2.0-2.5\Msun), the mass accretion rate scales as $t^{-1.95\pm0.49}$. 

\subsection{\textsf{Accretion onto Herbig Be stars}}
\label{sec:acchms}
The MA paradigm is shown not to work for the higher mass Herbig Be
stars, while evidence has been presented that these massive objects accrete matter in a
different manner than for T Tauri and Herbig Ae stars.  The natural
next question is how material is delivered to the massive Herbig Be 
stars.  Understanding accretion onto Herbig Be stars and their formation 
also addresses a wider issue.  The intermediate mass Herbig stars connect 
the low mass T Tauri stars and the most massive stars whose formation is 
a topic of intense debate \citep{krumholzbonnell}. Observational studies 
have not yet reached the same level of detail as for lower mass objects,  
as massive pre-Main Sequence stars  are often optically invisible; due to 
their rapid collapse, they are still enshrouded by dust from their natal 
clouds \citep{Zhang2018,Frost2019}. The Herbig stars therefore may offer 
valuable insights into the formation of the most massive stars.

Regarding the disk-to-star accretion on Herbig Be stars, we
should bear in mind a crucial piece of evidence provided by the linear
spectropolarimetric studies.  The data for the higher mass Herbig Be
stars is remarkably similar to those of the classical Be stars, which
are readily explained with disks that reach onto the stellar surface
\citep[e.g.][]{Poeckert1976,Poeckert1977}. 

Hence, it would appear that the Herbig Be stars accrete material from 
a disk that is in contact with the stellar photosphere. This situation 
is reminiscent of the so-called Boundary Layer (BL)
accretion that is a natural consequence of a viscous circumstellar
disk \citep{Lynden1974}. The BL is a thin annulus close to the star in
which matter slows down from its (Keplerian) rotation velocity to the slower stellar rotation. In this region, kinetic energy will be dissipated.  This model considers how material is transferred from a Keplerian
rotating disk to the, more slowly rotating, stellar surface. The
region where the disk material slows down and releases its kinetic
energy is the boundary layer. 

The BL paradigm is being applied to
many accreting astrophysical systems such as White Dwarfs and neutron
stars in binary systems \citep[e.g.][]{Tylenda1977,Romanova2012}. It had even been explored for T Tauri
stars before the MA scenario prevailed \citep{Bertout1988}. 
\citet{Blondel2006} applied this scenario to a sample of mainly later type Herbig
stars. The models could mostly
reproduce the UV emission and the authors derived accretion rates
using this approach.  The UV excess, and its associated effective
temperature, can be controlled by the projected size of the BL on the
sky, while the accretion luminosity depends on the amount of matter
accreted. The BL model requires higher accretion rates than the MA model
to reproduce similar accretion luminosities.  This is because the
kinetic energy from Keplerian rotating material is half that of the
gravitational potential energy. We would expect, therefore, that - all
other things, such as radiating area being equal - when applied to the
higher mass Herbig stars, the BL derived accretion rates would need to
be higher. This could well result in a steeper slope of
the $M_{acc} - M_*$ relation  
at the high end, yielding a universal accretion behaviour across the PMS mass range.

Although the BL accretion has been suggested to apply in the case of 
the more massive Herbig stars \citep[e.g.][]{Mendigutia2011b,Cauley2014}, 
it has never been tested nor investigated in detail. The time is ripe for 
dedicated magneto-hydro\-dyna\-mic models to be
applied to the full Herbig star mass range to investigate the
viability of the, surprisingly underinvestigated, BL model and provide
predictions of the resulting covering factors and mass accretion
rates to compare with observations.

\bigskip

\noindent \textit{Summary}: The magnetospheric paradigm for inferring the accretion properties of Herbig Ae stars does not apply to their more massive counterparts. It is possible that these more massive stars accrete via a boundary layer, but more this has not yet been modeled in detail. 

\begin{figure*}
\begin{center}
\includegraphics[width=0.99\textwidth]{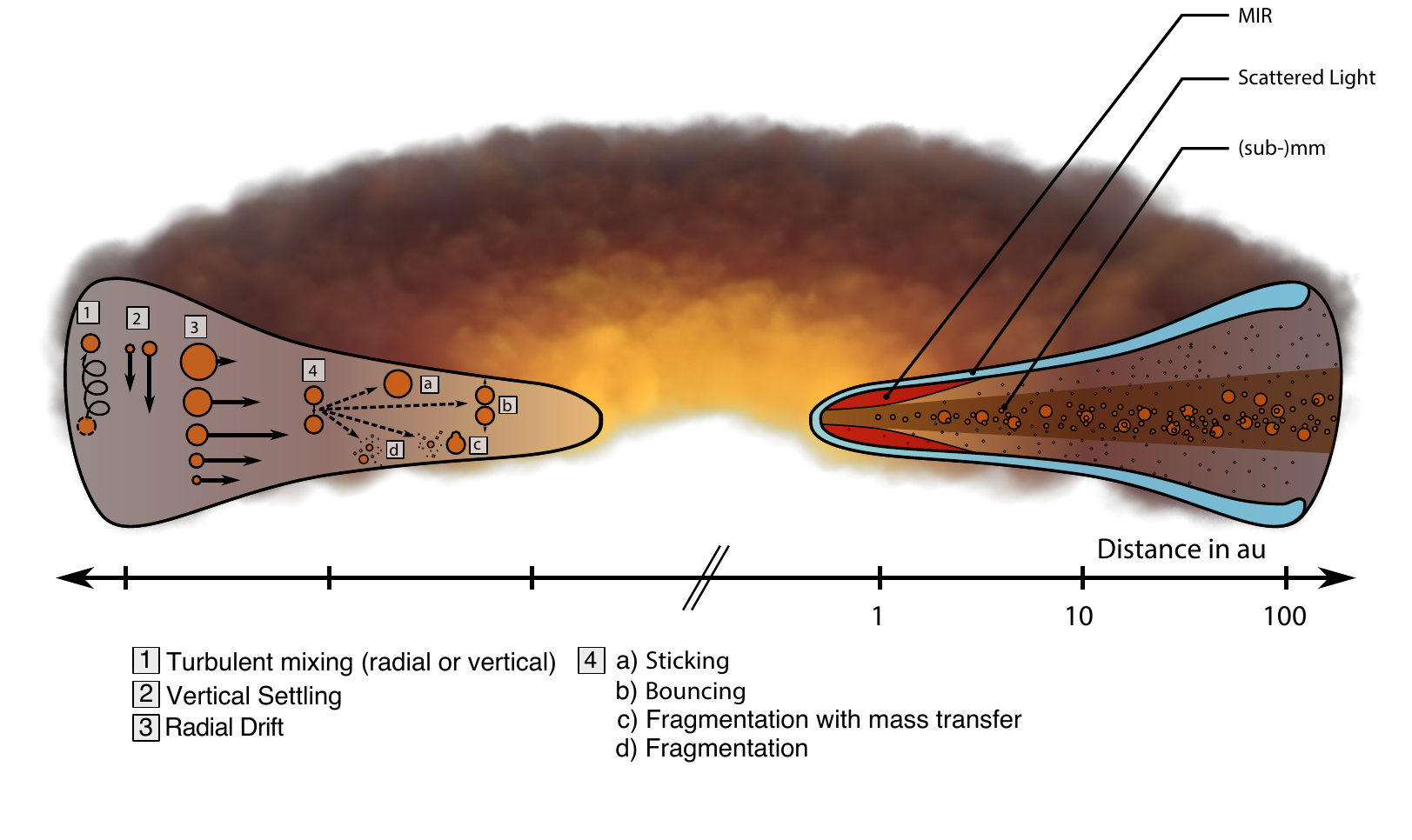}
\caption{Schematic of a disks around young star adapted from \citet{Testi2014}. This schematic highlights the principal processes that drive the evolution of dust (left) and the regions of the disk probed by various wavelengths (right).}
\label{fig_testi_disk}
\end{center}
\end{figure*}

\subsection{\textsf{Probing the accretion regions}}

To directly observe the accretion region we need to reach as close to the 
stellar surface as possible. With the exception of the UV excess, we have 
thus far discussed indirect probes of the accretion.  With regards to 
spatially resolved diagnostics, the smallest scales that can be probed 
directly are through milli-arcsecond optical and NIR interferometry 
of the H$\alpha$ and Br$\gamma$ lines. Over the years, such studies  have been 
conducted using facilities such as the VLTI, aimed at the Br$\gamma$ line  
\citep{Kraus2008,Mendigutia100546,Tam2016,Kreplin2018}   and CHARA, whose 
observations target H$\alpha$ \citep{Benisty2013, perraut2016,Mendigutia2017}. 

The picture that has emerged essentially confirmed the original findings from a small
sub-sample by \citet{Kraus2008}; there is a diversity in line properties, emission 
region sizes and line origins.  Perhaps most relevant to note here is that these
milli-arcsecond resolution interferometric optical and NIR studies of 
the hydrogen line emission do not necessarily identify the line emitting regions 
of Herbig stars with the magnetospheric accretion channels. 

For example \citet{Kraus2008,Mendigutia100546,Tam2016,Kreplin2018} and \citet{Mendigutia2017} 
report the line emission to be sometimes consistent with compact accretion channels and 
sometimes larger than that, making them more likely to be due to winds and outflows. 
When applying models to the data, it appears that, generally speaking, the line emission 
regions and properties are consistent with a disk-wind, and compatible with MHD driven 
winds \citep{Garcialopez2015,Tam2016,Caratti2015}. Although the lines themselves may not 
probe the accretion itself, the natural question as to how the emission line strengths 
allow us to determine the accretion rates can be answered fairly straightfowardly. As
outflows in young stars are accretion powered \citep[e.g.][]{hartmann2016}, a correlation between line luminosity
and accretion luminosity may therefore not be surprising after all. Indeed, it also explains why we 
can use the line luminosities for accretion rate determinations, and explains the observation that 
the line luminosities do not vary at the same timescales or amplitudes as the UV excess \citep{Mendigutia2013}. 

\bigskip

\noindent \textit{Summary}: There is compelling evidence that Herbig stars with spectral types later than $\sim$B7 accrete magnetosphericaly, and their H\textsc{i} and other atomic emission lines scale with the accretion luminosity though the origin of this relationship remains uncertain. The accretion rates of Herbig stars with masses $\lesssim 4 \rm \ M_{\odot}$ 
are ${\rm \ 10^{-8} \ M_{\odot}~yr^{-1}}\lesssim \dot{M}\lesssim10^{-6} \ \rm M_{\odot}~yr^{-1}$. The relationship between the stellar luminosity and the accretion luminosity is consistent $-2\lesssim \rm log({\it L_{*}}/L_{\odot})\lesssim2$. There is a break in this relationship for stars with higher luminosities (corresponding to $\sim$ 4\ \Msun). A study comparing the accretion rate of IMTT stars to Herbig stars will clarify how the stellar accretion rate evolves over time.

\section{\textsf{DUST DISKS AROUND HERBIG STARS}} 
\label{sec:dustdisks}
Our understanding of the properties and evolution of disks around young stars has advanced dramatically over the last 20 years. For a review about the evolution of dust in disks, we point the reader to a chapter from Protostars \& Planets VI by \citet{Testi2014} and the review by \citet{Andrews2020}. At the time of this writing, updated reviews are being prepared for Protostars \& Planets VII and should be available in 2023. In Fig.~\ref{fig_testi_disk}, we present a schematic that highlights the dominant processes thought to drive the evolution of dust in disks. In this section, we will summarize the advances in our understanding of dust in Herbig disks, while gas will be discussed in Sect.~\ref{sec:gasdisks}. 

That the SEDs of Herbig stars show an excess at infrared wavelengths was established in the early 1970's \citep[e.g.][and references therein]{Strom1972b}. Both disks and 
envelopes - some with a cavity carved by an outflow - were invoked to explain this excess 
\citep[e.g.][]{Gillett1971, Strom1972, Hillenbrand1992, berrilli1992, hartmann1993, mannings1994}. Unfortunately, due to the low spatial resolution of the observations (typically $\sim$20\arcsec \ at millimeter wavelengths, corresponding to 3000 au at 150 pc), it was not possible to distinguish which of the proposed morphologies fitting the SED was the correct one.  

The first spatially resolved observation of continuum emission around a Herbig star was obtained by \citet{Mannings1997b} using mm interferometry with the Owens Radio Valley Telescope (OVRO); they 
found extended emission up to 110 au for HD~163296, using beam sizes of 4-5\arcsec. These authors also detected Keplerian velocities of the CO gas, providing the first observational confirmation of a disk around a Herbig star (see Sect.~\ref{sec:gasdisks}).
The Hubble Space Telescope provided the first resolved scattered light images of Herbig disks \citep[e.g.,][]{weinberger99, Grady1999}, 
and in the last years, high contrast imagery with 8m class telescopes provide even greater detail of dust structures in the disk 
\citep[e.g.][]{garufi2017Msngr}. With the availability of the ALMA, disk emission can now routinely be spatially resolved for many Herbig stars thanks to the dramatically improved spatial resolution: typically 0.05-0.1\arcsec \ (7-15 au at a distance of 150 pc; \citealt{Dsharp1}). 

Before we look more into the detailed properties of the dust disks around Herbig stars, it is worth to consider the differences that are related to the spectral type of the central star. 

\subsection{\textsf{Herbig Be and Herbig Ae disks}}
\label{subsec:herbigbe}

Due to their rapid evolution towards the ZAMS, Herbig Be stars are typically surrounded by the cloud material from which they were formed, and this cold envelope will contribute to their continuum emission at longer wavelengths. For instance, a significant fraction of the Polycyclic Aromatic Hydrocarbon (PAH) emission is thought to originate from the surrounding envelope, instead of from the disk, as is the case in Herbig Ae stars \citep{Boersma2009,verhoeff2012}. 

Herbig Be stars have a much stronger UV radiation field than the cooler Herbig Ae stars: from 3450 L$_\odot$ for a B2V, 50 L$_\odot$ for a B8V and 7 L$_\odot$ for an A0V type star \citep{Jimenez-Donaire2017}. This translates to important observational differences, mainly in terms of gas emission lines (see Sect.~\ref{sec:disk-diff-AandB}) and disk properties. \citet{alonso-albi2009} compared the mm emission of Herbig Ae stars to Herbig Be stars, and found that, in general, the flux of the Herbig Be stars is 5-10 times lower, pointing to smaller disk masses. This can be attributed to photo-evaporation of the outer disk by UV photons on short timescales \citep[$\sim 10^5$ yr;][]{Gorti2009}. This is further supported by mid-infrared (MIR) observations, showing that the disks around Herbig Be stars are typically vertically flatter and smaller in size than those around Herbig Ae stars \citep{verhoeff2012}.

Furthermore, \citet{Vioque2018} derived the excess in the NIR (1.24-3.4~$\mu$m) and MIR (3.4-22~$\mu$m) for a large sample of Herbig stars. They found that the total infrared excess for objects more massive than 7~\Msun \, is typically smaller than that of the lower mass objects. This again points to important differences in disk properties depending on the stellar mass, and hence the stellar radiation field, although external evaporation due to nearby massive stars cannot be excluded.

Another important difference between Herbig Ae and Herbig Be stars is the typical distance to the objects. Herbig Be stars are typically located at $d \geq$ 1000 pc, while the best studied Herbig Ae stars are at distances of 100-200 pc, implying that the spatial resolution that can be achieved will differ by a factor of 5-10.

Due to the complications caused by the presence of envelope emission, in combination with the low spatial resolution when observing Herbig Be stars, we will now concentrate on the cooler Herbig Ae stars, thereby including objects from spectral type B9.5 up to $\sim$ F5.

\bigskip

\noindent \textit{Summary}: Herbig Be stars evolve toward the main sequence much faster than Herbig Ae stars, thus they tend to still be enshrouded in remnant cloud material. The NIR, MIR, and mm emission of Herbig Be stars is substantially lower than that of Herbig Ae stars. The faster disk dissipation timescale of Herbig Be stars is likely due to their higher UV luminosity and thus faster photoevaporation of the disk.

\subsection{\textsf{Early disk models: SED classification into group I and II}}
\label{s_diskmodels}

One of the first efforts to classify the SED of young stars were provided by \citet{Lada1984} and \citet{Lada1987}. These authors divided the SEDs of young stars into three classes, based on the slope of the SED between 2$\mu$m and 25$\mu$m given by,
\begin{equation}
    a=\frac{d{\rm log}(\lambda F_{\lambda})}{d{\rm log}(\lambda)}.
\end{equation}
Objects whose SED had a slope $0\leq a \lesssim 3$ where labeled Class I. Similarly, objects with $-2\lesssim a \lesssim 0$ where labeled Class II, and objects with $-3\lesssim a \lesssim -2$  were labeled Class III. Eventually this classification scheme was expanded to include Class 0 objects that were only detected at $\lambda \gtrsim 10\mu m$ \citep{Andre1993}. Most of the objects discussed in this review belong to class II.

The earliest disk models of Herbig stars required extremely high accretion rates ($> 10^{-6}$ \Msun/yr) and large optically thin inner holes to fit the SED near 3 $\mu$m \citep{Hillenbrand1992}. However,  \citet{hartmann1993} pointed out that, at the accretion rates necessary to fit the observed 3 $\mu$m excess, the inner region of the disk would be optically thick and thus emit too much radiation at short wavelengths to agree with the observations. Instead, they suggest that most Herbig stars do not harbor disks but envelopes. However, as noted above, mm interferometry and scattered light imagery established the presence of disks around Herbig stars just a few years later.

\citet{Chiang1997} modelled the excess emission observed in the SEDs of the lower-mass T Tauri stars with passive disks in hydrostatic radiative equilibrium. In those disks, the surface density goes as $\Sigma(r) \propto r^{-3/2}$. They distinguished between flat and flared disks: in a flat disk, the opening angle of the disk is constant, while in a flared disk it increases with distance. The optically thick disk is surrounded by an optically thin layer, the disk surface, that is directly heated by the star, and where the temperature is higher than that of the midplane. This surface layer will also heat the interior of the disk. Such 2-layered models can also be used for Herbig disks, with some modifications, as will be discussed in Sect.\ref{s_improv_diskmodel}. 

\begin{figure}
\begin{center}
\includegraphics[scale=0.3]{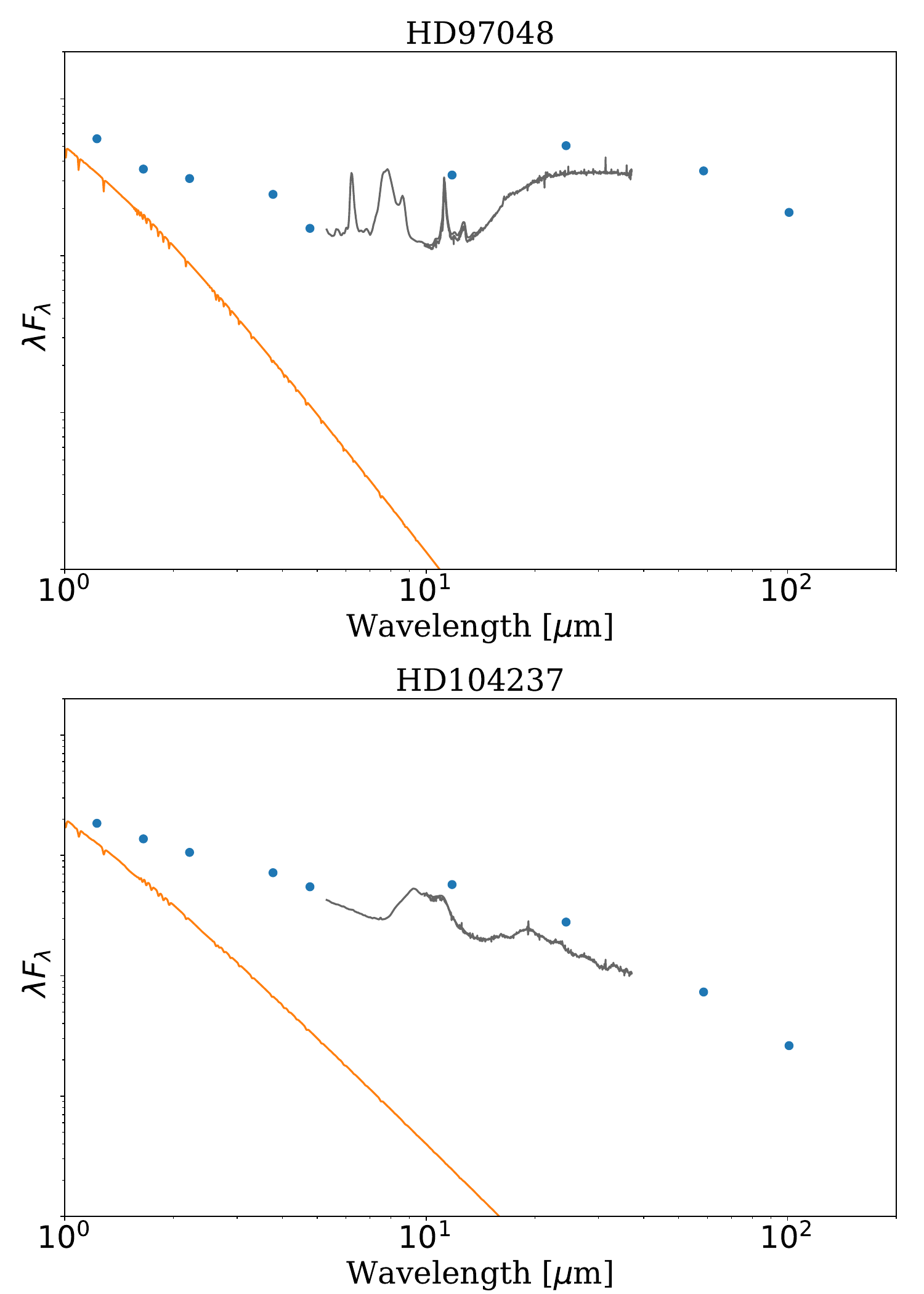}
\caption{IR SED of a group I (top) and a group II star (bottom). Blue dots: photometry, grey line: Spitzer IRS spectrum, orange line: Kurucz atmosphere model. After \citet{pascual2016}. }
\label{fig_GIGII}
\end{center}
\end{figure}

For a sample of 45 Herbig stars, the SED could be well-characterised in the infrared thanks to observations with IRAS \citep{malfait1998}. These authors proposed an evolutionary sequence where the initially continuous disk would evolve into a disk with a gap (a pre-transitional disk), 
after which the NIR excess would disappear (a transition disk), and finally a gas-poor debris disk would remain. A few years 
later, \citet{meeus2001} proposed a classification scheme based upon the shape of the IR SED\footnote{This classification scheme 
should not be confused with the one proposed by \citet{Hillenbrand1992} where group I sources 
have an infrared excess that scales as $\lambda^{-4/3}$ longward of 2 $\mu$m, group II sources 
have an SED with a positive slope in the infrared, and group III sources have a minimal infrared 
excess.}: Meeus group I (GI) objects have an SED where the continuum of the IR to sub-mm region 
can be reconstructed by the sum of a power-law and a black body, while Meeus group II (GII) 
objects have an SED where the continuum can be reconstructed by a power-law alone (see Fig.~\ref{fig_GIGII}). Possible 
locations for these components are an optically thick, geometrically thin disk (power-law 
component) and an optically thin flared region (black body). 

To facilitate the observational classification into GI and GII, \citet{vanboekel2005} introduced a criterion based on the NIR to total IR luminosity ratios and the IRAS 12$-$60 $\mu$m color: 
GI sources have $L_{\rm{NIR}}/L_{\rm{IR}} \leq (m_{\rm{12}} - m_{\rm{60}})$ + 1.5, while GII 
sources have $L_{\rm{NIR}}/L_{\rm{IR}} > (m_{\rm{12}} - m_{\rm{60}}) + 1.5$. A more recent (and equivalent)
criterion, based on an IR flux ratio that is easier to use was proposed by \citet{Khalafinejad2016}: 
sources with flux ratio F$_{30\mu \rm{m}}$/F$_{13.5\mu \rm{m}}$ $<2.1$ are GII, while those 
with a larger ratio belong to GI. 

\bigskip

\noindent \textit{Summary}: The disk of Herbig stars can be classified based upon the shape of their SED: to fit their excess, GI disks need a power-law + black body component, while GII disks only need a power-law. Criteria based on infrared photometry can be used to classify the disks.

\subsection{\textsf{Improvements to the early disk models}}
\label{s_improv_diskmodel}

For a long time, the structure of the disk remained a topic of debate due to the lack of spatial resolution, with both models of accretion disks with an optically thin hole and spherically symmetric envelopes matching the observed SEDs. 
However, \citet{Tuthill2001} were the first to spatially resolve the inner disk of a Herbig star, LkH$\alpha$~101, using interferometric data of Keck in the H and K band. They found a central cavity that was much larger (up to 10 times) than previously assumed from theoretical accretion disk models, and connected the location of the hot inner disk edge with the radius at which dust grains sublimate due to the stellar irradiation.

Disk models that could properly account for the NIR structure of the SED were developed by \citet{natta2001}. They proposed a model where the material inside the dust evaporation radius is optically thin gas and introduced the concept of a hot puffed-up inner wall of optically thick dust grains at the dust evaporation radius that could explain the NIR excess. \citet{Dullemond2001} gave an expression for the black body temperature of the rim,

\begin{equation} 
T_{\rm{rim}} = 
\left( \frac{L_*}{4\pi R_{\rm{rim}}^2 \sigma} \right)^{1/4} 
\left(1+ \frac{H_{\rm{rim}}}{R_{\rm{rim}}}\right)^{1/4}
\end{equation}

They further developed this idea with a semi-ana\-lytical model in which the hot puffed-up inner wall casts a shadow, obscuring parts of the flared disk. 
In some stars, even the entire disk might be shadowed. 

Following this work, \citet{dominik2003} modelled the observations with a disk model in hydrostatic equilibrium 
where the GI sources were modelled with a disk with varying surface densities, for some stars even increasing with radius. In this case most of the material is located in the outer regions of the 
disk, while the inner region is depleted. The GII sources, on the other hand, could more easily 
be modelled with a compact disk and/or a flattened outer region. 
However, the presence of the silicate feature in GII disks indicates that at least a small part of the disk intercepts stellar radiation \citep{dominik2003}.

\citet{dullemond2004self-shadow} used 2-D radiative transfer to model GI and GII disks 
and found that disks with a higher optical depth are flaring, while disks with a lower optical depth become self-shadowed. Building on this work, \citet{Dullemond2004settling} showed that the growth and subsequent settling of dust grains towards the 
midplane could enhance self-shadowing. 

This work provided a successful framework for interpreting the SEDs of Herbig stars in the evolutionary scheme proposed by \citet{malfait1998} that is intuitively compelling. Qualitatively 
the idea is that disks start out in a flared geometry (with envelopes). As small grains agglomerate to form larger 
grains, they settle toward the midplane resulting in flat, shadowed disks.
However, more recent work has challenged this picture, as we will discuss in the following section. 

\bigskip

\noindent \textit{Summary}: The disks around Herbig stars can be described with an optically thick midplane surrounded by an optically thin surface layer. The inner disk edge is puffed-up, and the innermost region is depleted in dust. GI sources emit stronger in the far-IR than GII sources.

\subsection{\textsf{Group I: disks with a cavity, Group II: self-shadowed disks} }
\label{sect_pretrans}

The paradigm of GI and GII being understood as an evolutionary sequence from young flared disks to settled flat disks stood for more than a decade until it was challenged by mid-IR spatially resolved studies. \citet{honda2012,honda2015} and \citet{maaskant2013} used MIR observations to conclude that several GI sources have large gaps in their disks, more specifically, a dust-depleted region between the inner and outer disk. Disks with gaps are
called pre-transitional disks, based on the assumption that as such disks further evolve and the inner disk dissipates, the SED would lack a near-IR excess (and become a transitional disk, see Sect.~\ref{sect_trans}). These authors suggested that GI disks could in fact be more evolved than GII disks, and proposed two distinct evolutionary pathways for disks. 

\begin{figure}[t]
\begin{center}
\includegraphics[scale=0.42]{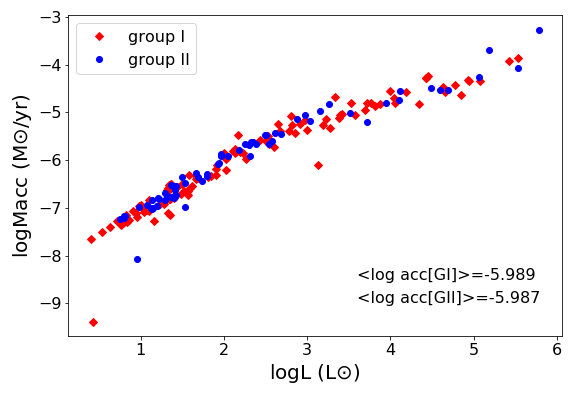}
\caption{Accretion rates for group I and II Herbigs, based on the HArchiBe catalogue  \citep{Guzman2021}.}
\label{fig_accGIGII}
\end{center}
\end{figure}

When comparing the accretion rate (determined from a variety of accretion tracers) in function of the stellar luminosity, no difference between GI and GII sources is observed (see Fig.~\ref{fig_accGIGII}). A similar behavior was found by \citet{grant2022} who compared the accretion rate, derived from the Br$\gamma$ line, for Group I and Group II sources and found a similar distribution, indicating that the accretion properties of the star is largely independent of the dust geometry.

In Table~\ref{tab_GIGII}, we summarize the main differences observed between GI and GII objects: 
selection criteria based on IR colors and fluxes, 
the mm flux scaled to a distance of 140 pc (`mm flux'), 
the slope $\alpha_{\rm{mm}}$ measured in the mm regions (`mm slope'), 
the extent of the dust continuum as observed at mm wavelengths
(`mm disk size'),
the gap size in the mm dust continuum (`gap size'), 
the disk substructure observed in mm continuum,
the extent of the small dust grains (`scattered light'),  
the appearance of the disk in scattered light, 
the extent of the disk seen in CO emission (`CO disk size'),
the inner radius where CO ro-vib lines are detected (`CO 
inner radius'), 
the emission transition from the highest rotational level in the ground vibrational state of CO observed (`CO transition'), the 10 $\mu$m silicate feature and finally 
the PAH strength.

\begin{table}
   \caption{Main observed differences between group I and II disks. References: mm disk size: \citet{stapper2022}; mm flux at 140pc, calculated from \citet{stapper2022}, mm slope: \citet{pascual2016}; scattered light, gap size, disk structure: \citet{Garufi2017} and references therein; CO disk size: Taun, MSc 2020; CO inner radius: \citet{Banzatti2018}; highest CO level detected:  \citet{Meeus2012,vanderWiel2014}; silicate feature: \citet{Juhasz2010}; PAHs: \citet{Acke2010}.}\begin{tabu}{XXX}
    \toprule

    Diagnostic     & Group I & Group II \\
    \hline
    $L_{\rm{NIR}}/L_{\rm{IR}} - 1.5$ & $\leq m_{\rm{12}} - m_{\rm{60}}$      & $> m_{\rm{12}} - m_{\rm{60}}$ \\
    F$_{30\mu m}$/F$_{13.5\mu m}$   & $>$ 2.2 & $<$ 2.2 \\
    $<$mm flux$>$ & 750 mJy & 220 mJy\\
    $<$mm slope$>$ & -3.41 & -3.05\\
    mm disk size & 40-250 au & $<$150 au\\
    mm gap size & large, 5-140 au & small, $<$5 au \\
    mm structure &cavity, rings & barely resolved\\
    scattered light & 100-600 au & elusive, $<$ 100 au\\
    scattered 
    
    light structure & cavity, rings, spirals & barely resolved\\
    CO disk size & 100-650 au & 80-200 au \\
    CO inner radius & 5-18 au & 3 au\\
    CO transition & J=36-35 & J=13-12\\
    silicate feature & sometimes absent & typically present \\
    PAHs & strong features & weak/absent \\
    \bottomrule
    \end{tabu}
 
    \label{tab_GIGII}
\end{table}

\begin{figure*}[t]
\begin{center}
\includegraphics[width=1.02\textwidth]{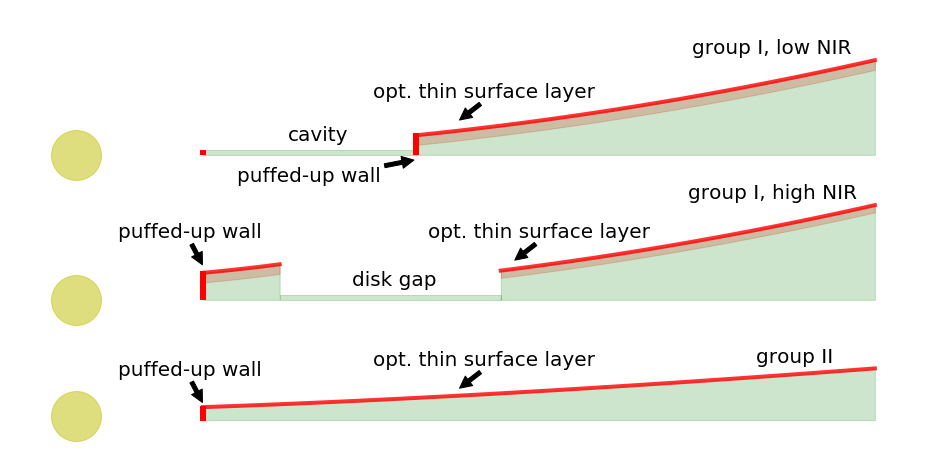}
\caption{The different disk structures proposed by \citet{Banzatti2018} for Herbig disks. Top: A schematic of a disk with an inner cavity and a flared outer disk. This geometry results in a GI SED with little to no excess NIR emission. Middle: A schematic of a disk with a gap that separates the inner and outer disk. This geometry also results in a GI SED, however, there is a substantial excess NIR emission. Bottom: A schematic of a continuous disk. This geometry results in a GII SED. Figure based on \citet{Bosman2019}.}
\label{fig_banzatti}
\end{center}
\end{figure*}

Recently, \citet{Banzatti2018} noted a dichotomy in the NIR excess of group I sources: it is either low ($<$ 10\%) or high ($>$ 25\%), while the group II sources have intermediate values. Furthermore, they found a connection between the strength of the NIR excess and the radius at which CO emission is detected: sources with a low NIR excess have CO emission starting at larger radii indicating that the inner regions are more depleted in these sources. We depict the morphologies of those 3 different disk types in Fig.~\ref{fig_banzatti}.
In a recent paper, \citet{garufi2022} studied GII disks in scattered light. They found an anti-correlation between the disk brightness in scattered light and the NIR excess, supporting the scenario where the inner wall casts a shadow on the disk behind it, as was proposed earlier by \citet{dullemond2004self-shadow}.

\bigskip

\noindent \textit{Summary}: GI disks have a large cavity, depleted in dust, that can explain the observed differences in flaring. 
GII disks, on the other hand, have either no cavity or a very small one and are self-shadowed, so they are fainter both at FIR wavelengths and in scattered light.

\subsection{\textsf{Dust properties in Herbig disks}}

The disk material from which planets eventually form originates in the interstellar medium (ISM). There, dust particles are either oxygen- (mainly silicates) or carbon-rich (mainly Polycyclic Aromatic Hydrocarbons - PAHs). 

Silicates can be divided into olivines (Mg$_{2x}$Fe$_{2-2x}$SiO$_4$) and pyroxenes (Mg$_{x}$Fe$_{1-x}$SiO$_3$). In the ISM, silicates are small ($<0.1~\mu$m) and amorphous, with a crystalline mass fraction $<$ 2.2\% \citep{kemper2004}, so that the detection of larger and/or crystalline silicates in Herbig disks would indicate dust processing. 
Dust grains can grow when the density is high and the collision velocity low enough \citep[e.g.][]{windmark2012}.
Crystallization of grains can occur either through thermal annealing of amorphous dust or condensation from the gas phase, both of which occur at high temperatures \citep[T $>$ 1000 K;][]{fabian2000}.
Our knowledge about dust in protoplanetary disks is limited by the fact that the bulk of the mass resides in the (partially) optically thick mid-plane, and that the emitting surface per unit mass decreases as grains grow. However, there are 3 portions of the dust population that can be observed through different methods: \\
1) submicron-sized dust grains in the disksurface scatter optical and near-IR light; \\
2) micron-sized dust grains, also located in the disksurface, emit thermally in the mid-IR when at the right temperature (T $\lesssim$ few hundreds of K), hence at a certain distance from the star. It is here that we can identify their main solid-state features;\\
3) mm-sized dust grains, located near the midplane at larger distances from the star, with T $\lesssim$ few tens of K, will emit thermally in the mm range, where the optical depth is lower.

We will now look at each of these methods and discuss the main results derived from them, with a focus on what we have learned from IR spectroscopy, revealing the properties of the dust grains. 

\subsubsection{\textsf{Scattering by micron-sized dust grains}}
Small grains or aggregates located in the surface layer of the disk can be traced through the light they scatter. If the disk is inclined, it is possible to measure the scattering efficiency at different angles (the phase function) of the dust grains.  The more inclined the disk is, the larger the phase angle range that can be observed (see Fig. 2 in \citealt{Benisty2022}). However, \citet{stolker2016} showed that it is important to take the flaring of the disk into account when deriving the phase function and interpreting the scattered light images, due to projection effects in the image plane. Several dust properties can be derived from the degree and intensity of polarization of the scattered light, as well as from the way these depend on the scattering angle. 

The phase function of a dust grain depends mainly on its size (but also on its shape, structure, and refractive index), with smaller dust grains ($a < \lambda_{\rm{obs}}$) having a more isotropic phase function, while the polarization by larger dust grains ($a > \lambda_{\rm{obs}}$) tends to peak towards small angles (0$\degree$, or forward scattering). Therefore, when comparing the ratio of forward and backward (180$\degree$) scattering intensity, the size of the aggregates can be estimated \citep[e.g.][]{tazaki2019}. Furthermore, for larger aggregates, the amount of polarization is related to the porosity of the material. Also the color of the aggregates can further reveal the porosity of the dust \citep{mulders2013}. In Herbig stars, polarized phase functions of both small and large aggregates have been observed in several objects \citep[e.g.][]{stolker2016,ginski2016}.
However, as these observations are very time consuming, due to the sensitivity and spatial resolution required, more progress in this field is expected to be made in the future \citep[see also][]{Benisty2022}.

\subsubsection{\textsf{Thermal emission of warm small dust grains}}

The largest advances so far in deriving detailed dust properties was done based on infrared spectroscopy. By comparing optical constants obtained in the lab, the dust responsible for the observed spectral features can be characterised in terms of its composition, degree of crystallisation, and grain size \citep[e.g.][]{henning2011}. Ground-based observations of silicates have mainly concentrated on the 10~$\mu$m window, where both amorphous and crystalline silicates can have strong features. However, for a more detailed study, space based observatories were necessary to obtain more sensitive and longer wavelength data. 

\begin{figure}[t]
\begin{center}
\includegraphics[width=0.75\textwidth]{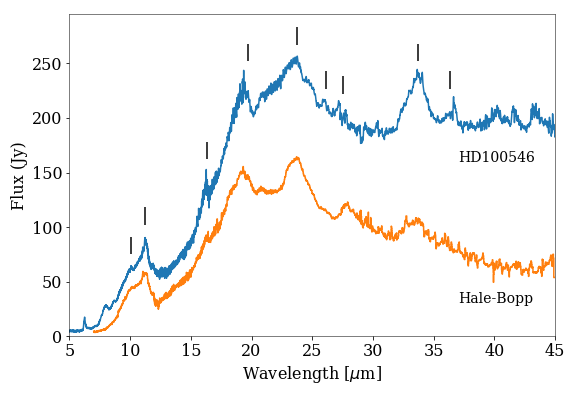}
\caption{The spectra of Herbig star HD~100546 and Solar System comet Hale-Bopp. With the vertical marks the positions of forsterite features are indicated. Figure based on \citet{malfait1998}.}
\label{fig_HaleBopp}
\end{center}
\end{figure}

Observations with ISO were crucial to make a first inventory of the IR spectral features in Herbig disks. A key result was the discovery of crystalline silicate features in the spectrum of HD~100546, reminiscent of features found in the Solar System comet Hale-Bopp 
\citep[see Fig~\ref{fig_HaleBopp};][]{malfait1998}. Furthermore, a study of 14 Herbig stars showed a wide variety in dust properties: from unprocessed silicates (mainly amorphous and small as in the ISM) to highly processed silicates (crystalline and large); with no relation to any of their stellar parameters \citep{bouwman2001}. Surprisingly, several GI sources do not show silicate emission, even though their SEDs were similar to other GI sources that did. This can be attributed to either a lack of warm and/or small silicates, what can be modelled by 1) removing the small grains and 2) increasing the height of the inner wall \citep{meeus2002,dominik2003}. 
However, both model options cannot explain the observations. Therefore, \citet{dominik2003} speculated that the absence of the silicate feature could be caused by a gap, removing those dust grains in the temperature range ($\sim$ 200-400 K) needed to emit in the MIR. Indeed, a decade later \citet{maaskant2013} showed with spatially resolved mid-IR observations that the absence of the 10~$\mu$m silicate feature in GI disks can be explained by certain gaps in the disk. 

\begin{figure}[t]
\begin{center}
\includegraphics[width=0.75\textwidth]{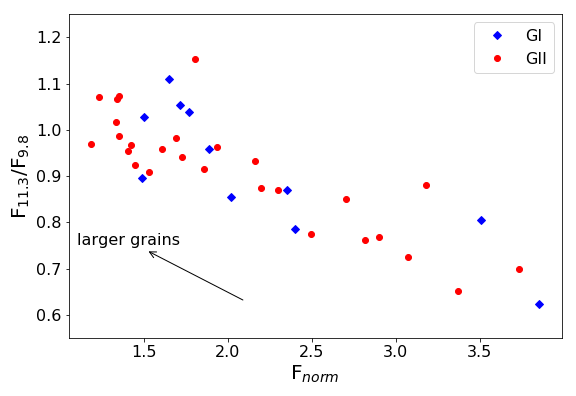}
\caption{The 11.3/9.8 $\mu$m ratio versus the peak to continuum ratio of the 10$\mu$m silicate feature for GI and GII sources, F$_{\rm{norm}}$ = 1 + (F$_{\rm{obs}}$ - F$_{\rm{cont}}$)/F$_{\rm{cont}}$. After \citet{Juhasz2010}.}
\label{fig_siliratios}
\end{center}
\end{figure}

To study the grain size, \citet{vanboekel2003} introduced an intuitive method, based on the change in strength and shape of the 10 $\mu$m silicate feature as grains grow: comparison of the 11.3/9.8 $\mu$m ratio of the silicate feature to its peak-to-continuum ratio. They showed that the observed band strength and shape are correlated, so that a weaker feature provides evidence for larger grains in the inner surface layers and cannot be attributed to a mere contrast effect.
This method indicates that there is no difference between GI disks and GII disks in terms of silicate grain growth (see Fig.~\ref{fig_siliratios}).
Studies of large samples with higher S/N spectra (taken with Spitzer) that were fit with a two-layer temperature distribution model confirm a variety of dust properties (dominated by amorphous olivines; \citealt{Juhasz2010}). These authors found that the crystallinity fraction of small dust graints in disks around Herbig stars ranged from 1\% to 30 \% and did not correlate with stellar or other disk properties.

The wavelength and band shape of the  69 $\mu$m feature emitted by forsterite at temperatures below 300 K are very sensitive tracers of the Fe content and temperature of the grains. This feature became accessible with Herschel/PACS \citep[e.g.][]{sturm2010,maaskant2015}. The analysis of 7 Herbig stars showed that most of the forsterite grains responsible for the 69 $\mu$m band have rather high temperatures of 100-200~K, and that the Fe content is less than 2\% \citep{sturm2013}.

It is important to keep in mind that the emission features seen in the IR spectra trace dust grains that are in the optically thin surface layer of the disk, while the bulk of the mass resides in the disk mid-plane. Also, those grains need to be at the right radial distance in order to be warm enough to emit at IR wavelengths \citep{kessler_silacci2007}. It is, therefore, uncertain how representative these dust grains are of the bulk of the disk. This will naturally depend on the amount of (vertical) turbulence in the disk, and the dynamical coupling between the gas and dust grains, which also depends on the grain size.

\subsubsection{\textsf{Spatial distribution of silicates in disks}}

\begin{figure}[t]
\begin{center}
\includegraphics[width=0.75\textwidth]{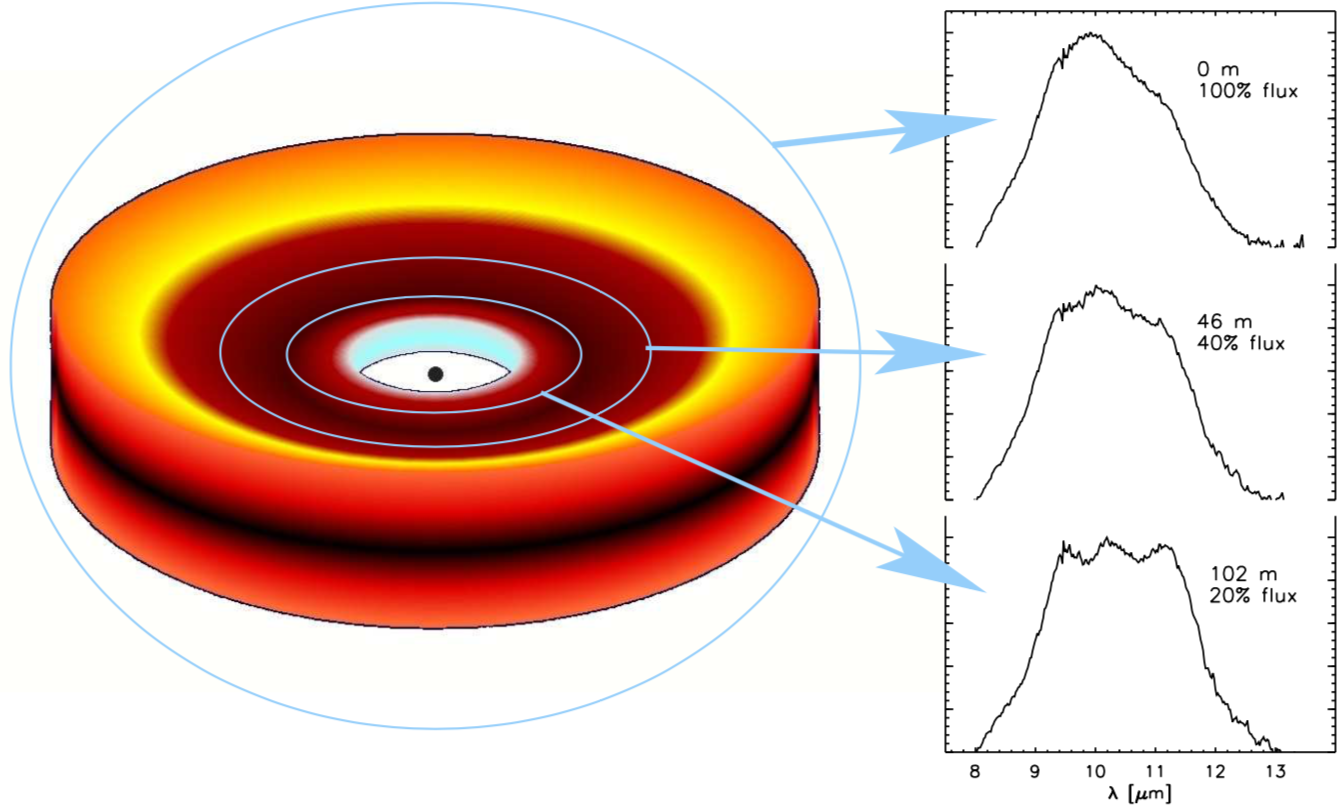}
\caption{The 10 $\mu$m spectra of HD~144432 taken with 3 different baselines, covering different radial distances in the disk. The dust in the innermost region is clearly more processed than that of the entire disk. Figure from \citet{vanboekel2006SPIE}. }
\label{fig_innerouter}
\end{center}
\end{figure}

The spatial distribution of silicates (in particular the crystalline silicates) can be measured with MIR interferometry, and indirectly in spatially unresolved spectra that cover a wide wavelength range and hence temperature range of the emitting dust. The \emph{MID-infrared Interferometer} at the VLTI \citep{leinert2003} was the first instrument that provided the angular resolution necessary to spatially resolve the silicate emission. It was shown that the abundance of crystalline silicates increases in the inner few au of Herbig disks \citep[see Fig.~\ref{fig_innerouter};][]{vanboekel2004}. Subsequent studies confirmed this result for other Herbig stars \citep[e.g.][]{menu2015} and T Tauri stars \citep{varga2018}. 

While more indirect, \citet{Juhasz2010} used the wide wavelength range (5-38 $\mu$m) of Spitzer/IRS to infer the radial distribution of the crystalline silicates in the disk. They split the spectra in 2 parts 
representing the warmer material in the inner disk and the colder material in the outer disk.  
They found forsterite at a wide range of radial distances, including regions far below the annealing temperature of silicates, suggesting they were formed through eruptive processes. In contrast, enstatite was preferentially (but not exclusively) found in the inner disk, so that the enstatite-to-forsterite mass ratio declines with distance from the star. For a discussion on possible formation scenarios, we refer to \citet{Juhasz2010}.

The abundance of forsterite can be locally enhanced under certain conditions. This is the case in HD~100546, a GI Herbig star with a bright disk wall at 13 au. \citet{Mulders2011a} found that most of the forsterite emission originates from the inner wall (at 13-20 au), with a dust temperature of 150 to 200 K. 
While there is a high forsterite abundance (40-60\%) in the wall, the observations are consistent with the absence of fosterite in the outer disk. 
Thus an observed strong feature does not necessarily mean a high average abundance in the disk, but can reflect a locally enhanced abundance in small crystalline silicate grains. Such local enhancements may be related to planet formation. Indeed, in the case of HD~100546, several planetary candidates have been proposed (see Sect.~\ref{s_planetformation}).
If present, such a massive planet may stir up a local population of larger bodies or pebbles in the exposed inner regions of the outer disk, and/or cause shocks that produce crystalline silicates. Hence, a direct relation between the planet formation process and the occurrence of crystalline silicates could explain the observed global lack of correlations with stellar or disk parameters. 

\subsubsection{\textsf{Thermal emission of cold large dust grains:  grain size and disk mass}}
\label{sect_GIGIIdust}

The thermal emission of mm-sized dust grains located closer to the mid-plane of the disk is expected to be optically thin at a distance $>$ 5-10 au. As their mm emission is in the Rayleigh-Jeans (R-J) limit, their flux is proportional to the dust opacity: $F_{\nu} \propto \kappa_{\nu} {\nu}^2$. Unfortunately, the optical properties of the dust are not known a priori. However, at mm wavelengths, $\kappa_{\nu}$ can be approximated by $\nu^{\beta}$, with $\beta$ the dust opacity index, such that: $F_{\nu} \propto {\nu}^{2+\beta}$. The dust opacity index $\beta$ is dependent on the dust properties, with dust grain size thought to be the most important - besides composition, shape and porosity. In the ISM, the value of $\beta$ is typically 1.7-2.0, and declines as the grains grow. Therefore, the observed spectral index, $\alpha$, of the SED ($F_{\nu} \propto \nu^\alpha$) can be used as a proxy for grain growth deeper in the disk, assuming optically thin emission in the R-J limit, so that $\alpha$ = $\beta$ + 2.

\citet{acke2004graingrowth} found that $\beta$ was lower for GII disks than for GI disks, suggesting enhanced grain growth in GII disks (see also \citealt{pascual2016}). However, these observations provide the radial average of $\beta$, while a  radial dependency of the spectral indices was found in several disks (e.g.  HD~163296 \citep{Guidi2016}, HD~141569A \citep{white2018}, and HD~100546 \citep{miley2019}), with a lower spectral index in the inner regions than the outer regions, pointing to enhanced grain growth in the inner disk. 

\citet{pinilla2014} found that $\beta$ is higher in sources with dust-depleted cavities (where grain growth is not favorable), and showed that $\beta$ can even be related to the cavity size. Therefore, the radially integrated value of $\beta$ can provide a misleading measure of the overall grain growth throughout the disk. Furthermore, \citet{Woitke2019} showed that while $\beta$ is thought to be dominated by the size of the grains, variation in the composition of the grains can lead to deviations in $\beta$ ranging from $\sim 0.2 - 0.7$ in the warm Herbig disks while cooler disks - such as around TTS - have even larger deviations.

\begin{figure}[t]
\begin{center}
\includegraphics[width=0.75\textwidth]{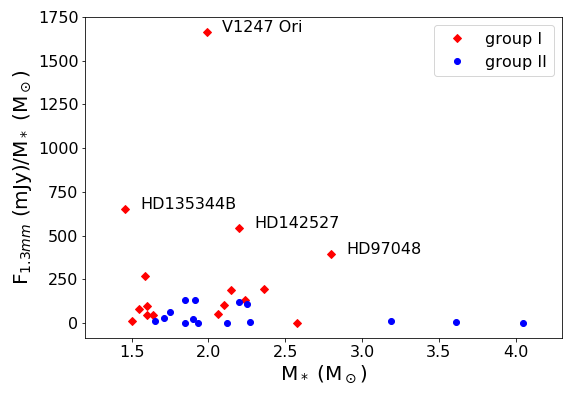}
\caption{The 1.3mm flux, scaled to 140pc and divided by M$_*$ versus M$_*$, a proxy for M$_{\rm{dust}}$/M$_*$, for stars with d $<$ 500pc. The largest fluxes are seen in GI sources, but many GI sources have similar values as GII. Based on data from \citet{stapper2022}.}
\label{fig_stappermass}
\end{center}
\end{figure}

The dust disk mass can be estimated from the observed flux at mm wavelengths, assuming that the emission is optically thin and that the dust opacity and temperature are known \citep[e.g.,][]{Beckwith1990, Henning1994}:

\begin{equation}
    \rm{M_{dust} = \frac{c^2 d^2 F_\nu}{2 \nu^2 k <T_{dust}> \kappa^{abs}_\nu}}
\end{equation}

\noindent with $c$ the speed of light, $d$ the distance, $F_\nu$ the flux, $\nu$ the frequency, $k$ the Boltzmann constant, $\rm{<T_{dust}>}$ the mass-averaged dust temperature and $\kappa^{abs}_\nu$ the absorption coefficient per unit dust mass (cm$^2$g$^{-1}$). 
\citet{pascual2016} found that, on average, GI sources have larger mm luminosities than GII sources; a similar result was found by \citet{stapper2022}; see Fig.~\ref{fig_stappermass}. Assuming that the temperature and dust opacity is the same in every disk, this would imply that GI disks have larger dust masses than GII disks. However, there is evidence that the disk temperatures and dust opacity of GI and GII sources are not the same \citep{Woitke2019}; also, the size of the largest dust grains that we can trace is similar to the longest wavelength we observe, thus large rocks etc.\ remain unaccounted for.  

\citet{Woitke2019} compared dust masses derived with 3 different methods: 1) the classical way, with a fixed dust opacity and dust temperature, 2) the classical way but with dust opacities and temperatures derived from SED modelling, and 3) dust mass directly derived from SED modelling. They point out that, 1) typically, the dust opacities are larger than the `canonical value of 3.5 cm$^2$g$^{-1}$, so that the actual dust masses are lower; 2) not all disks are optically thin at (sub)mm wavelengths; 3) the dust temperature can be higher in the case of a small disk, or lower in the case of an extended disk, leading to actual lower/higher dust masses for smaller/larger disk then when derived in the classical way. However, when comparing the results from method 1) and 3), \citet{Woitke2019} found that these effects tend to cancel out, so that only an uncertainty of 0.5 dex in dust mass remains. Additionally, as we will discuss in Sect. \ref{sec:gasmassevolution}, the gas to dust ratio in disks is poorly constrained.
While the use of the mm luminosity to estimate the disk mass is convenient, the poor characterization of the required parameters suggests that this approach should be taken with some caution. 

\bigskip

\noindent \textit{Summary}: Evidence for dust processing in terms of crystallization and grain growth has been found in the IR spectra of Herbig disks, but no connection between the amount of the dust processing and stellar or disk properties was found. However, the dust grains observed in the MIR represent only a small fraction of the dust present in the disk. 

The absence of the 10 $\mu$m silicate feature in GI disks can be explained by dust cavities at a distance where dust grains would obtain a temperature of 200-400K. 

The decrease in enstatite/forsterite ratio with distance from the star suggests that (long-lasting) eruptive processes are the dominant source of cold forsterite in the outer disk regions. On the other hand, the low Fe content observed in the 69 $\mu$m forsterite feature is consistent with silicate formation in chemical equilibrium at high temperatures. More detailed modeling of crystallization processes, taking into account radial transport, shock heating, and stellar eruptions is clearly needed. 

Under the (somewhat simplistic) assumption that the Herbig disks have similar temperatures and dust opacities, GI disks have, on average, larger dust masses than GII disks, as derived from their mm fluxes assuming optically thin emission. However, when ignoring radial variations of $\beta$, GI disks have, on average, higher $\beta$ values than GII, pointing to smaller grains, thus resulting in stronger emission for a similar mass. In addition, different dust compositions will result in different values of $\beta$. It is clear that, when deriving and comparing dust disk masses, more attention needs to be paid to dust opacities and temperatures. 

\bigskip

\subsection{\textsf{PAHs in Herbig disks}}
\label{sect_PAH}

While silicates are oxygen-rich materials, Polycyclic Aromatic Hydrocarbons (PAHs) are carbon-rich. They are actually large molecules rather than dust particles, so they are suspended in the disk atmosphere. Since PAHs emit by reprocessing UV photons, they trace the surface of the disk exposed to the radiation field of the star, and can thus be detected at larger distances from the star than silicates, which are thermally excited.

\citet{meeus2001} found that in GII sources, PAHs are absent or weak, while GI sources show stronger PAH emission features. 
This result was confirmed by \citet{Acke2010} for a sample of 53 Herbig Ae stars: the PAH-to-stellar luminosity ratio is higher in targets with a flared disk (GI). The observation that the PAH luminosity is stronger in GI than in GII disks, for a given stellar temperature, can be related to a larger amount of PAHs that are exposed to UV photons. Indeed, when a gap opens in the disk, a larger part of the disk is exposed to stellar photons that warm the disk and thus increase the flaring. 
The few GII sources where PAHs were detected \citep[e.g. HD~142666,][]{meeus2001} might have small gaps \citep{menu2015}. Alternatively, for some sources the gas in the disk could be flared, even when the dust has settled \citep{Acke2010}.

The variations seen in the positions of the features are mainly 
due to chemical differences of the PAHs induced by the stellar 
UV field \citep{Acke2010}: the C-C bonds at 6.2 and 7.8 $\mu$m shift to longer wavelengths with decreasing stellar effective temperature and is a measure for the aliphatic/aromatic content ratio of the hydrocarbon mixture.  Furthermore, \citet{maaskant2014} found that the ionization state of the 
PAHs (that can be deduced from the I$_{6.2}$/I$_{11.3}$ band ratio) critically depends on the 
optical thickness of the disk, with a higher ionization fraction in optically thinner disks. 
These authors also propose that PAHs are not only located in the disk surface as is generally 
assumed, but that they are also present in the more optically thin disk gaps. 

\bigskip

\noindent \textit{Summary}: PAHs trace the disk surface out to large distances. Their emission is stronger in flared disks where they can intercept more stellar photons. The observed chemical differences (aliphatic vs. aromatic) are due to differences in the stellar UV radiation, while the ionization state depends on the optical thickness of the disk. 

\subsection{\textsf{Ices in Herbig disks}}
\label{sect_ice}

Ices are thought to be an important component in the framework of planet formation, due to 1) their higher sticking coefficient, making it easier for icy grains to grow \footnote{However, experimental work by \cite{gundlach2018} did not confirm the theoretical prediction that ices have higher tensile strengths than silicates.},
and 2) the increase of the density in solids when dust grains are covered in ice. 
Therefore, the location of the ice line (where a particular molecule freezes out in a disk) determines the region where the growth of planetesimals could occur more easily.

Unfortunately, the spatial distribution of ices in Herbig disks is not well known, mainly due to a lack of spatial resolution at far-IR wavelengths, as well as due to the fact that most ices are located deeper in the disk, where the optical depth is too high for them to be seen. However, at 3.1 $\mu$m, water ice can be observed 
through scattered light spectra. With this method, a decrease in surface brightness at 3.1 $\mu$m was observed in HD~142527, beyond the cavity and inner wall of the outer disk \citep{Honda2009}. Furthermore, a crude spatial distribution of water ice was derived by \citet{Honda2016} for HD~100546 from the radial profile of the 3.1 $\mu$m opacity, but the results are inconclusive.

\begin{figure}[t]
\begin{center}
\includegraphics[width=0.75\textwidth]{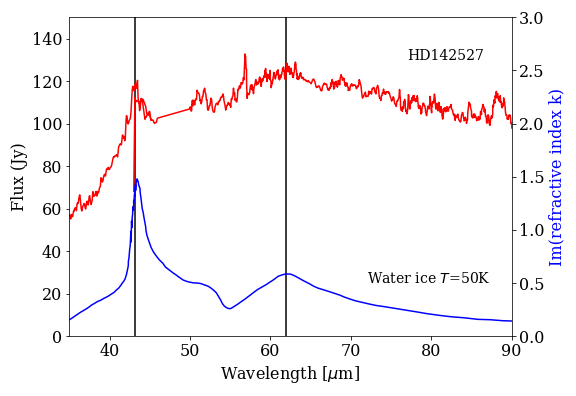}
\caption{The ISO SWS and LWS spectra of HD~142527, and the imaginary part of the refractive index of water ice at 50 K in blue. Indicated are water ice emission peaks at 43 and 62 $\mu$m. After \citet{Min2016b}.}
\label{fig_ices}
\end{center}
\end{figure}

In the far-IR, the water ice feature at 43 $\mu$m can be strong, but that wavelength unfortunately lies out of the wavelength covered by Spitzer and Herschel, but it was detected for a few sources with ISO/SWS \citep{Malfait1999}. Another ice 
feature at 62 $\mu$m has a large width (see Fig.~\ref{fig_ices}) that hinders straightforward identification as it is rather weak. Both features were identified in HD~142527 \citep[note that HD~142527 has a large dust gap ($>$ 140 au) with likely low gas density;][]{casassus2015}. 
\cite{Min2016b} modelled the far-IR spectra and determined that the emitting water ice resides in the outer disk of HD~142527 where the ratio of water ice to silicates is 1.6, showing that the density of solids clearly increases due to ice. They also concluded that 80\% of the oxygen in the outer regions resides in water ice, a similar amount as what is found in the outer solar system and in dense interstellar clouds. 

\cite{Min2016b} found the water ice in HD~142527 to be mainly crystalline, even though it is located in regions with temperatures for which the crystallisation timescale is too long to have occurred. Therefore, they considered various scenarios to form crystalline ice: 1) heating by a strong accretion burst, 2) formation in the inner disk and subsequent transport to the outer regions by radiation pressure, or alternatively, and 3) the break-up of larger icy bodies through collisions in which crystalline ice is preserved, again with subsequent transport outwards by radiation pressure. These authors considered the third scenario the most likely. 

The temperature at which water and CO freezes out under typical disk conditions is 128-155 K and 23-28 K, respectively \citep{zhang2015}. By deriving the disk temperature profile, the location of the ice line thus can be determined, as \citet{Isella2016} did for HD~163296 based on ALMA observations: for the midplane, $T_m (r)$ = 24K $(r/100au)^{-0.5}$ and for the surface, $T_s(r, z)$ = 68K$(\sqrt{r^2 + z^2}/100au)^{-0.6}$. This would place the water snow line at $\sim$ 3-4 au, too small to be resolved with ALMA, and the CO ice line at $\sim$ 75-110 au, well within reach with ALMA.

\bigskip

\noindent \textit{Summary}: 
Water ice has been detected at 3.1 $\mu$m in scattered light in a few Herbig stars. Despite water ice having strong features at 43 and 62 $\mu$m, the paucity of observations and detections at those wavelengths results in a poor understanding of water ice in Herbig disks. To detect the snow line, very high spatial resolution is required; so far water has not been detected in a disk around a Herbig star with ALMA. 
Soon, JWST will provide more sensitive measurements of water emission from disks, as MIRI (5-29 $\mu$m) covers the mid-IR lines of water with excitation temperatures of a few 1000 K with a spectral resolution of 3000, so water lines in Herbigs are expected to be detected. 

\subsection{\textsf{Transition Disks to Debris disks}}
\label{sect_trans}

\begin{table}[t]
    \centering
       \caption{Comparison of the properties of circumstellar disks at different evolutionary stages. Listed are typical values, for detected sources - many are not detected, especially the debris disks in CO. References: \citet{Meeus2012} and $^{(a)}$\citet{difolco2020}; $^{(b)}$\citet{hughes2018}, $^{(c)}$\citet{Moor2017}. CO and mm fluxes are normalised to a distance of 140 pc. }
    \begin{tabu}{XXXX}
    \toprule
    Disk type & Planet & Transitional & Debris  \\
     &  forming &  &   \\
    \hline
    Onset IR excess &NIR & MIR & FIR\\
    Gas content  &gas-rich & intermediate & gas-poor\\
    $\dot M_{\rm acc}$ (\Msun/yr) &7 $\times $10$^{-7}$ &2 $\times $10$^{-8}$&--\\
    L$_{\rm{IR-mm}}$/L$_{*}$ & 0.5 & 0.01 &$< 0.008^{(b)}$\\
    F$_{1.3\rm{mm}}$ (mJy) & 500 & 1.7-7$^{(a)}$ & 0.1-3$^{(c)}$ \\
    CO~2-1 (Jy km/s) &2-50&6.3$^{(a)}$& 0.05-3$^{(b)}$\\
    Prototype & AB Aur & HD~141569A & Vega \\
    \bottomrule
    \end{tabu}
 
    \label{t_trans}
\end{table}

Disks evolve from a massive primordial gas-rich disk into a gas-poor debris disk, after passing through the transition disk phase. The definition of a ``transition" disk is varied in the literature. This term is variously used to define disks whose near-IR excess is smaller than the median among disks in Taurus \citep{Calvet2005a}, disks possessing a cavity \citep{Espaillat2014}, and disks whose SED reveals no NIR excess and a significant excess at wavelengths beyond 10$\mu$m \citep{strom89}. In this review, we adopt the latter definition. The lack of a NIR excess can be attributed to the depletion of small grains in the inner disk \citep[e.g.,][]{strom89}, creating a cavity in the disk. This can be caused by clearing by a giant planet, but other scenarios, such as grain growth or photo-evaporation might also contribute \citep[e.g.][]{armitage2019}. In Table~\ref{t_trans}, 
we compare the main properties of planet forming, transitional and debris disks.

\begin{figure}[t]
\begin{center}
\includegraphics[width=0.75\textwidth]{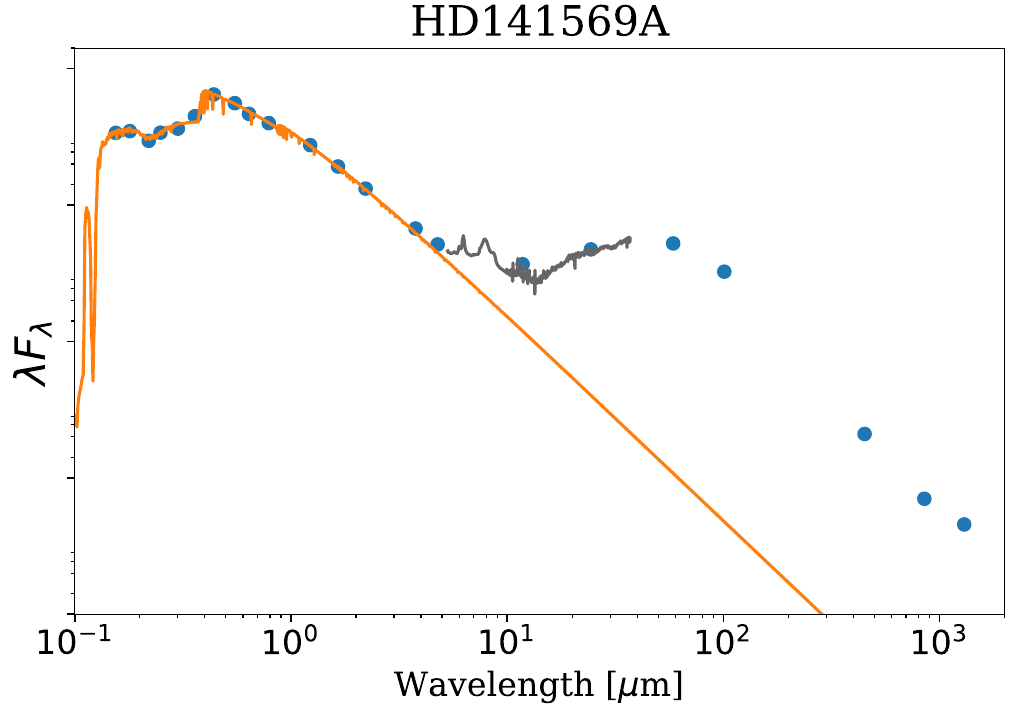}
\caption{SED of the transition disk HD~141569A. The excess in the NIR is absent, and only starts beyond 5~$\mu$m.}
\label{sed_HD141569}
\end{center}
\end{figure}

A survey of sub-mm emission for debris disks in combination with a literature compilation shows that the dust mass in debris disks is lower by about two orders of magnitude \citep[adopting a constant dust opacity and representative dust temperatures,][]{Panic2013}. However, there is some overlap between primordial disks and debris disks in the age range of a few~Myr to 20~Myr.

A prime example of a transition disk is the 9 $\pm$ 4.5 Myr A0 star HD~141569A. The infrared excess emerges near 5 $\mu$m (Fig.~\ref{sed_HD141569}), and its fractional IR 
luminosity is at least a factor 10 smaller than that of a typical Herbig Ae star \citep{pascual2016}. The H$\alpha$ emission line is double peaked and the accretion rate is $2\times10^{-8}$\Msun yr$^{-1}$ \citep{Fairlamb2015}. The presence of warm molecular gas was identified from ro-vibrational CO emission \citep{Brittain02} and the presence of cool molecular gas was identified from rotational CO emission \citep{zuckerman95}. The presence of gas in the disk is further confirmed through emission lines of [OI] and [CII] in the far-IR \citep{Thi2014}. 

\citet{difolco2020} compared spatially resolved images of CO emission and mm continuum acquired with ALMA with resolved scattered light imagery acquired with the \textit{Spectro-Polarimetric High-contrast Exoplanet REsearch} (SPHERE) on the VLT. They found large differences in the radial distribution of the material probed by these observations. The mm thermal continuum extends to $\sim$250 au while the disk is detected out to $\sim$400 au in scattered light, where several rings are seen. The CO gas 
was only detected out to the distance of the mm continuum (250 au). The flux of these pure rotational lines can be converted to a total gas mass of $\sim 70 M_{\oplus}$ ($\sim\!2 \times 10^{-4}$~M$_\odot$), on the lower end for Herbig stars, but 10 times higher than the most gas-rich debris disk known to date \citep{Moor2017}. Curiously, with the disk mass inferred from the CO lines, the stellar accretion rate could only be sustained for a few thousand years, suggesting that either the CO is severely depleted or the inference of the stellar accretion rate inferred from the veiling of the Balmer continuum is overestimated (see Sect.~\ref{sec:gasmassevolution}).

\bigskip

\noindent \textit{Summary}: A transition disk is a disk that is in transition between the protoplanetary and debris phase. In those disks, the infrared excess starts in the MIR. Gas is still present in these disks and the stars accrete. The total IR luminosity, CO luminosity and mm flux is in between that of Herbigs and debris disks.

\subsection{\textsf{Detailed disk morphology}}
\label{duststruct}

The MIR image of HD 97048 taken by \citet{Lagage2006} revealed for the first time a strongly flaring disk surface through the emission of PAHs located in the disk surface where they are transiently heated by stellar UV photons. This flaring surface was later confirmed with NIR scattered light observations \citep{ginski2016}.
On the other hand, under the assumption of optically thin emission, the surface density profile of the large grains can be derived from mm observations.  ALMA is perfectly suited for this purpose, as it traces the thermal emission of cold mm-sized dust grains with high spatial resolution. ALMA uncovered several large ($>$ 10 au) cavities in the dust continuum, the largest -140 au- seen in HD~142527 \citep{Perez2015}, and 80 au in HD~34282 \citep{vanderplas2017_HD34382}.
Subsequent progress in observational techniques at both shorter and longer wavelengths (increased sensitivity and spatial resolution) led to the discovery of substructures in the disks, as will be discussed below. But first we will summarize what is expected from theoretical disk models. 

\subsubsection{\textsf{Theoretical predictions for disk substructure}}
\label{sect_theo}

In a continuous disk, large mm-sized grains are expected to spiral inwards as they experience a headwind due to the surrounding gas, making them lose angular momentum, a process called radial drift. The smaller micron-sized dust grains, on the other hand, are coupled to the gas and do not experience such a headwind. Grains 
can also spread outwards to redistribute the angular momentum in the presence of (turbulent) viscosity \citep[e.g.][]{Birnstiel2010}.

\begin{figure}[t]
\begin{center}
\includegraphics[width=0.75\textwidth]{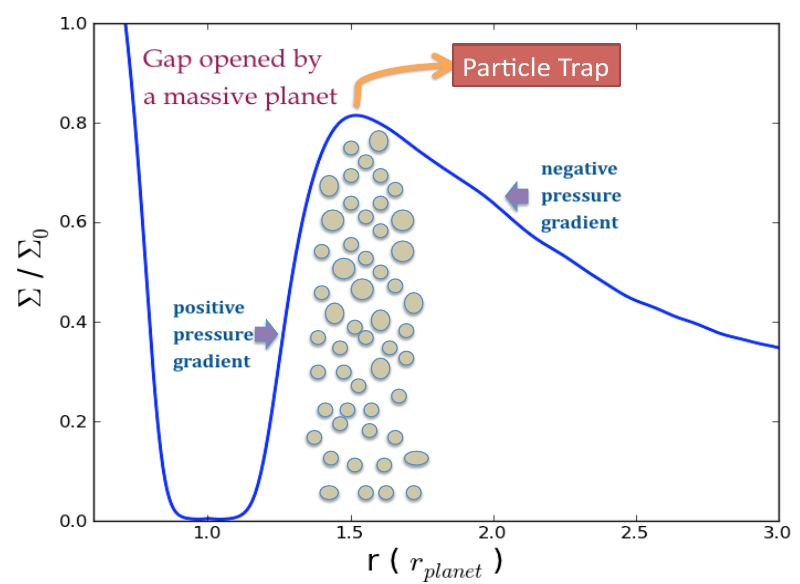}
\caption{Gas surface density in function of distance, illustrating particle trapping in a disk, the result of pressure bumps caused by a massive planet located at 10 au (Pinilla, private communication).}
\label{fig_trapping}
\end{center}
\end{figure}

Models predict that an embedded planet, brown dwarf, or low-mass stellar companion can dynamically clear its orbit and open a gap in a disk \citep{lin1979}. When a planet opens a gap in the disk, a local enhancement in gas pressure is created - a positive pressure gradient, trapping dust grains and thus creating a region with low collisional velocities in which the dust grains can grow more rapidly, as is illustrated in Fig.~\ref{fig_trapping} \citep{Pinilla2012,Pinilla2016}. 
Another scenario to create a gap is a dead zone, a region with a low ionization fraction, and by consequence a low turbulence level. At the edge of such a dead zone a pile-up of material is expected, creating a gap in the disk \citep[e.g.][]{regaly2012}.
For both scenarios, the viscosity is an important parameter. 

The efficiency of dust trapping depends on the grain size: larger grains are trapped in the pressure bump while smaller grains migrate inward - a process referred to as dust filtration \citep{rice2006, Zhu2012}. This can be verified with high spatial resolution observations: at mm wavelengths, tracing the larger grains, one would observe a ring-like structure with a large gap, while in the NIR scattered light, tracing smaller grains, this gap would either be smaller or even non-existent \citep[e.g.][]{dejuanovelar2013}. The width of the ring that is induced by a massive planet through pressure bumps will depend on the disk and stellar mass, as well as the location of the planet and the viscosity parameter $\alpha$ \citep{pinilla2018}. 

This process of dust filtration was confirmed by observations: the cavity size seen in NIR scattered light is often smaller than that in the mm, for instance for HD~135344B $\sim$ 28~au in the NIR versus $\sim$ 46 au at mm wavelengths \citep{Garufi2013,Muto2012}. To match the observations, \citet{Dong2012} used a disk model where inside the cavity, the surface density profile of the small grains is flat and can reproduce the observed NIR excess, while the larger dust grains are depleted.

\citet{Zhu2012} find that in the presence of a giant planet, grains $\rm \gtrsim 10-100~\mu m$ are trapped in the pressure bump and this accounts for roughly 99\% of the dust mass (assuming a grain size distribution $n(a) \propto a^{-3}$ in the outer disk, where grains have already grown to mm sizes, only 1\% of the dust mass is in small micron-sized grains). Thus the material that moves inward is heavily depleted in solids. 
Therefore, the relative abundance of refractory elements accreting onto the star should also be depleted relative to volatile elements. In the case of Herbig stars with radiative exteriors, this could result in a photospheric abundance pattern where refractory elements are depleted relative to volatile elements \citep[i.e., the $\lambda$ Bo\"{o} phenomenon,][ see Sect.~\ref{s_lamdaboo}]{kama2015}. 

Another feature that can be induced by a planet in a disk are spiral arms \citep[e.g.,][]{Kley2012}. If the planet is sufficiently massive, spiral arms can be observed in scattered light (planet to stellar mass ratio $q \rm \gtrsim~a~few~\times 10^{-3}$; \citealt{dong15spiral}). This is a kinematic effect that results in increasing the scale height of the disk above the spiral arm such that the disk intercepts more light. Furthermore, \citet{Dong2016} used hydrodynamical and radiative transfer simulations to determine how the disk appearance depends on its inclination and position angle. 
To summarize, a giant planet cannot only create a cavity but, with he right parameters ($\alpha$, h/r, $M_{\rm{planet}}$) also rings or spiral arms, or simply cause asymmetries. 

Turning to the inner disk, we find a puffed-up inner wall at the dust evaporation radius located behind an optically thin gas-rich region (Sect.~\ref{s_improv_diskmodel}, \citealt{natta2001}). This model was further refined by  \citet{Isella2005}, who realised that, as the dust evaporation temperature is density dependent, and there is a vertical density gradient in the disk, the inner wall will be curved, rather than have a sharp edge. Due to this curvature, the NIR excess will weakly depend on the inclination, while the observed shape will strongly depend on the inclination. Face-on disks will reveal rims that appear axisymmetric, while inclined disks will reveal rims that appear elliptic, with one side brighter than the other one. 
Furthermore, \citet{kama2009} showed that the rim morphology depends on the dust grain size, composition and the inner disk surface density.

In the following sections we will discuss how observations compare with these predictions. We will first look at the outer regions of the disk and then continue with the inner disk, before discussing the different substructures seen in the disk. 

\subsubsection{\textsf{Observations of the outer disk region}}

As we mentioned in Sect.~\ref{sec:dustdisks}, the first time a disk was resolved was through mm interferometry \citep{Mannings1997b}.
In a recent study, \citet{stapper2022} analyzed archival ALMA data of 36 Herbig stars within 450 pc. For the disks with d $<$ 225 pc, 15/25 `nearby' disks are resolved, but this sample was likely biased towards the brightest objects. However, given that there are only 31 Herbig stars within 225~pc, we can conclude that at least 50\% of them are resolved, all of which show substructure.

Another way to obtain high spatial resolution is to move to shorter wavelengths. Optical and NIR images trace light that is scattered off the disk surface by small, submicron-sized dust grains, but the stellar brightness limits how close to the star one can get (inner working angle). Instruments such as SPHERE and the \emph{Gemini Planet Imager} (GPI) use polarimetric differential imaging (PDI) techniques to remove the light of the central star and reveal the scattered light of the disk. What remains is light polarized by the dust grains in the disk. The inner working angle of PDI on 8m class telescopes is typically 0.1\arcsec, or 15 au at a distance of 150 pc.  

A comparison between ALMA and SPHERE observations of the outer disk has shown that 
the large and small grains are not always co-spatial, with the larger grains being more radially confined than the smaller ones \citep[e.g. the GI star HD~97048, large grains seen up to 350 au, small grains up to 640 au;][]{walsh2016}. On the other hand, the small grains are more or less co-spatial with the gas, indicating that they are dynamically coupled.

Another important result is that the extent of the disk in the mm continuum is often much smaller than that of the gas, derived from low-$J$ pure rotational CO transitions: e.g. 350 vs. 750 au in HD~97048 \citep{walsh2016}, and 375 vs. 1000 au for HD~34282 \citep{vanderplas2017_HD34382}. These results were confirmed in a larger study (Taun, MSc 2020; see Sect.~\ref{sec:radialgas}).

\begin{figure}[t]
\begin{center}
\includegraphics[width=0.75\textwidth]{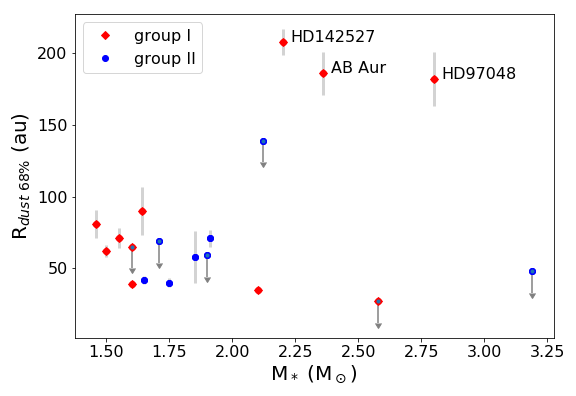}
\caption{The dust radius at 68\% of the light (ALMA band 6 or 7) versus stellar mass for stars with d $<$ 225pc. The error bars are indicated as well as the upper limits. The largest radii are seen in GI sources, but many GI sources have similar sizes as GII. Based on \citet{stapper2022}.}
\label{fig_stapper}
\end{center}
\end{figure}

The differences in disk sizes do not only depend on the material being traced (large or small dust grains, or gas), but also on the disk group considered. Based on SED modeling, \citet{dominik2003} proposed that GII disks are smaller than GI disks, an idea that now can be tested as we can often obtain high enough spatial resolution.
Early on, GI sources were routinely resolved with ALMA, while the few GII sources observed were either unresolved or small ($< 100$~au; e.g. \citealt{walsh2016,vandermarel2021}). 
\citet{stapper2022} further showed that, on average, GII disks are more compact than GI disks, although this trend is mainly caused by 3 outliers (AB Aur, HD 97048 and HD 142527; see Fig.~\ref{fig_stapper}), and bright disks such as the GI source HD100546 (68\% radius of 35 au) have similar sizes as the GII disks. 

In a study of Herbig disks with SPHERE, \citet{Garufi2017} found that GII disks, when detected in scattered light, are much fainter than GI disks which are commonly detected. Also their spatial extent is different: disks around GI sources are detected out to distances of 100-600 au, while disks around GII sources are usually smaller than 100~au.  Indeed, several nearby (d$\lesssim$ 200~pc) GII disks are not resolved at all, such as HD~150193A \citep{Garufi2014}, which is a binary with a nearby companion (a=1.1\arcsec ; \citealt{bouvier2001}) that likely has truncated the disk to give it its compact appearance. However, the disk of HD~100453, a GI star, is also truncated by a companion at 1.1\arcsec \, but here the companion is thought to give rise to the observed spiral arms reaching out to 42 au \citep{Wagner2018}. Similar behavior has been observed in the MIR with 8m telescopes \citep[e.g.][]{honda2015}. These differences can be attributed to larger gaps ($> 5$ au) and/or larger flaring angles in GI disks, while GII disks are either small or self-shadowed \citep{Garufi2017}. 

However, from a systematic comparison between ALMA and SPHERE data, \citet{garufi2022} found that \textit{the extent of GII disks in scattered light} is much smaller than that at mm wavelengths, due to the outer disk only being partially illuminated. This is in stark contrast with the observations of GI disks, where the extent in scattered light is larger than that in the mm, as seen in e.g. HD~97048 \citep{walsh2016}. This is because GII disks are flat and/or self-shadowed, while GI disks are flared, so that their disk surfaces can be traced much further out.

\subsubsection{\textsf{Observations of the inner disk region}}
\label{s_dust_innerdisc}

In a pioneering study of TTSs and Herbig stars with the IOTA interferometer, \citet{monnier2002} found a correlation between the `size of the inner disk', measured in the $H$- or $K$-band, and the stellar luminosity. 
They explained this finding with an inner rim at the dust evaporation radius, located behind an optically thin inner region. Later, \citet{monnier2005} measured more Herbig disks in the $K$-band with the Keck Interferometer and found sizes between 0.02 and 4~au, in agreement with emission at the dust evaporation radius, and showed that, for spectral types A to late B, the inner disk size can be related to the stellar luminosity as $R \propto \sqrt{L_*}$.

Another important clue about the inner disk region came from studying the brightness variations of UXors 
that show the blueing effect due to obscuration by dust in the line of sight (see Sect.~\ref{s_variability}). 
\citet{dullemond2003} realized that the variability timescale of weeks to months means that the obscuring cloud must be in the inner disk, at the location of a puffed-up inner rim. They also showed that this effect would only be seen in GII disks, where the line of sight towards the inner rim remains largely undisturbed. The blueing effect in UXors is thus caused by the clumpy nature of the inner disk.  

\citet{Lazareff2017} observed 51 Herbig stars with PIONIER, a NIR interferometric instrument that combines the 4 VLT telescopes. They found that the inner rims are smooth, radially extended, and consistent with axisymmetry. However, \citet{Kluska2020} further studied the inner disk rim morphology with PIONIER, reaching a spatial resolution of a few milli-arcseconds, and found evidence for a non-axisymmetric structure in 27\% of the objects. This could be due to warping of the inner disk or instabilities at the inner rim, potentially linked to the presence of a companion. MATISSE, also combining 4 telescopes but now in the L, M and N band \citep[][]{Lopez2022}, 
uncovered $L'$-band variable brightness asymmetries in the disk of HD~163296 at scales $<$~0.3~au \citep{varga2021}. Such variations are also detected in the H and K band, and persist over several years \citep{sanchez2021}. They reflect an orbiting, inhomogeneous dust distribution in the innermost disk regions, similar to what was already predicted from the UX Ori-like brightness and color variations.

\subsubsection{\textsf{Disk substructures}}
\label{s_diskdetails}

\begin{figure}[t]
\begin{center}
\includegraphics[width=1.01\textwidth]{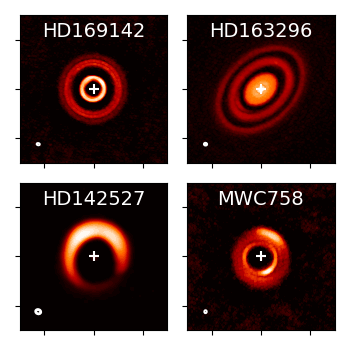}
\caption{ALMA continuum images illustrating the variety in structure present in the Herbig disks. The beam size is indicated in the lower left.  The tickmarks go from -1.5\arcsec \ to +1.5\arcsec for HD~169142, HD~163296 and MWC758, and from -3\arcsec \ to +3\arcsec for HD~142527. Figure from van der Marel, private communication, after \citet{vandermarel2021}}.
\label{fig_ALMA}
\end{center}
\end{figure}

More detailed ALMA observations of disks with a cavity reveal substructures \citep[for an overview see][]{Andrews2020,vandermarel2021}, some of which are shown in Fig.~\ref{fig_ALMA}. Multiple rings were found in e.g. HD~169142 \citep{Fedele2017} and HD~97048 \citep{walsh2016,Vanderplas2017_HD97048}.
Other objects, such as HD 142527, show asymmetries and/or shadows, which are referred to as horseshoes \citep{Casassus2013}. Finally, some objects show spiral arms: e.g. AB Aur \citep{Tang2017}. 

\citet{Zhang2018} created a 2-D hydrodynamical gas + dust model grid to study the effect the disk parameters viscosity, scale height 
and planet mass have on the presence of gaps in the disk. From a comparison with the DSHARP ALMA survey \citep{Andrews2018}, they deduced that the observed disk gaps could be caused by planets located beyond 10 au with masses between that of Neptune and Jupiter. More massive planets, the super-Jupiters, are capable of carving out larger ($>$ 20 au) cavities in a disk, as shown by hydrodynamical modelling of PDS~70 \citep{muley2019}.

With more and more objects being observed with ALMA, a reflection on the methods used for interpretation is needed. 
\citet{kim2019} performed a synthetic analysis to determine the best set of ALMA bands to derive the dust properties of the T Tauri star TW Hya 
and concluded that, in order to constrain the temperature profile properly, several ALMA bands are needed, preferably with the largest frequency intervals possible, and covering both optically thin and optically thick emission. Such an analysis likely also applies to Herbig stars, meaning that the ALMA bands with which the disk is observed should be carefully selected to include frequencies that trace both optically thin and optically thick emission.

\begin{figure}[t]
\begin{center}
\includegraphics[width=0.99\textwidth]{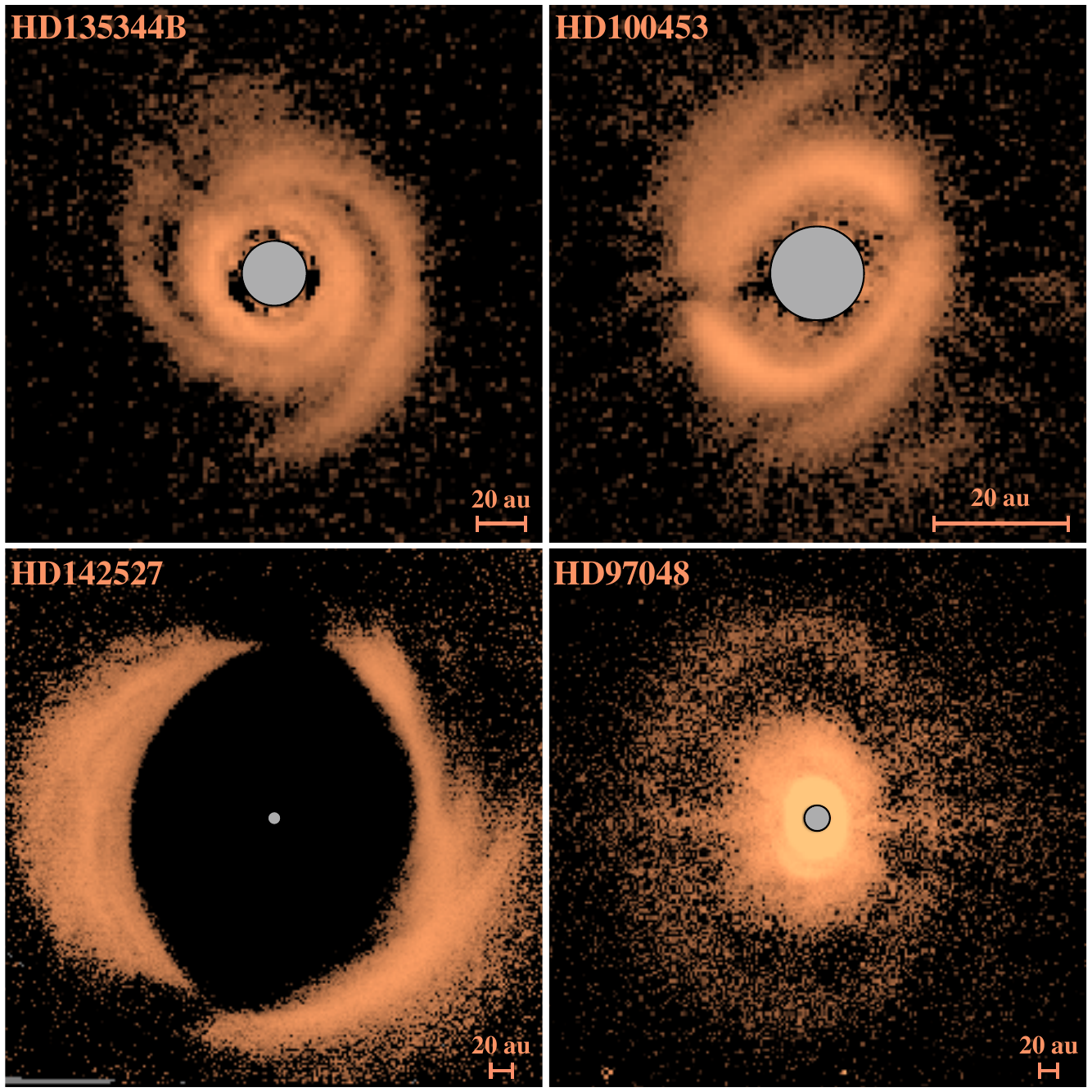}
\caption{Key examples of disks observed with SPHERE in PDI: spiral arms (HD~135344B and HD~100453), a large cavity (HD~142527) and rings (HD~97048). Note the large differences in scale between the different panels: HD~100453 is much smaller than HD~142527 and 
would fit 3-4 times inside its cavity. A. Garufi, private communication, after \citet{garufi2017Msngr}.}
\label{fig_sphere}
\end{center}
\end{figure}

Scattered-light observations probe small grains and thus can be used to constrain the flaring surface of a disk as was done for HD~100546 \citep{Avenhaus2014,stolker2016} and HD~97048 \citep{ginski2016}.
Large scattered light surveys of protoplanetary disks were made with GPI on Gemini South \citep[][]{macintosh2008}, \textit{High-Contrast Coronographic Imager for Adaptive Optics} (HiCiAO) on Subaru \citep[][]{tamura2009} and SPHERE on the VLT \citep[][]{Garufi2017}, uncovering a wide variety in disk substructures (see Fig.~\ref{fig_sphere}). Spirals were seen in e.g. HD~135344B \citep{Muto2012,Garufi2013} and HD~36112 
\citep[MWC758;][]{Grady2013,Benisty2015}, while rings were found in e.g. HD~169142 \citep{quanz2013} and HD~97048 \citep{ginski2016}. Other disks, such as those around AB Aur \citep{Hashimoto2011} and HD~142527 \citep{Avenhaus2014} show large, asymmetric structures that could be seen as parts of spiral arms or even rings. Indeed, \citet{Dong2016} studied the effect the inclination and position angle (PA) of the disk has on its observed appearance, and found that a disk with 2 spiral arms might be masked as an asymmetric structure with either only one trailing arm, or two arms on the same side of the star, possibly winding in different directions. The interpretation of inclined disks should, therefore, take into account this masking effect.  

\citet{garufi2018} organized the disks according to their appearance in scattered light and found that bright GI disks show spirals, rings, and bright rims at the edge of a cavity, while other disks (mainly GII) are faint and/or compact and show no such features. They related the disk features with the NIR excess, and found that sources with spirals or shadows have a high NIR excess, while sources with rings show a low NIR excess. Sources with a GII disk have an intermediate NIR excess. The high NIR excess could be attributed to the presence of a massive ($\geq 1 M_{Jup}$) planet perturbing the orientation 
and scale height of the inner disk and finally causing spiral waves.

\citet{bohn2022} compared the inclination of the inner ($<$ 1 au) and outer ($>$ 10 au) disks as derived from VLTI/GRAVITY and ALMA (CO) observations. When the inner and outer disks are misaligned, it is expected that the inner disk will cast a shadow on the outer disk. For those 3 GI sources where misalignment was observed, \citet{bohn2022} predicted the location of the shadow, and found this to agree well with scattered light observations. 

It is surprising to see that the substructures derived from scattered light and thermal emission do not always coincide. For instance, HD~100453 shows spirals in scattered light, but only a ring disk in thermal emission \citep{wagner2015,vanderplas2019}. In general, spirals are more often detected in scattered light than in thermal emission, indicating that the micron-sized grains experience more the effect of density waves in the disk. For example, HD~163296 has an SED that places it at the boundary of GI and GII disks. \citet{muro-arena2018} found that, while the mm continuum data show several rings, scattered light images only trace the innermost ring, leading the authors to conclude that the outer disk is either more depleted of small grains or more settled. However, in multi-epoch HST/STIS images, four  rings were detected \citep{rich2020}, with the fourth ring located at 330 au.  Similarly, \citet{Tang2017} found 2 CO spirals inside the large mm cavity of AB Aur, that were confirmed in scattered light \citep{boccaletti2020}. However, the spiral arms that were detected in CO in the outer disk are not seen in scattered light. 

\bigskip

\noindent \textit{Summary}:
The preferred scenario to form a gap in the disk is a pressure bump, either caused by a planet or a dead zone. While the large particles are trapped, the smaller particles can move inwards with the gas as they are dynamically coupled. Dust filtration might lead to selective accretion of gas rich in volatiles, so that the stellar photosphere is depleted in refractory elements. 

The inner rim is located at the dust evaporation radius and has a puffed-up curved inner wall whose properties depend on the grain size distribution and inclination. 
Once a disk is resolved, it often shows some substructure like rings, spirals or asymmetries. Small micron-sized grains are found both closer to and further from the star than the large mm-sized grains. These distinct locations can be explained by a combination of radial drift (large grains) and dust filtration (small grains). 
At mm wavelengths, some GI disks are very large, while many other GI and GII disks overlap in size; however, not that many GII sources have been observed with sufficient spatial resolution.  
In scattered light (optical/NIR), GI disks are routinely resolved while GII are either faint or undetected. Contrary to GI disks, GII disks are smaller in scattered light than in mm emission, due to self-shadowing. 

To conclude this dust section, the combination of improved spectral coverage in the IR with space-based observatories and sub-arcsecond resolution imaging at optical/NIR and mm wavelengths on the ground has revolutionized our understanding of the dust disks surrounding Herbig stars. It is clear that the IR-mm excess of Herbig stars arises from disks with remarkable structures including gaps, rings, and spiral arms. These disks also have a rich mineralogy indicative of significant dust processing and grain growth. There is strong evidence that the differences between GI and GII disks are tied to the presence of gaps in the disk. GI disks are flaring and have large gaps, while GII disks appear to be compact and/or shadowed. The old evolutionary scenario where disks evolve from GI to GII is not consistent with later studies. Instead, a large GII disk where no or only a small gap is present might open up a larger gap and become a GI disk 
while small GII disks may stay small and gradually lose their content as they evolve into debris disks. Alternative, GI and GII disks may follow independent evolutionary paths. The extent to which these differences reflect differences in the early evolution of the circumstellar disk and envelope or interactions with massive companions remains unclear. Even though the gaps in GI disks are commonly attributed to planets, the influence that stellar and planetary companions have on the evolution of the disk remains an important open question. 

\section{\textsf{GAS IN DISKS AROUND HERBIG STARS}} 
\label{sec:gasdisks}

Early interferometric observations with OVRO of $^{12}$CO $J\!=\!2-1$ 
revealed a rotating disk of gas around the star MWC\,480 \citep{Mannings1997}. 
Several pioneering surveys have since then systematically targeted the intrinsically 
strong CO (sub-)mm lines as key tracers for the disk bulk gas content \citep[e.g.][]{zuckerman95, Dent2005, Hales2014} 
and find that disks around many young Herbig stars contain traceable amounts of cold 
(a few~10~K) gas. Due to the large range of temperatures present --- a couple of 1000~K 
in the inner disk ($<$~few~au) to a few~10~K in the outer disk ($\gtrsim\!100$~au) --- 
the emission of gas is traced over a wide range of wavelengths. The cold outer disk is 
mostly traced at far-IR to mm wavelengths (e.g. low rotational lines of CO), while the 
inner disk is traced by near- to MIR wavelengths. Due to their brightness and warm 
temperatures (stellar luminosities heating the disk range from a few to a few 10~L$_\odot$), 
Herbig disks are generally easier to detect and to spatially resolve compared to T\,Tauri disks. 
This enables studying many processes related to disk evolution and planet formation in more detail. 
The main drawback here is the often small sample sizes and less well constrained ages of 
these objects (Sect.\ref{sec:properties}). In the following, we summarize our current understanding 
of the physics, chemistry and physical processes pertaining to the gas inside these disks 
that we obtained based on multi-wavelengths observations.

\subsection{\textsf{How and when do Herbig disks lose their gas?}}
\label{sec:gasmassevolution}

As summarized in section 4, the classification scheme of Herbigs is largely based on the distribution of dust in the disk, however, gas comprises 99\% of the disk mass. Ideally, the evolutionary state of Herbig disks would be informed by the evolution of the gas mass of the disk as they evolve from gas rich accretion disks to dusty debris disks. 

Several methods using far-IR and sub-mm line emission have been proposed to study disk gas masses directly: 
(1) The CO sub-mm line emission \citep{Thi2001}, (2) the [O\,{\sc i}]\,63~$\mu$m line together with the CO 
sub-mm line \citep{Kamp2011}, (3) HD lines \citep{Bergin2013}, and (4) the CO isotopologue sub-mm lines 
\citep{Williams2014,Miotello2016}. Below, we present an overview of the results from such observational attempts.

\citet{Dent2005} conducted a northern survey of the $^{12}$CO\,$J=3-2$ line in bright Herbig disks using 
the JCMT. Following this up, \citet{Hales2014} carried out an APEX/ASTE survey targeting the $^{12}$CO\,$J=3-2$ line in 52 southern Herbig disks. They report that $\sim\!45$\,\% of the IR bright disks ($L_{\rm IR}/L_\ast\!\geq\!0.01$) 
--- 21 out of 46 in the combined sample --- are not detected in CO, the pre-ALMA, sensitivity limit being 
$\sim\!10^4$\,Jy\,km/s\,pc$^2$. At the faint end of IR excesses this has been continued using ALMA by 
\citet{Moor2017} to detect the $^{12}$CO\,$J=2-1$ line in young A-type debris disks ($10-50$\,Myr); 
they find that CO emission does not correlate with IR excess ($10^4-10^6$\,Jy\,km/s\,pc$^2$ for objects with $L_{\rm IR}/L_\ast\!<\!0.01$). 
\citet{Pericaud2017} did a similar study focusing on debris disks around A-type stars younger than 100\,Myr 
using APEX and IRAM, targeting the $^{12}$CO\,$J=2-1$ and $3-2$ lines. Folding in literature data  on T\,Tauri and Herbig 
disks, they identify a general correlation between CO line flux and mm(sub-mm) continuum flux; however, this correlation is strongly driven by the T Tauri disks. They also find a subclass of disks around A-type stars which have CO line fluxes brighter than the above correlation predicts based on their sub-mm continuum flux and termed them ``hybrid" disks (HIP\,76310, HIP\,84881, HD\,21997, HD\,131835, 49\,Cet, HD\,141569).
Some of these sources show indications of inner dust cavities (a few to several 10\,au) 
\citep[HIP\,76310, HIP\,84881, HD\,21997, HD\,131835,][]{Lieman-Sifry2016, Kospal2013, Hung2015} 
and/or rings \citep[HD\,141569, 49\,Cet,][]{Augereau1999, Wahhaj2007, Biller2015}. This places them into the category of transitional disks (see Sect.~\ref{sect_trans}), but the two classifications (based on either dust or gas observables) do not necessarily agree for all objects.

\citet{Williams2014} proposed to combine the two less abundant isotopologues $^{13}$CO and C$^{18}$O to estimate disk gas masses. This method has been refined using thermo-chemical disk models by \citet{Miotello2016} including isotope selective photodissociation for the regime of TTauri and Herbig disks. This technique  produces in general gas-to-dust mass ratios in disks that are well below the canonical one of 100 \citep[e.g.][]{Ansdell2016,Miotello2017}. Since this result is hard to reconcile with the estimated mass accretion rates, carbon element depletion is often invoked as an explanation.

Thermo-chemical disk models have shown that the [O\,{\sc i}]\,63~$\mu$m emission requires excitation temperatures of several 
10~K and typically originates from inside 300\,au \citep{Kamp2010}. The [O\,{\sc i}]/CO line ratio serves here as 
a temperature proxy and the [O\,{\sc i}]\,63~$\mu$m line emitting region shifts with disk mass, indicating that the combination of the 
line ratio and the [O\,{\sc i}]\,63~$\mu$m line flux could be a reliable tool to estimate the total disk gas mass. \citet{Meeus2012} applied 
this method to the GASPS\footnote{GASPS (Gas Survey of Protoplanetary Systems has been an Open Time Herschel Key 
Program led by Bill Dent \citep{Dent2013}} sample of Herbig disks deriving disk gas mass estimates in the range 
$2.4 \times 10^{-4}$ (HD\,36112) to $2.5 \times 10^{-2}$~M$_\odot$ (HD\,163296), spanning two orders of magnitude.
However, using the ages recently determined from Gaia DR2 \citep{Vioque2018}, there is no clear trend of gas mass 
with age between 5 and 15\,Myr. Likely other disk parameters, such as size play a more dominant role. We 
will come back to this in Sect.~\ref{sec:radialgas}.

The lowest rotational lines of HD ($J\!=\!1-0$ at 112~$\mu$m and $J\!=\!2-1$ at 56~$\mu$m) have not been detected 
with Herschel in the disks around Herbig stars \citep{Kama2020}. Upper limits on the HD $J\!=\!1-0$ line at this 
stage can only rule out very massive disks ($10^{-1}-10^{-2}$~M$_\odot$) around most sources. Interestingly, HD\,163296 
yields an HD upper gas mass limit of $6.7 \times 10^{-2}$~M$_\odot$, compatible with the estimate based on [OI]. 

For HD\,163296, many alternative techniques have been more recently applied to estimate the gas mass 
\citep[full thermo-chemical disk modeling, dust radial drift, the very rare isotopologue 
$^{13}$C$^{17}$O,][]{Rab2020, Woitke2019, Powell2019, Booth2019}, 
all of them now converging on a fairly high disk gas mass of $\sim\!0.2$~M$_\odot$. \citet{Kamp2011} 
have shown from thermo-chemical modeling that optical depth effects turn disk gas mass estimates via 
the [O\,{\sc i}]/CO method into lower limits for gas masses larger than $10^{-3}$~M$_\odot$. So, the 
[O\,{\sc i}]/CO disk gas mass lower estimate is consistent with the more recent studies. The HD upper gas 
mass limit is lower than many of the estimates derived using other techniques. Since the method of \citet{Powell2019} using dust radial drift\footnote{This method uses dust radial drift in a gas-rich disk to estimate gas masses. To first order, the spatial extent of the dust emission at different sub-mm wavelength is affected by the efficiency of dust drift and hence amount of gas in the disk; strictly, this works only for disks that do not show substructure.} is independent of element abundances, the most probable reason for this gas mass discrepancy could be in the gas temperatures underlying the HD estimate. 
The conversion from line flux to gas mass is very sensitive to the gas temperature \citep{Bergin2013} 
and if 2D thermo-chemical disk models are used, the model gas temperature depends strongly on the 
properties of the dust (e.g., grain size distribution, composition, and settling). 

The masses of disks are in tension with the stellar accretion rates around Herbig stars (Sect.~\ref{sec:AccretionRates}). The typical accretion rate of Herbig stars is $\sim$ a few 
$\rm \times 10^{-7}$ \Msun yr$^{-1}$. 
For such rates to be sustained 
for $\sim$1~Myr would require disk masses approaching 0.1M$_{\star}$ - much higher than the typical disk masses inferred from the techniques 
described above. 

\bigskip
\noindent \textit{Summary}: Gas masses in Herbig disks ($5-15$\,Myr) span about three orders of magnitude 
with no clear trend with age. There is a clear correlation between 
the strength of sub-mm gas emission and sub-mm continuum emission. This correlation becomes less clear for older 
objects ($\gg 10$\,Myr), where in several objects significant levels of CO gas are detected while the dust is 
clearly of secondary (debris) origin (termed ``hybrid'' disks or ``gas-rich'' debris disks). 
So, while continuum observations seem to indicate a clear dichotomy between primordial and debris dust masses 
around $\sim\,5-20$\,Myr (see Sect.~\ref{sect_trans}), it is not clear that the gas follows a similar evolution.

\subsection{\textsf{Turbulence in Herbig disks}}
\label{sec:turbulence}

The exquisite sensitivity and spectral resolution of ALMA allows now the precise measurement of the width of spectral lines. If the gas probed by ALMA is turbulent, then the lines should be broader than predicted by the sound speed of the gas. Early pre-ALMA estimates of gas turbulence were limited by spatial and spectral resolution as well 
as the S/N of the channel maps \citep{Dartois2003, Pietu2003, Pietu2005, Pietu2007, Hughes2011}. 
ALMA has significantly improved our ability to estimate the gas turbulence in the outer disk since its high 
quality data sets (bright sources) allow the separation of the front and back side of the disk and so a 
better disentangling of gas temperature and turbulence. Early measurements were strictly applicable 
only to the outer disk (100~au scale), while ALMA can now also probe regions closer to the star. 
\citet{Rosenfeld2013} used ALMA science verification data to demonstrate the high potential of the 
gain in spatial and spectral resolution; they demonstrate the presence of a vertical temperature 
and radial pressure gradient for the disk around HD\,163296. Estimates of gas turbulence in disks 
keep pushing its upper limits to smaller values, typically less than 5-10\% of the sound speed at 
30-few 100~au (see Table~\ref{tab:turbulence} and references therein). This has strong implications 
for the process of angular momentum transport in these disk, since it implies very low $\alpha$ 
viscosity values for the outer disk \citep[$\alpha\!<\!10^{-2}$,][]{Flaherty2020}. 

Another method for inferring supra-thermal line broadening of gas is through spectral synthesis of the CO overtone bandheads \citep{Carr04}. The rotational-vibrational transitions of the R-branch grow closer together with increasing rotational level and eventually pile-up at the blue-end of the spectrum before turning around moving to redder wavelengths. This pile-up is referred to as the bandhead. Because the transitions near the bandhead are very closely spaced, the instrinsic width of the lines affects the opacity of the bandhead (i.e., the shape of the lines determines the opacity of the pseudo continuum). Thus even without spectrally resolving the individual lines comprising the bandhead, the intrinsic line broadening of the lines can be inferred from the emergent flux. Because of the large number of transitions involved, the temperature of the gas is well constrained. Further, the shape of the bandhead is determined by the bulk motion of the gas allowing for the determination of the radial extent of the emission if the stellar mass and disk inclination are known. This method has been used to measure the turbulent velocity of gas in the disk atmosphere within the inner $\sim$1~au of intermediate mass stars such as WL 16 (where the turbulent broadening is twice the sound speed; \citealt{Najita1996}), SVS 13 (1.5-3 times the sound speed; \citealt{Carr04}), V1331 Cyg ($\sim$~the sound speed; \citealt{Najita2009, Doppmann2011}), and HD 101412 ($\lesssim$~the sound speed; \citealt{Adams2019}). However, there is no direct tension with the (sub)mm measurements that probe the turbulence in the disk at much larger distances from the star. 

\begin{table*}[t]
    \caption{Overview of turbulence measurements for Herbig Ae disks.}
    \centering
    \begin{tabu}{lllll}
        \toprule
        object & tracer & method & $v_{\rm turb}$  & reference \\
         &  &  &  [$c_s$] & \\
        \hline
        \multicolumn{5}{c}{inner disk (few au)}\\
        \hline
        V1331\,Cyg & NIR CO bandhead, water & spectrum & $\sim\,1$ & \citet{Najita2009}\\
        HD\,101412 & NIR CO bandhead, water & spectrum & $\sim\,1$ & \citet{Adams2019}\\
        \hline
        \multicolumn{5}{c}{outer disk}\\
        \hline
        HD\,34282 & \multirow[c]{3}{*}{CO sub-mm isotopologues}  & \multirow[c]{3}{*}{visibilities} & \multirow[c]{3}{*}{$<\,1$} & \multirow[c]{3}{*}{\citet{Pietu2003,Pietu2005,Pietu2007}}\\
        AB\,Aur & &  &  & \\
        MWC\,480 & &  &  & \\
        \hline
        HD\,163296 & CO sub-mm & channel maps & $\sim\,0.4$ & \citet{Hughes2011}\\
        HD\,163296 & CO sub-mm isotopologues & channel maps & $<\,0.03$ & \citet{Flaherty2015}\\
        HD\,163296 & DCO$^+$ & channel maps & $<\,0.04$ & \citet{Flaherty2017}\\
        MWC\,480 & CO sub-mm & channel maps & $<\,0.08$ & \citet{Flaherty2020}\\
        \bottomrule
    \end{tabu}
    \label{tab:turbulence}
\end{table*}

There are alternative methods suggested to estimate the gas turbulence in disks around Herbig stars and to understand 
the underlying driving mechanism: smoothed out snow lines \citep[concentration gradients,][]{Owen2014}, 
dust settling \citep[mm-dust grains confined to a geometrically thin midplane, ][]{Pinte2016}, vortices in disks 
\citep[stability,][]{Zhu2014}, planet induced structures \citep[morphology,][]{Dejuanovelar2016}, and outer 
gas disk shape \citep[sharpness,][]{Facchini2017}. SED shapes contain information on dust settling, though 
the problem of fitting them is highly degenerate. \citet{Mulders2012} show that the median SED shapes of 
brown dwarf, T Tauri and Herbig disks are consistent with a low turbulent $\alpha$ value, not depending 
on the (sub-)stellar mass. Comparing ALMA dust and gas observations with 2D hydrodynamical models shows 
that the turbulence affects the sharpness of planet-induced gaps, ring-gap separation, and the gas-to-dust 
mass ratio inside \citep{Lin1993, Long2018, Liu2018}. For HD\,163296, \citet{Liu2018} deduce a smaller 
$\alpha$ in the inner disk ($\sim\!60$~au) compared to the outer disk ($>\!100$~au); however, $\alpha$ 
stays well below $10^{-2}$. 

Given the observed mass accretion rates (Sect.~\ref{sec:AccretionRates}), it remains unclear what the 
underlying driving mechanism is and if the turbulence depends strongly on radial distance from the star. 
Non-ideal MHD effects causing disk winds could offer a solution here \citep{Bai2013, Riols2018}. In their 
review, \citet{hartmann2016} suggested that different mechanisms might cause turbulence in the different 
disk regions (thermally and non-thermally activated MRI, winds, non-ideal MHD effects). 

\bigskip

\noindent \textit{Summary}: ALMA has enabled the measurements on non-thermal broadening of gas lines which place upper limits on the turbulence of the gas in the outer disk at $\lesssim5-10\%$ of the sound speed. For a hand full of objects with CO bandhead emission, high resolution NIR spectroscopy has been used to measure non-thermal broadening of the gas in the innermost disk that is comparable to the sound speed. Other observational evidence points to minimal turbulence in the outer disk including smoothed out snow lines, dust settling, vorticies in disks, planet induced structures, and the sharpness of the boundary of the outer disk. 

\subsection{\textsf{Are Herbig disk surfaces flaring or flat?}}
\label{sec:GasFlaring}

Scattered light observations of small dust grains can trace the shape of the disk surface 
(see Sect.~\ref{duststruct}). Direct observations of gas emission can also 
reveal how the stellar luminosity is reprocessed by the disk and hence provide indirect 
evidence of the degree to which the disk surface is flaring.

\begin{figure}[t]
    \centering
    \includegraphics[width=0.85\textwidth]{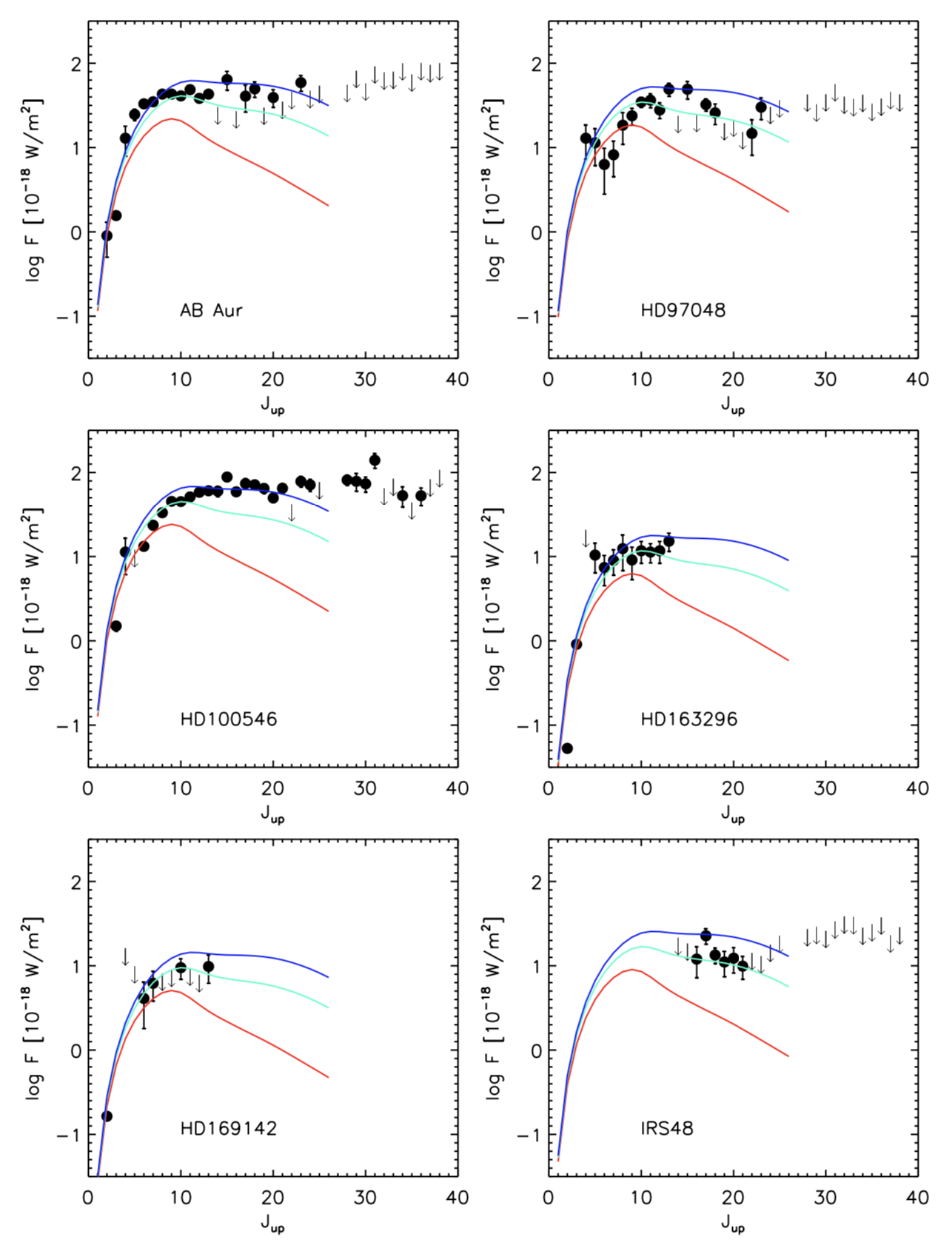}
    \caption{Compiled CO ladders for six Herbig disks. Overplotted are the normalized modelled CO ladders for a typical Herbig disk by \citet{Fedele2016}. The three lines represent disks with different flaring angles, $\beta=1.05$ (red), 1.15 (green), 1.25 (blue). The models have been normalized to the observed J$=\!10-9$ line flux; for IRS48 we used the first detected J line.}
    \label{fig:COladder}
\end{figure}

An example is the Herschel survey of the [O\,{\sc i}]\,63\,$\mu$m line 
\citep[Open Time Key Program GASPS,][]{Dent2013}. Comparing the data with a large grid of 
thermo-chemical disk models \citep{Woitke2010, Kamp2011}, \citet{Pinte2010} show that the 
brightness of this cooling line agrees with a stellar UV heating mechanism. \citet{Meeus2012} 
confirm this by demonstrating a clear correlation between the UV stellar flux and the line 
luminosity for the GASPS sample of 20 Herbig disks. However, they do not find a correlation 
between the SED slope (IR-mm) and the strength of the [O\,{\sc i}] cooling line. In contrast, the CO 
high-J rotational lines are only detected in disks that show a shallower SED slope 
\citep[group I,][]{Meeus2013}. However, Sect.~\ref{sec:dustdisks} clarifies that the SED slope 
is not necessarily a direct reflection of the amount of disk flaring. The CO ladders for the 
few GI Herbig disks that show line detections with sufficient coverage in $J$ stay flat 
between $J\!=\!10$ and 20 (see Fig.~\ref{fig:COladder}). According to thermo-chemical models, 
this is a trend seen in moderate to strongly flaring disks such that $\beta\!\geq \!1.15$, where $\beta$ is the power law exponent of the radial dependence of the gas scale height \citep[i.e., $H(r) \propto r^{\beta}$;][]{Woitke2010}.

The [O\,{\sc i}] fine structure lines react strongly to the amount of disk flaring \citep{Woitke2010}, 
but several other parameters play an important role as well such as inner gaps, dust settling and the 
stellar heating. The CO ladder also reacts to flaring \citep{Bruderer2012, Fedele2016}
--- the ladder drops steeply 
beyond $J=10$ for flat disks due to the altered gas temperature profile --- and here the gas-to-dust 
mass ratio is the most cofounding other variable. The Herbig disks differ substantially in their geometry, making a simple "grid" approach for the interpretation of observational data 
rather difficult. Recently, the DIANA project \citep[DIsc ANAlysis;][]{Woitke2019} modeled six 
of these Herbig disks consistently using multi-wavelength dust and gas observations with a 
unified thermo-chemical modeling approach. The observational data includes the far-IR [O\,{\sc i}] 
and CO rotational lines. For all except one of the disks (AB\,Aur), they find the need for a flared 
outer gas disk $1.19\!\geq\!\beta\!\geq\!1.07$, a range that is well below the maximum Chiang and 
Goldreich solution \citep[$\beta\!=\!9/7$,][]{Chiang1997}.

The geometry of the upper tenuous layers of the inner disk is much better probed using lines that 
predominantly originate there, i.e.\ the forbidden [O\,{\sc i}]\,6300\,\AA \, line 
\citep{Finkenzeller1985, Acke2005, vanderplas2008}. The excitation mechanism for the line can be thermal 
(gas temperature of a few 1000~K or non-thermal \citep[photodissociation of OH,][]{Stoerzer2000}. Due to 
the limited spectral resolution in earlier work, the origin of this line emission remained disputed: infall, 
wind or disk origin. \citet{Acke2005} found that the majority of Herbig disks show narrow symmetric profiles 
(FWHM$\sim\!50$~km/s), with the line profiles observed towards GII sources being on average $\sim\,20$~km/s broader
than the line profiles observed toward GI sources. Also, the detection rate of this line is much lower in GII disks than in GI disks. 
Several of the line profiles display the characteristic double peak suggesting an origin within a rotating 
disk. \citet{vanderplas2008} investigate the radial intensity profiles of the [O\,{\sc i}] line and find 
that it can be related to the inner disk geometry (puffed up regions casting shadows on regions further out). 
Interestingly, HD\,101412 shows two emission peaks, possibly indicating that there is a second 'puffed' up 
region at larger distances (few au); this is a feature seen in hydrostatic disk models 
\citep[e.g.][a second shadow for the tenuous surface layers]{Woitke2009, Min2009, Meijerink2012}. In some cases 
\citep[e.g.\ the Herbig Be star MWC~147,][]{Bagnoli2010}, the line emission could even originate from inside the 
dust sublimation radius. 

\begin{figure}[t]
    \centering
    \vspace*{-10mm}
    \hspace*{-2mm}
    \includegraphics[width=0.75\textwidth]{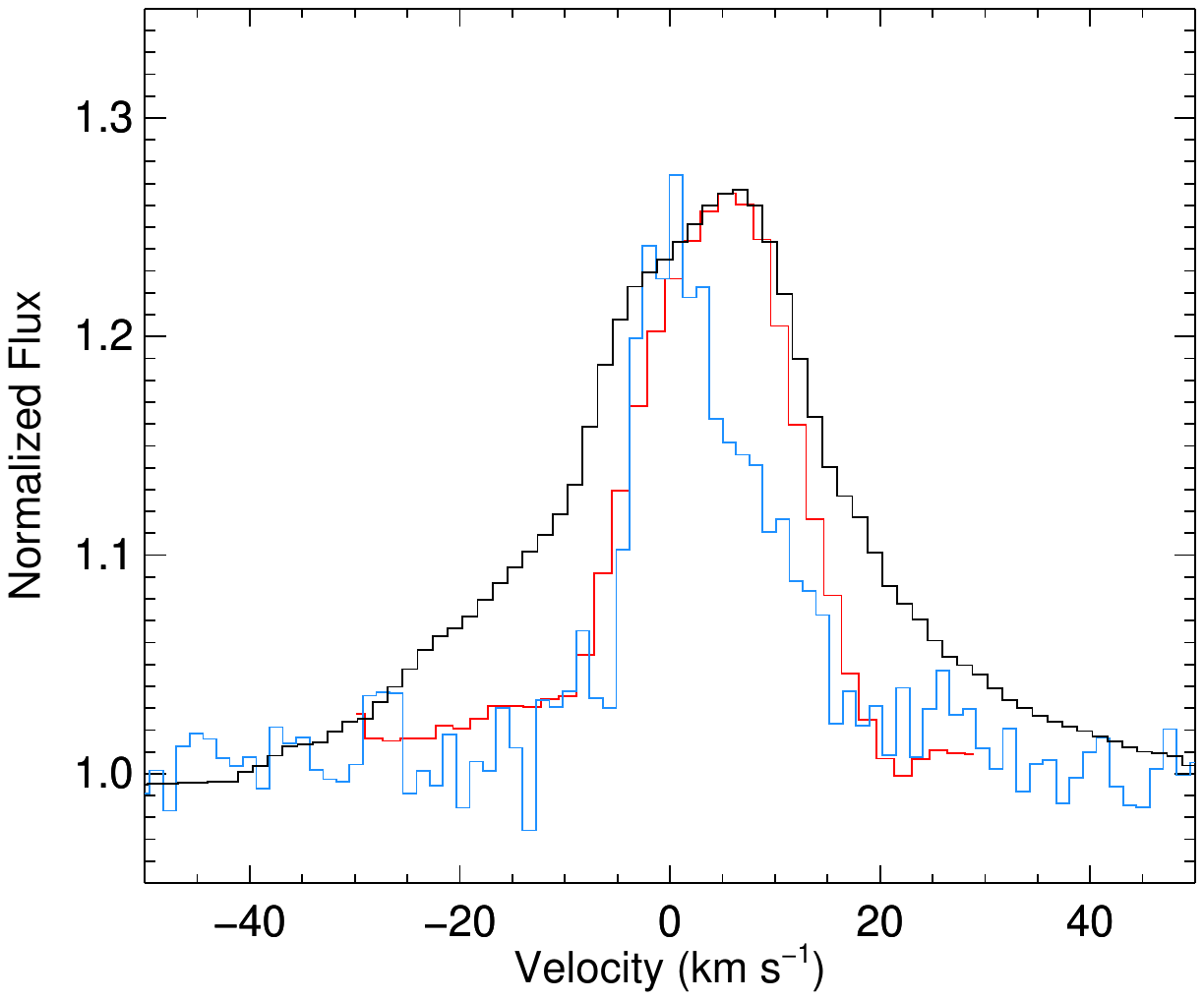}
    \vspace*{-8mm}
    \caption{Comparison of the [O\,{\sc i}]\,6300\,\AA (black), $v=1-0$~OH P4.5 (blue) and the $v=1-0$ P26 CO (red) line profiles from the disk around HD\,100546.}
    \label{fig:HD100546lineprofiles}
\end{figure}

The OH P4.5 doublet (2.934~$\mu$m) was been detected in 4/5 GI disks and 0/6 GII disks by \citet{Fedele2011}, 
sometimes showing very complex profiles suggesting multiple radial emission zones.  
\citet{Brittain2016} detected the OH P4.5 doublet in about 50\% of the Herbig disks 
in their sample; again, the detection rate is much higher for the GI disks (61\%) than GII disks (25\%). 
The line profile of the high J CO line and the OH doublet are consistent (see Fig.~\ref{fig:HD100546lineprofiles}). The fact that OH is more readily observed among 
GI disks indicates that that OH photodissociation could contribute to the excitation of the 
[O\,{\sc i}]\,6300~\AA\ line \citep{Fedele2011}.

Also [Ne\,{\sc ii}] emission at 12.81~$\mu$m is often used to probe jets, outflows and disk winds. 
\citet{BaldovinSaavedra2012} detected this line for the first time in a disk around a Herbig Be star (V892 Tau); 
its centrally peaked narrow line profile is consistent with a photoevaporative wind. Since this line 
requires the presence of X-rays, most of the work so far has been focused on T Tauri disks 
\citep[e.g.][]{Pascucci2011, BaldovinSaavedra2012, Sacco2012}. A more systematic study combining 
high spectral resolution ($3-10$~km/s) optical ([O\,{\sc i}]\,6300~\AA) and IR spectra (OH and CO ro-vib, like shown in Fig.~\ref{fig:HD100546lineprofiles}) 
with dust interferometry could allow a detailed study of the inner disk gas and dust geometry and the 
potential existence of weak disk winds, something that has been piloted by \citet{Fedele2008} for three sources. 

\bigskip

\noindent \textit{Summary}: Gas studies suggest that the outer disks of Herbigs are moderately flaring (below the theoretical maximum of $9/7$). However, the inner disks show more complex geometries required to interpret gas line profiles, such as shadows and disk winds. A systematic study of combined near-IR dust and gas observations for a large sample of Herbig disks would help here to disentangle the processes shaping the inner disks.

\subsection{\textsf{Gas temperatures and the position of icelines}}
\label{sec:GasTemperature}

\begin{figure}[t]
    \hspace*{-4mm}
    \begin{center}
    \includegraphics[width=0.99\textwidth]{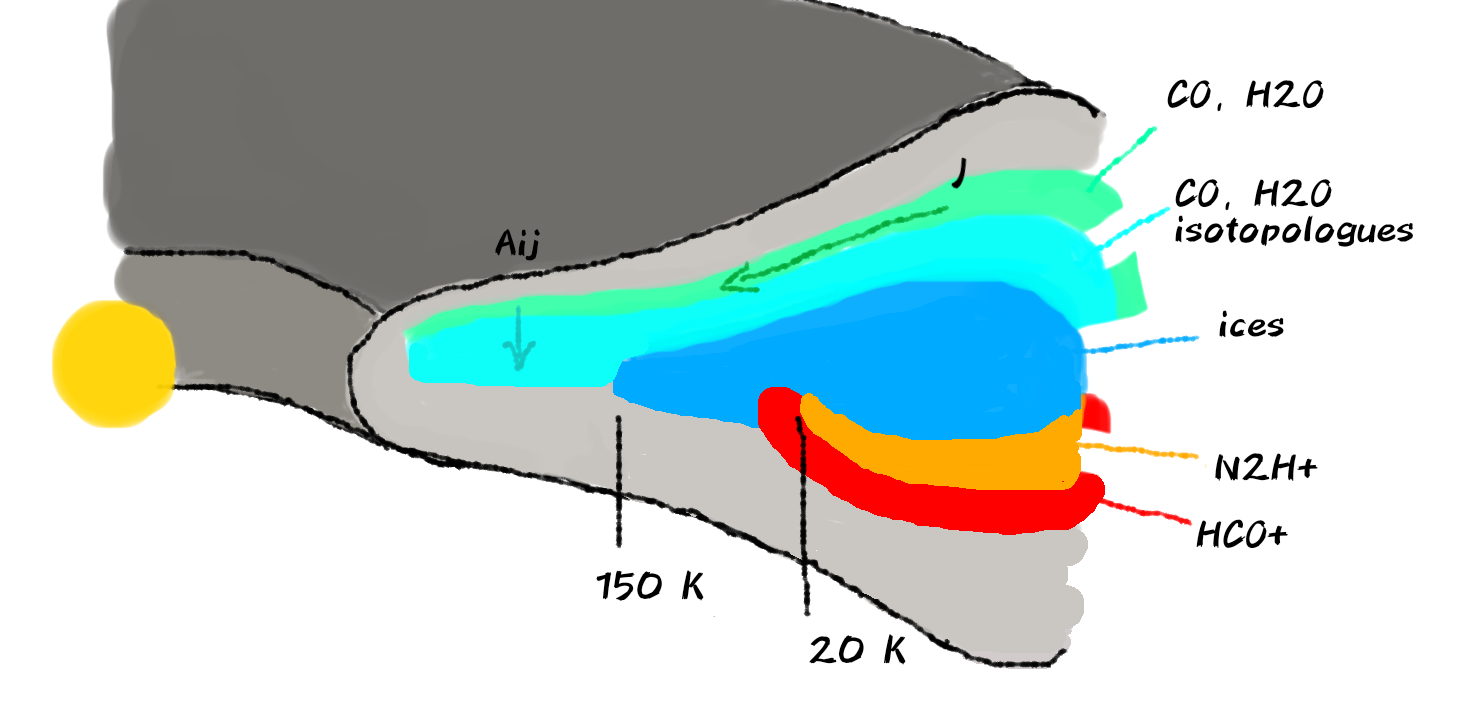}
    \caption{Schematic of various methods to assess the disk gas temperature profiles and icelines (radially and vertically). Indicated are the key molecular tracers used to derive the temperature profiles. Arrows indicate how the line emission region shifts with increasing rotational quantum number $J$ or decreasing Einstein coefficient $A_{ij}$.}
   \label{fig:icelines-methods}
   \end{center}
\end{figure}

The thermal structure of the disks is a key element in understanding planet formation as it determines 
the stability of disks (Toomre~Q parameter\footnote{This parameter depends on distance from the star 
$r$ and is defined as $Q(r)=c_s(r)\Omega/\pi G \Sigma_g(r)$, with $c_s$ the sound speed, $\Omega$ the 
angular velocity, $G$ the gravitational constant and $\Sigma_g$ the disks gas mass surface density.}) 
as well as the composition of the material that is forming the planets (icelines). The midplane of 
Herbig disks is typically optically thick, in the inner disk ($<\!30$~au) often up to sub-mm wavelengths. Figure~\ref{fig:icelines-methods} illustrates how we can 
either use gas emission lines to infer radial and vertical temperature profiles of disks or revert 
to indirect tracers such as the spatial distribution of molecules associated with specific icelines 
(e.g.\ N$_2$H$^+$, DCO$^+$, water).

The temperature profiles for the gas can be derived from molecular emission line studies in several 
ways: (1) interferometric gas line observation (spatial and spectral resolution), (2) from a suite 
of gas line profiles (spectral resolution), (3) from a suite of unresolved emission line fluxes from 
lines with different excitation temperatures (e.g.\ CO ladder). We discuss each of these in the 
following and discuss to which extent the conclusions from them agree and also how they compare to 
the dust temperature studies. However, as a matter of fact, the continuum optical depth at any wavelength 
limits the vertical depth down to which we can probe. For the inner disk ($<\,30$~au), this limits the 
information we can get since these regions are often optically thick even at sub-mm wavelengths. Exceptions are disks where this region has been partly cleared.

\begin{figure}
    \centering
    \includegraphics[width=0.85\textwidth]{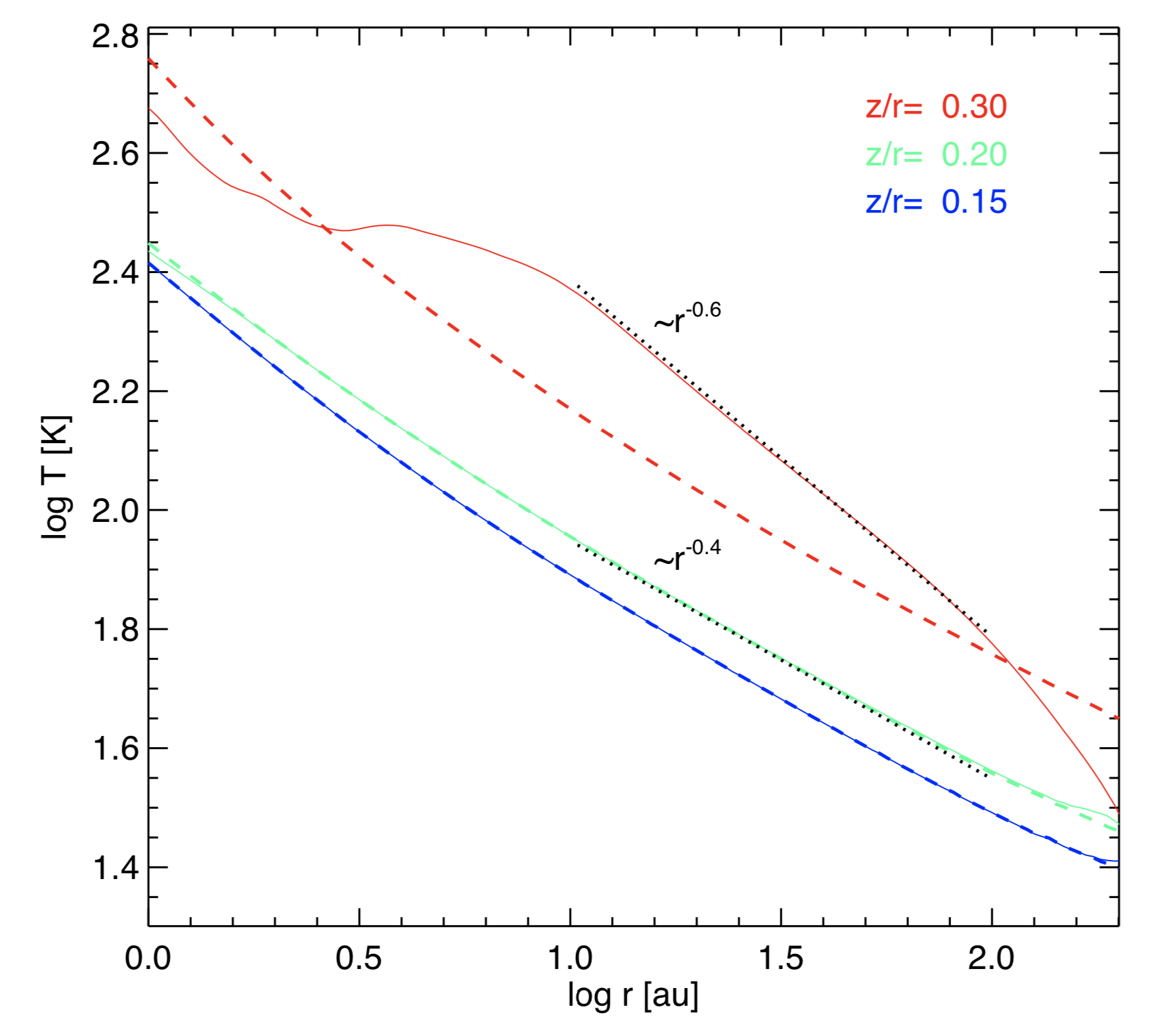}
    \caption{Gas temperature profile in a typical Herbig disk model extracted at three different 
    heights in the disk (solid lines). Dashed lines show the corresponding dust temperature profiles. Black dotted lines show two typical power laws with exponents -0.6 and -0.4.}
    \label{fig:Tgasprofiles}
\end{figure}

\paragraph{Interferometric gas line observations}
\citet{Pietu2007} derived gas temperature profiles using IRAM/PdB interferometric data for a suite of CO 
and HCO$^+$ lines. Fitting visibilities with simple power law disk models ($T\!\propto\!r^q$) for MWC\,480, 
they find gas temperature profiles of the molecular emitting region with power law exponents of $q\!=\!0.65$ 
for $^{12}$CO and shallower ($q\!=\!0.37-0.28$) for $^{13}$CO. They also note a clear vertical temperature 
gradient of several 10~K based on $^{12}$CO $J\!=\!2-1$, $^{13}$CO $J\!=\!2-1$ and $J\!=\!1-0$ lines which 
originate from different heights in the disk. Such values agree quite well with typical thermo-chemical disk models \citep[Fig.~\ref{fig:Tgasprofiles}, showing extracted radial temperature profiles from the MWC\,480 model,][]{Woitke2019}. 
\citet{Flaherty2015} follow a similar approach but use a full 
2D power-law disk model. They find fairly shallow temperature gradients for all CO isotopologues 
($q\!=\!0.216-0.278$) in the disk around HD\,163296, while \citet{Isella2016} find a steeper profile 
($q\!=\!0.6$) for the molecular layer. However, these results depend to some extent on the assumed model 
parametrization; for example \citet{Flaherty2015} assume the same power law exponent for the molecular 
layer and midplane temperatures and we know that the dust distribution in this disk shows prominent rings 
inside 200~au and thus deviates from a simple power law. So, there is a strong push to develop methods that 
are independent of model assumptions. 

Channel maps with high spatial and spectral resolution provide a 
promising alternative. They can be used to reconstruct the temperature profile directly from the CO brightness 
temperature if the dust is optically thin \citep[assuming optically thick line emission,][]{Pinte2018, Dullemond2020}. 
If substantial CO freeze-out occurs, this method measures the temperature of the CO ice surface, otherwise, it 
can be used to estimate the midplane temperature. For HD\,163296, this method finds a very shallow temperature 
gradient ($q=0.14$). There is considerable agreement for the midplane temperature profiles for HD\,163296 with 
different methods (e.g.\ estimated temperatures at 100-150~au differ by less than 3~K), but the underlying 
assumptions need to be evaluated on a case-by-case basis. For example, detailed thermo-chemical models matching 
the DSHARP data (dust and gas) of HD\,163296 show significant freeze-out of CO in the midplane at 100-150~au \citep{Rab2020}.

\paragraph{Gas line profiles}
\citet{Fedele2013} use HD\,100546 to show that the CO line profiles of CO $J\!=\!16-15$, $10-9$ (Herschel/HIFI) 
and $3-2$ (APEX) become successively narrower, thus confirming the shift of their radial origin to increasingly 
larger distance from the star. This method had already been suggested by \citet{Bruderer2012} based on 
thermo-chemical disk modeling. The resulting best fit temperature profile from simple disk power law models has 
a radial slope of $0.85\pm0.1$ (between $\sim\!30$ and 300~au), steeper than that found for MWC\,480 and also 
steeper than what is typically found for the dust (Sect.~\ref{s_diskmodels} and \ref{s_improv_diskmodel}). 

\paragraph{Emission line flux ladder}
In the case of spectrally and spatially unresolved lines, we use the fact that molecular emission lines trace 
the gas temperature in the vertical layer and radial regime over which they emit. Interesting tracers in that 
context are molecules that possess a suite of lines with a large range in excitation energy and Einstein A 
coefficients. The CO ladder, where the excitation energy ($T_{\rm ex}$) increases with the rotational quantum 
number ($T_{\rm ex}(J\!=\!4)\!=55.3\!$~K to $T_{\rm ex}(J\!=\!36)\!=\!3669$~K), scans the disk surface 
with higher rotational lines originating ever closer to the star \citep{vanderWiel2014,Fedele2016}. The same 
holds for the series of OH doublets ($T_{\rm ex}$ ranging between 120 and 875~K). Also water lines span a huge 
range of excitation conditions.
The second dimension, the vertical depth, is probed by studying isotopologues (e.g.\ $^{13}$CO, C$^{17}$O, 
C$^{18}$O, $^{13}$C$^{17}$O), and/or by choosing lines with low Einstein A coefficients \citep[e.g.\ water;][]{Notsu2017}. 
This method to extract radial temperature profiles works particularly well at mid- to far-IR wavelengths and 
has been discussed by \citet{Bruderer2012, vanderWiel2014} based on Herschel/PACS and SPIRE CO ladders. 
\citet{Bruderer2012} demonstrate that the gas temperature in the disk surface (atmosphere) has to be larger 
than that of the dust in order to explain the CO ladder.
This is confirmation of earlier work, both theory \citep{Kamp2004, Jonkheid2004} and observations 
\citep[e.g.\ H$_2$ MIR observations]{Carmona2008}. \citet{Fedele2016} applied simple power law and 
thermo-chemical disk models to FIR data of four Herbig disks. The power law temperature gradients 
derived from simple models fall in the range $0.5-1.0$. The observed CO ladders stay flat in many cases 
out to high rotational quantum numbers ($J\!\sim\!16-20$) \citep[][see Fig.~\ref{fig:COladder}]{Meeus2012, Meeus2013, vanderWiel2014}, 
a behavior that requires a steep temperature gradient in the power law disk models, something that 
can be achieved with a high flaring angle in the thermo-chemical disk models \citep[][]{Fedele2016}. 

\paragraph{Icelines}
More indirectly, specific molecules are associated with specific phase transitions in the disk, 
i.e.\ icelines. Even though they are pressure dependent, they serve as important calibration 
points for the gas temperature profiles. Water at the high pressure conditions in 
the midplane freezes out at temperatures below $\sim\!150$~K, and CO typically freezes out at around 20~K. 
The term `iceline' can be misleading because the phase transition of a molecule does not occur at a single distance from the star. Rather 
it bends towards larger radii as the density decreases towards the disk surface. In addition, non-thermal desorption 
processes such as photo-desorption and cosmic ray (CR) desorption cause the iceline to eventually bend back 
towards the midplane at larger radii where disks become more optically thin to UV radiation 
(stellar and interstellar).

Icelines can be traced directly using the molecule in question and spatially and/or spectrally 
resolving its emission in optically thin lines. Examples of this are $^{13}$CO observations of 
HD\,163296 \citep{Qi2011} putting the CO iceline at 155~au and low Einstein A water lines 
constraining the snowline in HD\,163296 to between 8 and 20~au \citep{Notsu2019}. 
This latter work modeled the water line emission to determine which isotopes 
and transitions are most suited to determine the location of the water snowline. 
Unfortunately, ALMA did not detect any of those selected transitions in HD~163296, 
with 3$\sigma$ upper limits for ortho-H$_2 ^{16}$O at 321 Ghz $< 5.3 \times 10^{-21}$
W~m$^{-2}$, and for para-H$_2 ^{18}$O at 322 GHz $< 8.5 \times 10^{-21}$ W~m$^{-2}$. 
So the water snow line position remains difficult to measure even in the brightest 
Herbig disks.

The water snowline can also be traced indirectly by molecules that correlate with its location. 
\citet{leemker2021A&A...646A...3L} studied the use of 
HCO$^+$ as a chemical tracer of the water snow line. The abundance of 
HCO$^+$ and H$\rm_{2}$O are anti-correlated due to the reaction 
HCO$^+$ + H$\rm_{2}$O $\rightarrow{}$ CO + H$_3$O$^+$. 
However, they concluded that, due to degeneracies complicating the interpretation, 
HCO$^+$ is not a good tracer of the snowline in Herbig disks.

Similarly, N$_2$H$^+$ and DCO$^+$ trace the CO iceline indirectly because both of these molecules are 
tightly connected to the gas phase CO. If CO is present, it will compete with N$_2$ in reactions 
with H$_3^+$, diminishing the production of N$_2$H$^+$; at the same time, N$_2$H$^+$ is also destroyed 
in reactions with CO, leading to the formation of HCO$^+$ \citep[for a detailed discussion see][]{vantHoff2017}. 
A consequence of this is that HCO$^+$ will be abundant just above the CO iceline. The low 
temperatures there ($\sim\!20$~K) are conducive to deuteration, which could result in DCO$^+$ 
to peak in abundance just above/inside of the CO iceline --- a method that has been applied 
earlier to pre-stellar cores \citep{Caselli1999, Pagani2012}. Observing DCO$^+$, \citet{Mathews2013} confirm the 
CO iceline location in HD\,163296 found by \citet{Qi2011}.

\bigskip

\noindent \textit{Summary}: There is convincing evidence that the gas temperatures in the disk atmosphere are higher than 
the dust temperatures. However, our understanding of the 2D temperature structure in Herbig disks 
remains incomplete. The resulting temperature profiles depend strongly on which tracer and method 
is chosen to extract them. HD\,163296 is one of the disks that has been studied using almost every 
available method. A trend emerges from the observations that shows a very steep radial temperature 
gradient for the uppermost layers close to the central star (CO high rotational lines) and a 
flattening of that gradient as one moves to tracers that probe layers closer to the midplane and 
further out (CO isotopologues). This agrees qualitatively well with 2D thermo-chemical disk models. The CO iceline estimate from CO and DCO$^+$ suggest a temperature of 
$\sim\!20$~K at 155~au, which agrees very well with the midplane temperature profiles derived from 
interferometric data. The snowline estimate from the ALMA water line data suggests $\sim\!150$~K at 
8-20~au, which is a higher temperature than what is inferred from the extrapolation of the midplane power law temperature profiles 
from interferometry. Given the intricate dust substructure seen inside 50~au \citep{Andrews2018}, 
such a simple extrapolation is however questionable. The presence of gaps and rings will alter the 
midplane temperature profile \citep[e.g.][]{Pinte2016, Rab2020}.

\subsection{\textsf{The radial distribution of gas}}
\label{sec:radialgas}

In the context of planet formation, it is of immediate interest how the gas mass is distributed 
within the disk, especially in the planet forming regions inside 50~au. For the understanding of 
the mechanisms creating gaps and rings observed in mm-sized dust, it is crucial to know whether 
gas and dust are spatially related (i.e., whether dust gaps also imply gas gaps). A key question is how 
the gas in the inner disk evolves during the planet formation era. Does it follow the dust 
behavior (gap/hole formation) or does it decouple from the dust? Also, the outer edge of the 
gas potentially carries an imprint of either viscous spreading or dynamic encounters truncating 
the disk. 

\begin{figure}
    \centering
    \includegraphics[width=0.85\textwidth]{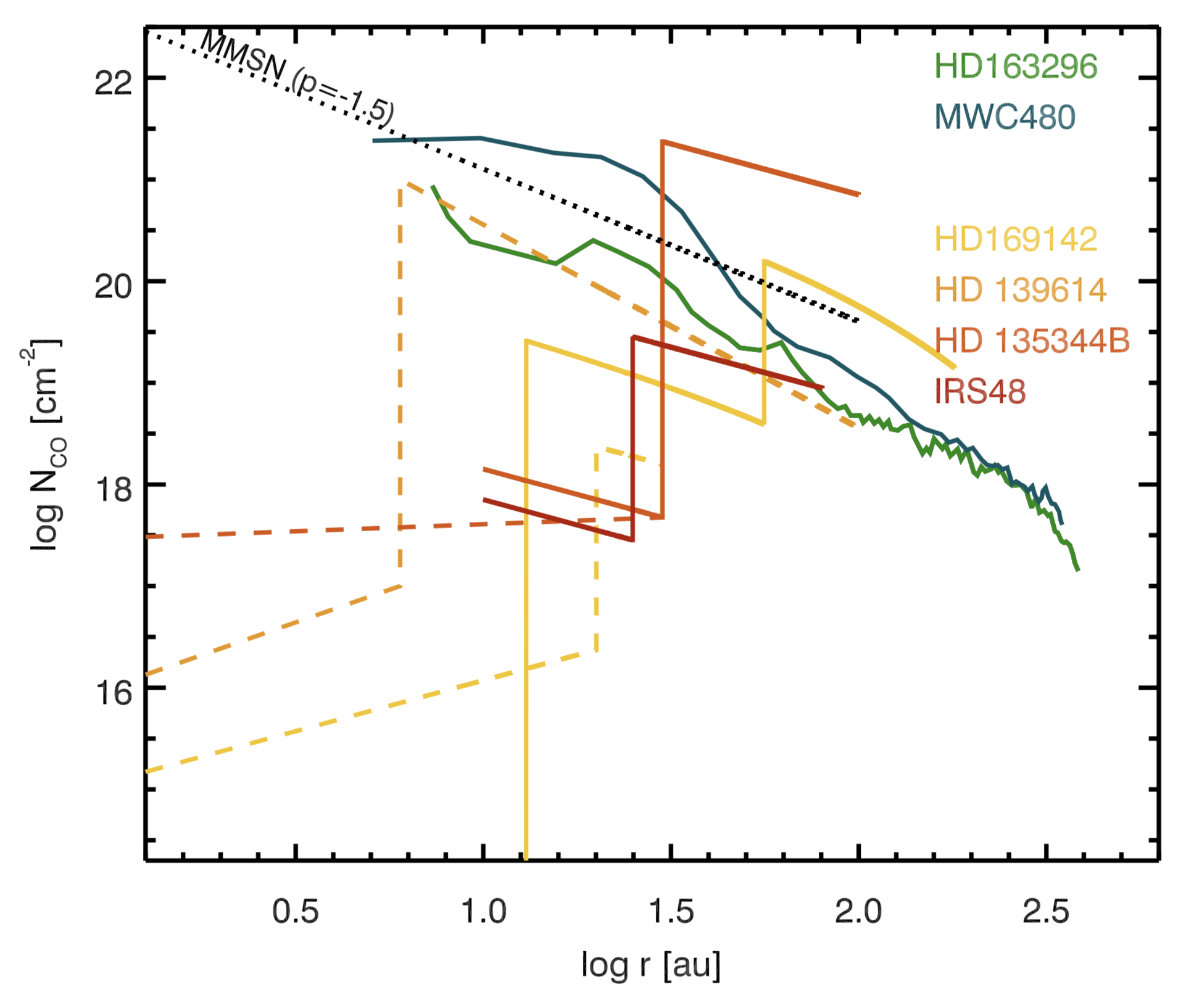}
    \caption{Compiled CO surface density profiles from the literature \citep{Zhang2021,Fedele2017,Carmona2014,Carmona2017,Carmona2021,vanderMarel2016}. We used a constant CO abundance of $10^{-4}$ if required to convert from total gas surface densities. The solid lines are derived from ALMA data, while the dashed lines are from CRIRES data. The black dotted line shows the Minimum Mass Solar Nebula. The blue/green colors indicate normal class {\sc ii} disks, while the yellow/orange/red colors indicate transitional disks.}
    \label{fig:COsurfdens}
\end{figure}

Studies of the CO ro-vibrational lines profiles show that in many cases, the onset of CO emission 
coincides with the dust inner radius \citep{goto2006,Brittain2007a,salyk2009,hein2014,vdp2015,hein2016}.
However, in some cases the authors find evidence for wide CO ro-vib line profiles in GI disks 
which host dust cavities. For example, \citet{salyk2009} 
modeled the CO ro-vib inner emission radius assuming a power law intensity profile and found for LkH$\alpha$330 (IMTTS) 
that the CO gas resides inside the dust cavity (50~au). Several more cases are shown by \citet{hein2016}. In those cases, gas and dust could be spatially de-coupled. 
\citet{banzatti2015} show that the CO gas in systems with large inner cavities (as deduced from the CO 
line profile) is vibrationally `hot'. One explanation of this could be UV fluorescence reaching the 
outer disk because the inner disk is also devoid of dust. Following up on the powerful CO ro-vib line 
diagnostic, \citet{Banzatti2017} showed that water line detections in T Tauri disks (2.9-33~$\mu$m) correlate with gas 
gap sizes deduced from the CO lines. In disks with small gaps, the highest water excitation lines are 
no longer detected and with increasing gap size, these non-detections expand to lower excitation water 
lines. \citet{Antonellini2015} showed that water lines with lower excitation temperatures tend to originate 
from increasingly larger distances from the star. Detailed thermo-chemical disk modeling of CO ro-vibrational 
lines by \citet{Bosman2019} and \citet{Antonellini2020} shows that the observations of Herbig disks may point 
to a more complex inner disk structure, possibly de-coupling of dust and gas and possibly positive density 
gradients consistent with gaps carved by giant planets \citep[e.g.][]{Bryden1999, Lubow2006a}. 
\citet{Carmona2014} used the CO ro-vib line profiles 
combined with a large set of multi-wavelength data (photometry, line fluxes, and images) for the disk around 
HD\,135344B to deduce the shape of the gas surface density profile inside $\sim\!50$~au. From the CO v=1-0 
line profile they find a positive surface density profile and also evidence for dust depletion with respect to 
the canonical gas-to-dust ratio of 100 inside the gap of this pre-transitional disk (GI, see Fig.~\ref{fig:COsurfdens}). Also fitting three CO isotopologues in the 
pre-transitional disk around HD\,139614 indicates a positive surface density profile inside 6~au, and even possibly 
a deep narrow gap that could point to a giant planet around 4~au \citep{Carmona2017}. Using the same method, the 
CO isotopologue lines in HD\,169142 are consistent with a flat or increasing gas surface density profile inside 
$\sim\!20$~au (Carmona et al. 2021; Fig.~\ref{fig:COsurfdens}).

\begin{figure*}[t]
    \centering
    \includegraphics[width=0.99\textwidth]{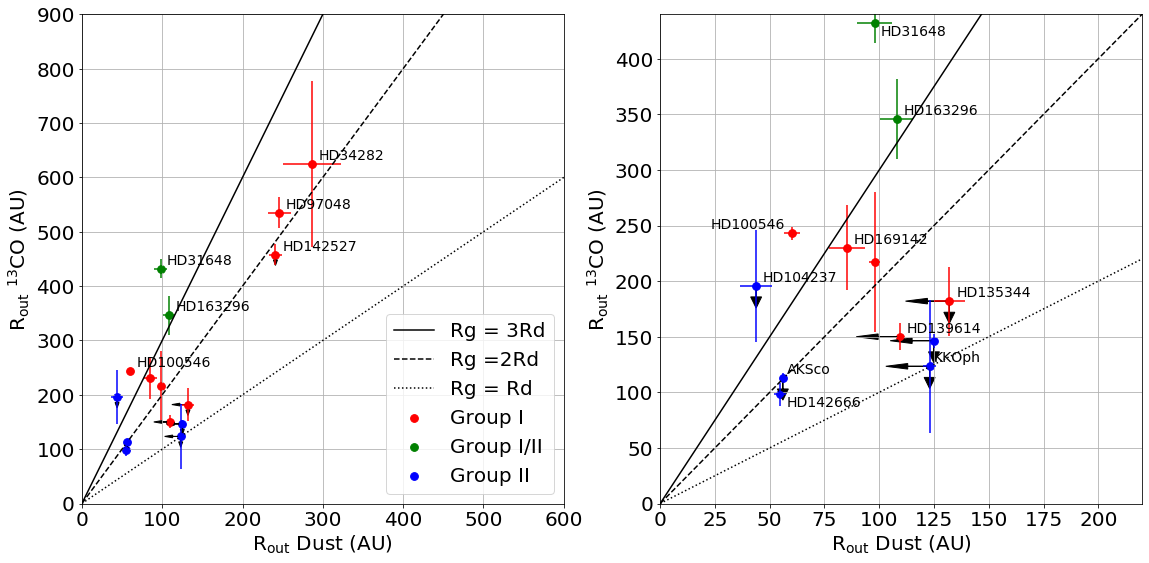}
    \caption{Gas outer radii of Herbig disks derived from ALMA $^{13}$CO archival data versus dust outer radii (Taun, MSc thesis). 
    The right panel zooms in excluding the three largest disks. }
    \label{fig:ALMAdisk-sizes}
\end{figure*}

Given the high spatial resolution that ALMA offers now, the CO pure rotational lines can also be used to probe 
the inner disk regions. The caveat here is that the typical gas temperatures inside 10~au are above a few 100~K 
and so the low rotational levels are not maximally populated. In a pioneering survey of disks with inner dust 
 gaps (pre-transitional disks\footnote{Note that the term transitional disk is here reserved to disks that have no near-IR excess; HD135344B and IRS48 do have such an excess.}), \citet{vanderMarel2016} characterized the depth and shape of such  gaps for 
HD\,135344B and IRS~48. They find that the sizes of gas gaps are typically smaller by a factor two compared 
to the dust gaps and that for HD\,135344B the data at this stage cannot yet distinguish between a smooth 
gas surface density profile with a negative gradient and a step-like positive one (different gas depletion 
levels); notably the CO ro-vib data is more constraining for this object (see above). IRS48 requires a sharp 
edge in the gas surface density profile at $\sim\!25$~au (see Fig.~\ref{fig:COsurfdens}). The DIANA models \citep{Woitke2018} show that 
the observational data are consistent with an outer gas surface density gradient smaller than $-1.5$ 
\citep[Minimum Mass Solar Nebula,][]{Weidenschilling1977}, a result previously also reported by 
\citet{Williams2011}. 

Concerning the outer disk, the DSHARP program \citep{Andrews2018} obtained high spatial resolution ALMA data (30-60~mas) for a few Herbig Ae disks, HD\,163296, HD\,142666 and the IMTTS HD\,143006. From this program, only one source, HD\,163296, has a high quality dataset that allows the determination of a detailed gas surface density 
profile (see Fig.~\ref{fig:COsurfdens}). Since the dust rings are not optically thin, derivation of the gas surface density profile requires 
simultaneous modeling of dust and gas. While \citet{Isella2016} inferred the presence of gas gaps for the inner two dust gaps at 60 and 100~au from power law (plus gaps) disk modeling of ALMA CO data, more detailed 
thermo-chemical dust+gas modeling using the DSHARP data does not confirm this \citep{Rab2020}. The observations 
of the high S/N high spatial resolution $^{12}$CO data (DSHARP) are still consistent with no gas depletion 
inside the dust gaps. 

In a sample of 22 disks from a pre-ALMA northern survey of the $^{12}$CO~$J=3-2$ line \citep{Dent2005}, \citet{Panic2009} find from disk modeling (power-law disk models without chemistry) that 75\,\% of this Herbig sample are smaller than 200\,au; of course, pre-ALMA data is likely biased towards large/bright disks and could underestimate this percentage. Unfortunately, in the epoch of ALMA, there does not exist a homogeneous large survey for disk sizes at the time of this writing. 
Taun et al.\ (MSc thesis, paper in preparation)
collected 
archival data of Herbig disks in band 6 (all three CO isotopologues) with spatial resolution of 20-100~au \citep[e.g. data from][]{miley2019}. They 
added Herbig disks observed only in band 7 ($^{13}$CO and C$^{18}$O) to enhance the sample. This led to 17 Herbig 
disks, ten GI, five GII and two intermediate disks (Fig. \ref{fig:ALMAdisk-sizes}). The dust and gas radii were derived using the cumulative 
fluxes (from elliptical apertures) and defining the outer radius when 90\% of the total flux was reached 
\citep[see][for details of the method]{Trapman2019}. In general, GI disks span a large range in 
dust (90$\sim$370 AU) and gas ($\rm 180\!\sim\!850\, AU$) outer disk radii while the five GII disks are small, with four 
unresolved dust disks and three unresolved gas disks ($^{13}$CO). The GI/II disks have similar sub-mm properties 
(gas outer radii, continuum and CO fluxes) to GI disks. Like in the case of the lower mass counterparts 
(T Tauri disks), the average ratio between gas and dust outer radii is $\sim\!2$ with a large spread \citep[e.g.][]{Ansdell2018}. \citet{Facchini2017} and \citet{Trapman2019} find that radial drift of dust grains has a profound effect on the outer disk thermal gas structure and the ratio between inferred dust and gas radii; specifically, radial differences of a factor four or more are clear signatures of dust evolution and radial drift.

Individual disks have been investigated using high spatial resolution ALMA data and find similar results: a factor 2.4 for 
HD\,97048 \citep{Vanderplas2017_HD97048}, more than a factor 2 for HD\,163296, \citep{Degregoriomonsalvo2013}, and more than 
a factor 5.8 for HD\,100546 \citep{Pineda2014}. For the disks around HD\,163296 and HD\,100546, the authors claim that the 
dust disk has a sharper cut-off compared to the gas, possibly indicating signatures of grain growth and radial dust migration. 
The DIANA disk modeling results show that for five Herbig disks, the outer disk edge is consistent with a soft edge 
($\gamma\!=0.5-1.0$ 
\footnote{The surface density is here assumed as $\Sigma (r) \propto r^{- \epsilon} \exp \left( -\left(r \over R\rm_{tap} \right)^{2-\gamma} \right)$ with the radius $r$ and the taper radius $R\rm _{tap}$.}
), and only for one object, 
the sharp edge previously reported is recovered 
\citep[HD\,163296, $\gamma\!=\!0.2$][]{Woitke2019}. 
\citet{Panic2021} investigated two binaries of intermediate separation (HD\,144668 and KK\,Oph) where both components are 
surrounded by dust-rich (gas-to-dust mass ratio $\leqslant \!2.4$) planet forming disks. The respective disk sizes are consistent with tidal truncation in these systems 
and they appear to have lower gas masses ($<\!0.1$~M$_{\rm Jupiter}$) than the average single Herbig disks; so it is unlikely that gas giant planets can still form around these two stars. 

\paragraph{Summary}
Based on NIR gas observations of Herbig disks with dust cavities (up to several 10~au), there is evidence for 
positive surface density gradients inside those cavities. In some cases, the gas clearly extends inside the dust cavity 
and in most cases, the gas-to-dust ratio is high inside the cavities. Broadly speaking, this is consistent with giant 
planet formation models. The gas surface density gradient in the outer disk (beyond several 10~au) is often more 
shallow than the minimum mass solar nebula (-1.5, see Fig.~\ref{fig:COsurfdens}). We have in a few cases high spatial resolution ALMA observations 
indicating that growth and radial migration of dust affect the outer edges of disks around Herbig Ae stars. Intermediate 
separation binaries also suggest that tidal truncation operates, maybe in tandem with promoting more efficient dust evolution.

\subsection{\textsf{Chemistry in Herbig disks}}
\label{sec:chemistry}

To understand what type of planetary systems can emerge from the disks around Herbig stars, we also need to address the question of how chemically rich these disks are. (1) Which level of molecular complexity do we find 
compared to disks around low mass stars? (2) Are the differences primarily due to differences in the physical and thermal disk structure (e.g.\ level of UV and X-ray irradiation) or do they reflect true differences in the gas chemistry? And, most fundamentally, (3) do Herbig disks contain significant a\-mounts of water, a tracer often  invoked to assess the possibility of forming habitable planets. (4) What is the C/O ratio in Herbig disks?

\begin{table*}[t]
    \caption{Detected molecules (transitions) in Herbig disks with ALMA beyond CO. References: \citet[][Bo19]{Bosman2019}, \citet[][vdP17]{Vanderplas2017_HD97048}, \citet[][A21]{Aikawa2021}, \citet[][B19]{Booth2019}, \citet[][B21]{Bergner2021}, \citet[][C18]{Carney2018}, \citet[][C21]{Cataldi2021}, \citet[][F17]{Fedele2017}, \citet[][G21]{Guzman2021}, \citet[][H17]{Huang2017}, \citet[][I21]{Ilee2021}, \citet[][L20]{Loomis2020}, \citet[][LG21]{LeGal2021}, \citet[][M13]{Mathews2013}, \citet[][M17]{Macias2017}, \citet[][O15]{Oeberg2015}, \citet[][P20]{Pegues2020}, \citet[][Q15]{Qi2015}, \citet[][Z21]{Zhang2021}.}
    \centering
    \begin{tabu}{ll|lllll}
    \toprule
   line & ident & HD\,97048  & MWC\,480 & HD\,163296 & HD\,143006 & HD\,169142 \\
    \hline
$^{12}$CO       & J=1-0             & vdP17      &          &             &     & \\
$^{13}$CO       & J=1-0             &            & Z21      & Z21         &     & \\
C$^{18}$O       & J=1-0             &            & Z21      & Z21         &     & \\
C$^{17}$O       & J=1-0             &            & Z21      & Z21         &     & \\
$^{12}$CO       & J=2-1             &            & Z21      & Z21         &     & F17 \\
$^{13}$CO       & J=2-1             &            & Z21      & Z21         &     & F17 \\
C$^{18}$O       & J=2-1             &            & B19,Z21  & Q15,B19,Z21 & B19 & F17 \\
$^{12}$CO       & J=3-2             & vdP17,Bo19 &          &             &     &     \\
HCO$^+$         & J=1-0             &            & A21      & A21         &     & \\
HCO$^+$         & J=4-3             & vdP17,Bo19 &          & M13         &     & \\
DCO$^+$         & J=3-2             &            &          &             &     & M17,C18 \\
DCO$^+$         & J=5-4             &            &          & M13         &     & M17,C18 \\
H$^{13}$CO$^+$  & J=1-0             &            & A21      & A21         &     & \\
H$^{13}$CO$^+$  & J=3-2             &            & H17      & H17         &     & \\
H$^{13}$CO$^+$  & J=4-3             & Bo19       &          & M13         &     & \\
H$_2$CO         & J=3-2             &            & P20,G21  & P20,G21     & P20 & \\ 
H$_2$CO         & J=4-3             &            & P20      & P20         &     & \\ 
\hline
HCN             & J=1-0             &            & G21      & G21         &     & \\
HCN             & J=3-2             &            & B19,G21  & B19,G21     & B19 & \\ 
H$^{13}$CN      & J=3-2             &            & O15,H17  & H17         &     & \\ 
HC$^{15}$N      & J=4-3             & Bo19       &          &             &     & \\ 
DCN             & J=3-2             &            & C21      & C21         &     & \\ 
CN              & N=1-0             &            & B21      & B21         &     & \\ 
CN              & N=2-1             &            &          & B21         &     & \\ 
N$_2$H$^+$      & J=3-2             &            & L20      & Q15         &     & \\ 
N$_2$D$^+$      & J=3-2             &            & C21      & C21         &     & \\ 
HC$_3$N         & 29-28             &            & I21      & I21         &     & \\ 
CH$_3$CN        & 12$_0$-11$_0$     &            & I21      & I21         &     & \\
\hline
C$_2$H          & N=3-2             &            & B19,G21  & B19,G21     &     & \\ 
C$_2$H          & N=1-0             &            & G21      & G21         &     & \\ 
c-C$_3$H$_2$    & 7$_{07}$-6$_{16}$ &            & I21      & I21         &     & \\ 
SO$_2$          &                   &            & LG21     &             &     & \\
$^{12}$CS       & J=2-1             &            & LG21     & LG21        &     & \\ 
$^{12}$CS       & J=5-4             &            & LG21     &             &     & \\ 
$^{12}$CS       & J=6-5             &            & L20      &             &     & \\ 
C$^{34}$S       & J=5-4             &            & LG21     &             &     & \\
H$_2$CS         &                   &            & L20,LG21 &             &     & \\
\bottomrule
    \end{tabu}
    \label{tab:ALMA-molecules-Herbigs}
\end{table*}

\paragraph{Level of molecular complexity:} A systematic SMA survey by \citet{Oeberg2010,Oeberg2011} finds a lack of cold chemistry tracers in Herbig disks, i.e.\ N$_2$H$^+$, DCO$^+$, DCN, H$_2$CO, compared to T\,Tauri disks. These early studies focused on the brightest (and so largest) disks in each  category. With ALMA's increasing spatial resolution and sensitivity, the studies of the chemical composition of Herbig disks has gained momentum. \citet{Oeberg2015} showed that the bright Herbig disk around MWC\,480 hosts a number of complex nitriles (HC$_3$N and CH$_3$CN) next to the commonly detected HCN and its isotopologue. Overall, the number of Herbig disks in which ALMA has searched for molecules is still too small and the data is too inhomogeneous (spatial resolution, transitions) to have conclusive answers as to how different Herbig disks are amongst themselves (see Table~\ref{tab:ALMA-molecules-Herbigs} for sources that have molecular detections beyond the CO molecule); this is not changing with the MAPS survey \citep{Oeberg2021}, since it includes only two of the already well characterized Herbig disks MWC\,480 and HD\,163296. In the inner disk, Spitzer surveys have shown that Herbig disks also lack the richness in molecular lines typically found in T\,Tauri disks \citep{Pontoppidan2010,salyk2011}. The Herbig disks show for example no detections of HCN, C$_2$H$_2$, CO$_2$ and OH; only HD\,101412 is detected in CO$_2$. Warm water ($\sim\!300-700$~K) is only detected at longer wavelengths (e.g.\ $29.85~\mu$m, \citealt{Pontoppidan2010}) among several Herbig disks (4/25). HD\,163296 is also clearly detected at 33\,$\mu$m \citep{Banzatti2017}. Given the low spectral resolution of Spitzer, this can be a result of limited sensitivity since models predict that the line flux scales weaker than linear with the central star luminosity \citep{Antonellini2016}. 

\paragraph{Origin of differences between T\,Tauri and Herbig disks:} Single dish sub-mm observations focused initially on detecting molecules in the cold outer disk and analysing integrated line ratios. \citet{Thi2004} presented a first comparative study of the sub-mm lines of CO, HCO$^+$, CN, HCN, and H$_2$CO in T\,Tauri stars (LkCa\,15 and TW\,Hya) and Herbig stars (HD\,163296 and MWC\,480); they found that the CN/HCN ratio is higher in Herbig disks than in T\,Tauri disks and attributed this to either differences in the radiation field (Ly\,$\alpha$ or X-rays) or a thermal effect (the two molecules have different freeze-out temperatures). Using the PdB Interferometer, \citet{Henning2010} found that C$_2$H emission is lower in the Herbig disk around MWC\,480 than in the T\,Tauri disks around DM\,Tau and LkCa\,15; again, they explained this by the presence of strong UV and lack of X-rays in the case of the Herbig disk. The SMA survey by \citet{Oeberg2010,Oeberg2011} shows no clear difference in the CN/HCN ratio between T\,Tauri and Herbig disks. Subsequent large surveys of T\,Tauri and Herbig disks using IRAM also show no systematic difference in CN emission between the two type of sources \citep{Chapillon2012,Guilloteau2013}. However, MWC\,480 and HD\,163296 show strong CN emission, while AB\,Aur, MWC\,758, CQ\,Tau and SU\,Aur show only upper limits or weak emission. Notably the strong CN disks are both intermediate GI/GII disks; also Table~\ref{tab:ALMA-molecules-Herbigs} shows that these are also to-date the two best studied disks. Contrary to the above studies, \citet{Chapillon2012} find that the CN/HCN line ratio is higher in Herbig disks compared to T\,Tauri disks --- however, the sample for which both lines have been observed with IRAM is small. 

A comprehensive ALMA line survey of LkCa\,15 (a T\,Tauri star) and MWC\,480 finds clear differences in the cold chemistry tracers N$_2$H$^+$, and DCN, and the nitriles HC$_3$N, CH$_3$CN. The latter are brighter in the Herbig disk, maybe due to its instrinsically higher temperatures \citep{Loomis2020}. \citet{LeGal2019} focused on the sulphur bearing molecules in disks including again MWC\,480. They find no difference in CS between T\,Tauris and this Herbig disk. On the other hand, a much larger sample studied by \citet{Pegues2020} shows that H$_2$CO (formaldehyde) is less abundant in Herbig disks compared to T\,Tauri disks. The observations suggest that both chemical channels, gas-phase and grain-surface, likely operate to explain the formaldehyde observations in disks; hydrogenation of CO ice can lead to efficient H$_2$CO formation in cold T\,Tauri disks. The general idea is that formaldehyde can be further hydrogenated to methanol \citep[e.g.][]{Hiraoka1994,Watanabe2002} and simulations support that this works under the conditions found in cold dense cores or cold YSO envelopes \citep{Cuppen2009}. \citet{Carney2019} carried out a deep search for methanol in HD\,163296 with no detection. Compared to TW\,Hya, the methanol to formaldehyde ratio is much lower. However, still multiple explanations are possible given that we still have an incomplete picture of the role of surface chemistry versus gas phase chemistry and also given uncertainties in the desorption efficiencies of these two molecules. 

The cold chemistry in the outer disk also drives deuteration of molecules. \citet{Huang2017} find no significant difference in the deuteration between the T\,Tauri and Herbig disks (MWC\,480 and HD\,163296). This is surprising given the earlier finding of a comparative lack/lower emission level of cold chemistry tracers in the outer disk of Herbig stars. 

Disk modeling studies \citep{vanZadelhoff2003,Cazzoletti2018} indeed suggest that higher FUV irradiation enhances the CN emission. \citet{Walsh2015,Antonellini2016,Agundez2018} used thermo-chemical models to investigate the difference in chemical composition between disks of various spectral types. Using the same generic disk structure and exchanging the central star, \citet{Agundez2018} do not find striking differences in the outer disk molecular reservoir between T\,Tauri and Herbig disks. \cite{Walsh2015} use the same approach and focus on the inner 10~au. They do not find significant differences between Herbig and T\,Tauri disks in key molecules such as HCN and C$_2$H$_2$; if at all, the Herbig disks have higher column densities. It is important to note that these works assumed high abundances of relatively  small grains in the disk surface (up to 1~$\mu$m size). This causes the H/H$_2$ transition to reside very high in the disk at warm temperatures (several 100~K), thus promoting a very efficient neutral chemistry \citep{Walsh2015}. This is not the case in disk models that assume a wider grain size distribution (up to mm-size) and dust settling; the small  grains are then indeed left in the disk surface, but their fractional mass is much smaller, preventing high abundances  of H$_2$O and C$_2$H$_2$ above $\tau\!\sim\!1$ in the disk surface \citep{Woitke2018, Greenwood2019}. 

\paragraph{Water in Herbig disks:} The lowest excitation water lines from the outer disk have been detected with Herschel/HIFI in one out of four Herbig stars \citep[HD\,100546 detection versus HD\,163296, MWC\,758, MWC\,480 non detections][]{vanDishoeck2021}. Also, water ice has been unambigiously detected in the Herbig disks HD\,142527 \citep{Min2016a} and HD\,100546 \citep{Honda2016}, see Sect.~\ref{sect_ice}. 

Moving slowly to warmer disk regions closer to the star, warm water ($\sim\!200-300$~K) has been detected in the far-IR with Herschel in HD\,163296 \citep{Meeus2012}, and through line stacking also in HD\,104237 and HD\,142527 \citep{Fedele2013}. Contrary to the fundamental water lines, HD\,100546 is not detected in warm water with PACS; the original claim by \citet{sturm2010} has been retracted \citep{Meeus2012}. Also warm OH ($\sim\!100-500$~K) has been detected in disks, both in GI and GII \citep{Meeus2012,Fedele2012,Fedele2013}. Higher OH/H$_2$O abundance ratios are found in Herbig disks compared to T\,Tauri disks based on slab models that assume the same spatial origin for both molecules. However, thermo-chemical models of T\,Tauri disks show that these lines originate from very different radial and vertical disk layers \citep{Woitke2018, Greenwood2019}.  

At NIR wavelengths, ground based studies have been used to search for hot water (few 1000~K) and OH in the inner disks around Herbig stars \citep{Fedele2011,Brittain2016,Adams2019}. The NIR detections of OH are more common among GI disks \citep{Brittain2016} in contrast to the far-IR OH detections. This could relate to the inner disk architecture where dust cavities/gaps (more common among GI disks) create better excitation conditions, e.g.\ at the inner edge of the outer disk. \citet{Brittain2016} find that the high J CO and OH P4.5 doublet line ratio is roughly constant ($\sim\!10$) in Herbig disks; T Tauri disks show a lower ratio between the CO P10 line and the OH doublet at 2.9\,$\mu$m \citep{Banzatti2017}. Subsequently, \citet{Adams2019} found that the H$_2$O to OH line ratio in Herbig disks \citep[based on HD\,101412 and upper limits from][]{Fedele2011} is systematically lower than for T Tauri disks \citep[][]{Banzatti2017}. 

\paragraph{The C/O ratio in Herbig disks:}
A key controversy at this stage is the global C/O ratio in disks, because it relates very closely to the composition of gas giant planets and planetary atmospheres forming within these disks. \citet{Bruderer2012, Kama2016a, Kama2016b} use a combination of atomic fine structure lines ([O\,{\sc i})], [C\,{\sc i}], [C\,{\sc ii}]) and CO sub-mm lines to investigate the carbon depletion in the upper layers of the outer disk; the [C\,{\sc i}] data originates from APEX surveys. They find that the volatile carbon abundance could be a factor $5-20$ lower than the solar one \citep[$2.7\times 10^{-4}$ relative to hydrogen,][]{Asplund2009}; however to derive robust conclusions, [C\,{\sc i}] detections are essential, thus warranting the sensitivity of ALMA. The recent MAPS project found no difference in the amount of CO depletion in three T\,Tauri and two Herbig disks \citep{Zhang2021}. Subsequent 2D thermo-chemical disk modeling of the five disks \citep{Bosman2021} shows that the C/O ratio varies strongly within each disk, irrespective of spectral type.

\bigskip

\noindent \textit{Summary}: Surveys of Herbig disks have shown less chemical richness compared to their lower mass counterparts, the T\,Tauri disks. The to-date richest Herbig disks are the two GI/GII disks HD\,163296 and MWC\,480; however, this could be a selection bias since few comprehensive Herbig line surveys have been done with ALMA. Despite thermo-chemical modeling, it remains unclear the extent to which this is due to differences in disk structure or chemistry. A systematic survey of OH, water and CO would again be key (see also Sect.~\ref{sec:GasFlaring}) in disentangling the excitation and abundance question for these molecules and understanding the differences between T\,Tauri and Herbig disks. If elemental carbon depletion is invoked to explain the weak CO lines in the submm, factors of up to several 10 (relative to solar) are found in Herbig disks; the C/O ratio could be as high as 2.

\subsection{\textsf{Differences between Herbig Ae and Be disks}}
\label{sec:disk-diff-AandB}

Most of the discussion above pertains to disks around Herbig Ae stars. Their B-type counterparts have much higher 
luminosities and they often reside in more distant star forming regions (see also Sect.~\ref{subsec:herbigbe}). The larger distance and their embedded nature makes them difficult targets for directly resolving their disks. However, they have been studied spectroscopically from optical to far-IR wavelengths.

\citet{Bik2004} detected the CO bandhead emission in a young B-type star (IRAS08576-4334, $M_\ast\!=\!6$~M$_\odot$) and 
showed that it is fully consistent with originating from a small (few au sized) disk in Keplerian rotation. Subsequent 
X-shooter and SINFONI data \citep{Ellerbroek2011} suggest that the system also features a jet and would thus be still 
accreting with rates of the order of $10^{-5}-10^{-6}$~M$_\odot$/yr. \citet{Ramirez-Tannus2017} classified and 
investigated young stellar objects in M17 ($d\!=\!1.98$~kpc) and found five of them having a spectral type B 
(B243, B268, B275, B331, B337) and at least two of the attributes: NIR excess, CO bandhead emission, double-peaked emission 
lines of hydrogen, the Ca~triplet, [O\,{\sc i}]\,6300\AA; B275 had been previously confirmed to be a young pre-main sequence 
B-type star by \citet{ochsendorf2011}. A detailed analysis of the line profiles shows that the gaseous disks are again very 
small (0.5-5~au scale) and that the hydrogen emission originates from further out compared to [O\,{\sc i}] and the Ca~triplet. 
\citet{Ilee2013, Ilee2014} detected the CO overtone emission at 2.3~$\mu$m with CRIRES and X-shooter in a sample of 6 out of 91 
Herbig disks (4 Be disks and 2 Ae disks); the detection rate is higher among the Be disks. Fitting the CO overtone line profile 
again suggests an origin within a small gaseous disk ($<$~few au). For the B-type stars, the CO emitting region is well inside 
the estimated dust sublimation radius, but well beyond the corotation radius. So, the picture that emerges is that these Be-type 
stars have a NIR excess and host small gaseous disks, likely inside the dust sublimation radius. 

At far-IR wavelength, \citet{Jimenez-Donaire2017} compared the Herschel far-IR spectra of two Herbig Be stars (R\,Mon with 
spectral type B8, 0.8~kpc, and PDS\,27 with spectral type B2, 1.25~kpc). While the spectrum of the B8 type star is rich in 
emission lines of [O\,{\sc i}], H$_2$O, OH and CO up to $J\!=\!34-33$, the B2 star barely shows any emission lines, besides 
CO up to $J\!=\!11-10$. However, due to the limited spatial and spectral resolution, the disk and outflow (shock) contribution 
to these lines cannot be disentangled. Further observations at far-IR and sub-mm wavelengths (either with high spatial 
or spectral resolution), are required to find out whether the more massive HBe stars possess outer gaseous disks 
similar to the Herbig Ae stars or not. 

\bigskip

\noindent \textit{Summary:} Over the past 25 years it has become clear that Herbig stars possess rotationally supported gas disks. Spectroscopic capabilities across the electromagnetic spectrum have revealed the presence of many molecules in these disks. Thermo-chemical models of disks can reproduce the main trends. The mass of disks remains highly uncertain and model dependent. There is a discrepancy between the masses of disks inferred from various indirect tracers and the stellar accretion rate that points to either missing physics in our interpretation of stellar accretion rates and/or the interpretations of tracers used to infer the mass of disks. While there is evidence that Herbig Be stars possess disks, these are much smaller and less chemically rich than their Herbig Ae counterparts. 

\subsection{\textsf{The relative radial distribution of gas and dust in disks around Herbig stars}}

At large spatial scales (few 10's of~au), the substructures frequently seen in the dust do not always have a corresponding gas structure and vice versa. For example, the clear dust gaps in HD\,163296 have a correspondence in the gas \citep{Isella2016}, but this can be a pure temperature and/or opacity effect \citep{vanderMarel2018, Rab2020}. The pre-transitional disk around HD\,142527 shows a clear dust trap in 1.3~mm emission, but the CO emission is smoother and extends much further inwards \citep{Boehler2017}. HCN and CS emission in this source is offset from the dust trap by almost 180$^{\circ}$ \citep{vanderPlas2014}; again, the explanation could be either a temperature or an opacity effect. AB\,Aur shows two prominent $^{12}$CO gas spiral arms  inside the mm dust cavity \citep{Tang2017}. 

A radial dependence on the vertical dust settling has been measured in the disk around HD\,163296. At the inner dust ring ($\sim\!70$~au), dust and gas remain vertically well mixed. At the outer dust ring ($\sim\!100$~au), the scale height of the dust is roughly ten times smaller than the scale height of the gas \citep{Doi2021}. This is consistent with only the inner ring being detected in scattered light \citep{muro-arena2018}.

At the spatial scales of the inner disk, interferometric studies of the dust and optical and near-IR spectroscopy of the [OI] and CO ro-vibrational lines can reveal whether or not gas and dust are co-spatial. \citet{vanderplas2009} inferred that CO is absent in the dust gaps ($<\!10$~au) of HD~97048 and HD~100546; however, [OI]\,6300~\AA\  emission extends well inside the dust gaps, thus showing the presence of gas. \citet{vdp2015} then compared the inner radius derived from CO emission with that of the dust and other inner disk gas tracers ([OI], PAHs) for a larger sample of Herbig stars. The GI disks show systematically larger CO inner radii than GII disks; in both cases, the [OI] emission extends further inward than the CO. \citet{banzatti2015, Banzatti17} compared the CO and water emission lines to measurements of the dust inner holes. They confirm that inner dust gaps are indeed depleted in molecular line emission (water and CO). \citet{hein2016} added to the CO diagnostic by introducing the line FWHM as function of upper level quantum number $J$ as a new diagnostic: (1) a constant FWHM versus $J$ can be used to infer the presence of a dust gap, (2) the presence of line wings can indicate the presence of gas inside a dust gap, and (3) a strongly decreasing line flux versus $J$ behaviour can indicate a gas depleted region (Hein Bertelsen 2015, PhD thesis). \citet{Carmona2014, Carmona2017} studied two pre-transitional disks (HD\,135344B, HD\,139614) with very deep CRIRES observations and find that for these two disks, the dust gaps (30 and 6~au respectively) are partially filled with molecular gas emission. This agrees with the findings from ALMA CO submm observations by \citet{vanderMarel2016}.

\bigskip

\noindent \textit{Summary}: The structures observed in gas lines and in the continuum do not necessarily correspond. This may be due to either temperature or opacity effects. The vertical mixing of the gas and dust has been shown to vary for at least one source which may explain differences in what is observed in the mm continuum and in scattered light. In the inner disk, the inward extent of the molecular gas tends to follow the inward extent of the dust, though there are exceptions. Atomic gas (as traced by [OI] for example) generally extends inward of the molecular gas. Thermo-chemical disk modeling work \citep[][Hein Bertelsen 2015, PhD thesis]{Bruderer2013, Bosman2019, vanderMarel2018} has revealed the complex interplay between gas and dust in the inner disk. The local gas-to-dust ratio, dust opacities, disk scale height, presence of dust traps and gas excitation mechanism all play a role in interpreting the observations.

\section{\textsf{PLANET FORMATION AROUND HERBIG STARS} }
\label{s_planetformation}
\subsection{\textsf{Planet occurrence rates among intermediate mass stars}}

The occurrence rate of super-Earths within $\sim$1 au of their host star peaks at 0.5M$_{\odot}$ (\citealt{Mulders2021}, see also \citealt{Howard2012, Mulders2015b}). On the other hand, the occurrence rate of supra-Jovian mass planets with periods  $\leq$4 years increases with stellar mass reaching a maximum at $\sim$2 \Msun and then rapidly declines (\citealt{Reffert2015}, see also \citealt{Luhn2019,Johnson2010}). The most massive star around which a planet has been detected thus far is a binary with a system mass of 6-10\Msun \citep{janson2021A&A...646A.164J}. The planet was detected by direct imaging and has a semi-major axis of 560~au. Thus it is clear that planets can form even around fairly massive Herbig stars.

Transit searches for planets find systematically lower occurrence rates for planets than radial velocity searches that is largely driven by selection effects of the surveys \citep{Moe2021}. These authors note that this driven by the fact that binaries with intermediate separations (a=0.5 -- 200 au) suppress planet formation and that radial velocity searches systematically exclude binaries while transit searches do not. \citet{Moe2021} argue that half of the dependence of the superEarth occurrence rate on stellar mass can be accounted for by the effect of binaries. The occurrence rate of supra-Jovian mass planets within 1 au around single AF stars is $8.6\pm2.3\%$ \citep{Moe2021} - consistent with the estimate by \citet{Reffert2015} for solar metalicity stars with masses ranging from 1.8-2.6M$_{\odot}$. It is not clear how the occurrence rate of supra-Jovian mass planets scales with orbital separation. 

\begin{figure}[t]
\begin{center}
\includegraphics[width=0.85\textwidth]{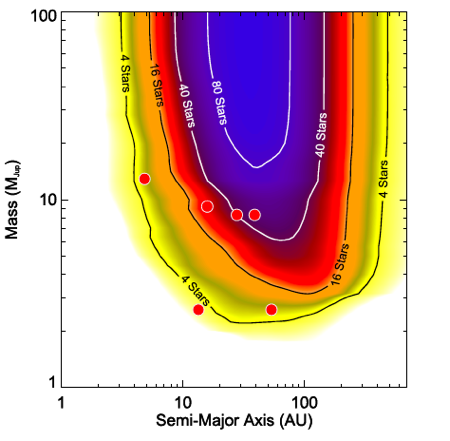}
\caption{Depth of search for the intermediate mass stars in the GPIES sample \citep{Nielsen2019}. Of the 123 intermediate 
mass stars observed thus far, four reveal planets: $\beta$~Pic, 51~Eri, HD~95806, and HR~8799. Three of the six imaged companions orbit HR~8799. }
\label{fig_nielsen}
\end{center}
\end{figure}

Indeed, direct images of planets around intermediate mass stars are rare. 
\citet{Bowler2016} find that high mass planets (5--13\,$M_J$) 
are detected at 30--300\,au orbital separation 
in only a small fraction of young A-stars (2.8\%). Similarly, the \textit{Gemini Planet Imager Exoplanet Survey} (GPIES) reports a modestly higher occurrence 
rate for the same mass range of supra-Jovian planets (9$^{+5}_{-4}$\%) from 10--100\,au (Figure \ref{fig_nielsen}; \citealt{Nielsen2019}). 
Assuming that this reflects the actual occurrence rate of massive companions orbiting 
intermediate mass stars (i.e., the planets haven't migrated inward of $\sim$10-30\,au), one might conclude that only a few percent of Herbig stars 
should reveal signatures of ongoing supra-Jovian gas giant planet formation at distances $\gtrsim$ 10 au. However, 
that does not appear to be the case.  Here we summarize the evidence for ongoing planet formation in Herbig disks and discuss the 
challenges to detecting forming planets in these systems.

\subsection{\textsf{Sign-posts of planets in disks}}
\label{sect_signposts}
There are several indirect signatures of the presence of planets in disks: the $\lambda$ Bo\"{o} phenomenon, falling evaporative bodies, gaps/rings in disks, kinematic signatures in the orbital motion of the gas in the vicinity of the orbit of a gas giant planet, and spiral arms. Here we describe what each of these signposts tell us about the presence of planets in disks around Herbig stars.

\subsubsection {The $\lambda$~Bo\"{o}tis phenomenon }
\label{s_lamdaboo}

The identification of the $\lambda$ Bo\"{o}tis phenomenon among Herbig stars has been interpreted as a signpost of a gap-opening
planet in the disk \citep[e.g.,][]{Acke2004lambdaboo, Folsom2012a, kama2015, Jermyn2018}.  
The $\lambda$ Bo\"{o}tis  phenomenon was first identified by 
\citet{Morgan1943} who noted a depletion of Mg and Ca in the photosphere of $\lambda$ Bo\"{o}tis. The definition of the 
class of $\lambda$ Bo\"{o} stars has been refined and is now understood to comprise late-B through early-F stars 
(10,500~K $\rm \leq T_{eff} \leq$ 6,500~K) that possess a depletion of refractory elements and near solar abundance of 
volatile species \citep{Paunzen2002}. Such stars comprise about 2\% of main-sequence field stars in this spectral type range. 

There are two related leading hypotheses to account for this anomalous abundance pattern. \citet{Waters1992} propose 
that $\lambda$~Bo\"{o} stars continue to accrete residual material from the circumstellar environment. The refractory 
elements are disproportionately found in dust that is blown away more efficiently than gas resulting in this abundance 
pattern. \citet{Kamp2002}  note that there is no correlation between the $\lambda$~Bo\"{o} phenomenon with stellar 
age spanning several billion years leading them to suggest that the anomalous abundance pattern arises from the passage 
of these stars through the diffuse ISM. In this scenario gas is accreted selectively relative to the dust as the dust is 
blown away by the radiation pressure of the star. 

The $\lambda$~Bo\"{o} abundance pattern observed among main sequence B-F stars has also been observed among Herbig stars 
\citep{Acke2004lambdaboo, Bruderer2012, Folsom2012a}.  In contrast to $\sim$2\% of field stars that show this abundance 
pattern, \citet{Folsom2012a} found that at least 33\% of Herbig stars show this effect.
Building on this work, \citet{kama2015} 
examined the relationship between the abundance of GI Herbig stars and GII Herbig stars  (see also \citealt{Kama2016a, Kama2016b, Jermyn2018, Kama2019, Castelli2020}).  
\citet{kama2015} find that while the abundance of volatiles among GI and GII Herbig stars are equivalent, GI 
Herbig stars are depleted in refractory elements by $\sim$0.5~dex. These authors note that all known disks with cavities among 
Herbig stars are GI sources. They note that pressure bumps in the disk at the boundary of gaps (perhaps due to planet-disk tidal interactions)  
trap and hold back grains (cf. Sect. \ref{sect_theo}), leading to accretion onto the star from an inner gas rich disk depleted in refractory elements. 
Based on their analysis they conclude that 
$\sim$1/3 of Herbig stars could possess a planet with sufficient mass to open a gap that results in a dust filtration. 

\subsubsection {Falling Evaporative Bodies}
\label{sec:feb}

As mentioned in Sec.~\ref{s_variability}, the narrow, intermittent components of metallic absorption lines in the spectra of Herbig stars have been interpreted as due to infalling, evaporating material such as exo-comets.

In fact, the presence of these infalling comets is another indirect signpost of planets in disks around Herbig stars. These were first observed towards the young debris disk $\beta$~Pic by \citet{Ferlet1987}. \citet{Beust1991} suggest that the presence of these infalling solid bodies result from stirring by a planetary mass body.  \citet{Grady1996} observed similar behaviour towards a large sample of Herbig stars and  suggest that this phenomenon is evidence that planets have already formed in the disk and thus perturb planetesimals onto eccentric orbits that bring them close to the star. However, the gas in the inner disk of Herbig stars should damp the eccentricity of these orbits. Further study of the association between the occurrence rate of red shifted absorption features and clearing of the inner disk (or perhaps the accretion rate) may clarify whether this mechanism is plausible. 

In the largest survey of this phenomenon to date, \citet{Rebollido2020} studied a sample of 117 main sequence stars with spectral types spanning G8 - B8. Their sample was selected on the basis of having either 1) previous evidence of $\beta$ Pic like phenomena, 2) an edge-on debris disk, 3) a debris disk with cold gas, 4) an infrared excess, 5) membership in a young association, 6) shell stars, or 6) a $\lambda$ Bo\"{o} abundance pattern.  Among this sample, 16 of the stars showed variable red or blue shifted features that could be attributed to infalling evaporative bodies (i.e., exocomets) all of which were earlier than A9. They note that this could be due to the difficulty of detecting these signatures against the structure in the cooler photospheres. An unbiased survey of young A and B stars may shed additional light on the presence of planets around these stars necessary for stirring exocomets and thus the occurrence of planets in disks around Herbig stars.

\subsubsection{\textsf{Rings and gaps in disks}}
\label{s_rings}

The presence of a planet in a disk also has other effects on its structure, as we saw in section~\ref{sect_theo}.
To summarize, the radial drift that would naturally occur in disks can only be halted by strong enough pressure bumps, that in turn are caused by objects of certain mass. Pressure bumps due to massive planets will lead to disk structures like cavities or rings, while the absence of strong pressure bumps will not be able to halt radial drift of the dust grains efficiently, making the disk more compact over time. In these disks, less massive planets such as super-earths could still form. Therefore, for a 1.5 \Msun \ star, disks with large cavities have massive planets ($> 200$ M$_{\bigoplus}$, while ring-like disks host somewhat lighter planets (M $\sim$ 70-200 M$_{\bigoplus}$), and lastly compact disks have planets with a mass $<$ 10 M$_{\bigoplus}$ \citep{vdm2021}.  

From a comparison of exoplanet statistics with a large ALMA disk survey, not only including Herbig stars but also lower-mass T Tauri stars, \citet{vdm2021} show that the occurrence rate of exoplanets inferred from radial velocity and transit surveys among stars in different mass bins is in agreement with occurrence rate of disk structures (cavity, rings). The authors also conclude that the mass of a planet present in the disk is crucial in determining the disk structure and size.

\subsubsection{\textsf{Kinematic Planetary Signatures}}
\label{s_kinks}

When planets in a disk open a gap (cf. Sect.~\ref{sect_theo}), the resulting pressure gradient will affect the orbital velocity of the gas \citep{Perez2015, Perez2018, Pinte2019, Armitage2020}. The gas exterior to the gap will experience a boost and the gas interior to the orbit of the planet will be slowed \citep[see for example][]{Armitage2007}. In the spiral structures emanating from the planet, the deviation from a Keplerian orbit reaches a maximum. This results in a `Doppler flip' \citep[for a review of this phenomenon see ][and references therein]{Armitage2020}. These authors propose a set of criteria for confirming that such kinematic signatures of planets (KSPs) are not artefacts due to processing and discuss other effects that can mimic this phenomenon. 

With the unparalleled sensitivity, spatial resolution, and spectral resolution of ALMA, several authors have presented evidence of KSPs in disks around Herbig stars: AB Aur \citep{Tang2017}, HD 163296 \citep{Pinte2018b, Teague2018, Teague2021}, HD 100546 \citep{Walsh2017, Casassus2019, Perez2020}, HD 97048 \citep{Pinte2019}, and MWC 480 \citep{Teague2021}. While these observations are too expensive to have conducted an unbiased survey of nearby Herbig stars, the results are suggestive. Application of this technique to more sources will elucidate the frequency of gas giant planet formation in disks. 

\subsubsection{\textsf{Spiral arms in disks}}
\label{s_imagerydisks}
\begin{figure*}[t]
\begin{center}
\includegraphics[width=0.99\textwidth]{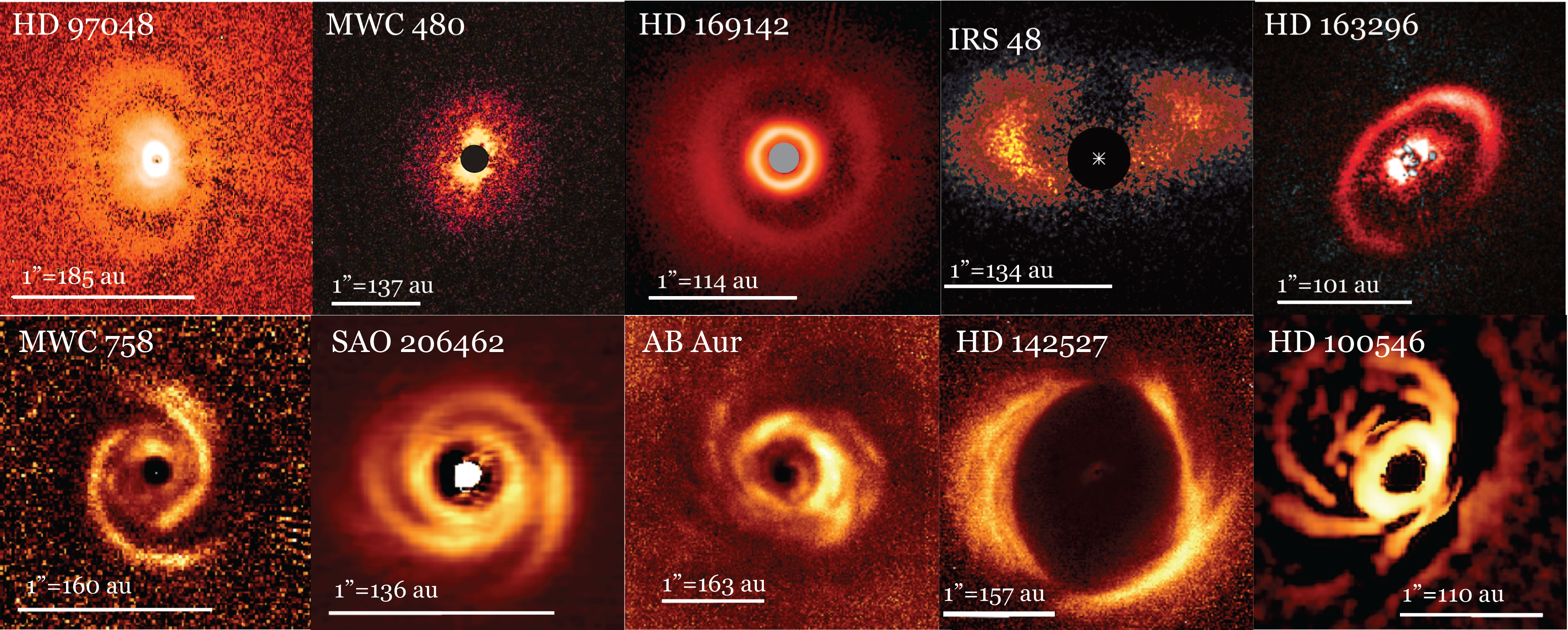}
\caption{10 single Herbig stars within 200pc that have been imaged with high contrast imagery \citep[based on figures from][]{ginski2016, Kusakabe2012, Pohl2017, vandermarel2021, muro-arena2018, Benisty2015, Garufi2013, Hashimoto2011, Avenhaus2017, Follette2017}. Of these 10 Herbig stars, five show spiral structure indicative of the presence of a supra-Jovian mass planet \citep{Dong2018d}.} 
\label{fig_dong}
\end{center}
\end{figure*}

High-resolution, high-contrast observations of Herbig disks with 8m class telescopes have produced exquisite images. 
Indeed, \citet{Dong2018b} noted that of the 10 Herbig stars within 200~pc that had been imaged and lack a stellar companion from 0.3\arcsec-5\arcsec \ that could drive spiral arms in the outer disk,
five show spiral structure (Fig. \ref{fig_dong}). Since then additional sources have been imaged such as HD~139614 \citep{Muro2020} which shows multiple rings and significant shadowing indicative of warped disk. 
 There are at least two mechanisms that can give rise to spiral structure: gravitational 
instabilities or planet-disk interactions with a supra-Jovian mass planet. A gravitationally unstable disk will form 
spiral structures (e.g., \citealt{hall2019}). While disk masses of order $M_D\sim0.5 M_\star$ are necessary to drive 
two-armed spirals and only last for thousands of years, multi-armed spirals can form with disk masses as low as 
$0.1 M_\star$ and persist for Myrs.  Are such masses representative of disks 
around Herbig stars? 

Typical dust masses in GI Herbig stars are around $1-3 \times 10^{-4}$ \Msun \citep{garufi2018}, so  assuming d/g = 0.01, disk masses would be typically $1-3 \times 10^{-2}$ \Msun. However, measuring disk mass is fraught with uncertainty. Estimates of disk mass from CO isotopologues, mm-continuum, and disk accretion rates span roughly two orders of magnitude (Sect.~\ref{sec:gasmassevolution}). 
As disk mass estimates from stellar accretion rates are often
consistent with $\rm M_{disk} \sim 0.1M_{\star}$, multi-armed spirals observed around Herbig stars could be the 
consequence of moderately gravitationally unstable disks. On the other hand, models of the upper limits on flux of HD  
suggest that most Herbig stars' disks may not be gravitationally unstable \citep{Kama2020}. Whatever the case, two-armed spirals
require much higher disk masses and survive in that state for 
very short periods of time, so it is unlikely that gravitational instability accounts for these. 

Alternatively, a massive planet can account for the spiral structure observed in these disks 
\citep[e.g.,][]{Fung2015, Dong2017b, Dong2018b}. Thus it appears that the population of supra-Jovian mass planets in the 
outer disk ($\gtrsim 30~au$) of young intermediate mass stars could be $\sim$20-50\% which is consistent with the occurrence rate inferred 
from \citet{kama2015} from the abundance patterns of Herbig stars.  However, at this time there has only been one confirmed detection of a gas giant 
planet orbiting a Herbig star (AB Aur b; \citealt{Currie2022}; see also \citealt{Zhou2022}). While the direct detection of young massive companions orbiting intermediate mass stars 
remains quite low \citep{Bowler2016, Nielsen2019}, indirect signposts of forming massive companions are quite common.
The disparity between the detection rate of supra-Jovian mass planets beyond $\sim$30~au and the frequency of indirect 
signatures of planet formation in this range is a puzzle.

\bigskip

\noindent \textit{Summary}: There are a number of indirect pieces of evidence that point to the ubiquity of planet formation 
in the disks of Herbig stars including the $\lambda$ Bo\"{o} phenomenon, the presence of falling evaporative bodies, rings and gaps in disks, KPSs, and spiral structure. While the exoplanet statistics and occurrence rate of gaps/rings in disks are fairly consistent \citep{vandermarel2021}, there is tension between the detection rate of supra-Jovian mass planets from 30-300au and the occurrence rate of spiral structures pointing to the presence of supraJovian gas giant planets in this region \citep{Dong2018b}. Validation of these indirect signatures by connecting them to detections of planets is crucial.

\subsection{\textsf{Detecting forming planets in disks}}
The detectability of gas giant planets depends on their formation pathway. The two limits are the hot start and cold start births \citep[e.g.,][]{marley2007, fortney2008, Spiegel2012}. In the case of the hot start, the accreting material goes into heating the planet making it more luminous. In the case of the cold start, the accreting material radiates its energy away resulting in a colder, less luminous young planet. For the first 100~Myr of the planet's life, this makes a significant difference in the luminosity of the planet. In the case of the young MS stars HR~8799 and $\beta$~Pic, it appears that the orbiting companions began their life as hot start planets \citep{Brandt2021a, Brandt2021b}. If this is the typical formation pathway for Jovian mass planets, then it sharpens the discrepancy between the evident signposts of forming massive companions and the direct detection of such companions orbiting young intermediate mass stars. 

One possibility is that the initial conditions of forming planets span the whole cold start -- hot start continuum \citep{Spiegel2012}. If there is a significant number of planets that begin their life as cold start objects, then the escaping accretion energy from the circumplanetary disk should be readily detectable \citep{brittain2020a}. Several complementary techniques have been employed to detect the presence of forming planets in disks such as sparse aperture masking (e.g., \citealt{Huelamo2011}), angular differential imaging (ADI; e.g., \citealt{Marois2006}), H$\alpha$ imaging (e.g., \citealt{Close2014b}), and deep searches with ALMA \citep[e.g.,][]{Andrews2021}. There have been several intriguing hints of planets imaged in disks around Herbig stars (e.g., HD 100546-\citealt{Quanz2013a, Currie2014, quanz2015ApJ, currie2015, Follette2017, Rameau2017, Currie2017}, MWC 758-\citealt{Reggiani2018k, Wagner2019, boccaletti2021}, HD 169142-\citealt{Reggiani2014, Ligi2018, Gratton2019}, and AB Aur \citealt{Currie2022}); however, AB Aur is the only Herbig star with a confirmed planet detected in the disk thus far. This low detection rate is similar to that of T Tauri stars where PDS 70 is the only one for which forming planets have been imaged \citep{keppler18, Muller2018, haffert2019, Isella2019, Zurlo2020, Benisty2021}.

It could be that planets in disks accrete episodically, and our sample is too small to have detected a forming planet in its accretion outburst phase \citep{brittain2020a}. Another possibility is that emission from circumplanetary disks peaks in the MIR and current NIR surveys for forming planets lack the requisite sensitivity to detect a forming planet (e.g., \citealt{Szulagyi2019}). An alternative approach to direct imaging of a forming planet is spectroscopic monitoring of warm gas emission lines arising from the circumplanetary disk. 

\citet{Rab2019} used the thermochemical code, ProDiMo \citep{Woitke2009}, to calculate the gas and dust temperature of a circumplanetary disk. They find that the gas is much hotter than the dust. In their reference model, gas at a temperature of $\gtrsim$1500~K extends to 0.3 au before dropping to 500~K at 0.5 au and 150~K at 1~au. The temperature depends on the interstellar background radiation, the accretion luminosity of the planet, and shock heating as gas enters the circumplanetary disk. They showed that it is plausible to detect rotational lines of CO from a circumplanetary disk with ALMA. The spatial and spectral resolution of ALMA will provide useful insight to the circumplanetary structure of material accreting onto forming planets. 

While the circumplanetary disk is not resolvable with 8m class telescopes, high-resolution spectroscopy 
serves as a surrogate for spatial resolution. This technique was pioneered in the study of classical T 
Tauri stars  and has been applied extensively to
the study of warm CO and OH emission in disks around Herbig stars \ref{sec:gasdisks}.  Such observations may also provide the
means to detect circumplanetary disks \citep{Brittain2019a}. 

Circumplanetary envelopes have complex structures that likely include a rotationally supported 
circumplanetary disk whose outer extent may range from one-third to the full 
extent of the Hill sphere \citep[e.g.,][]{quillen2004, Ayliffe2009-acc, Ayliffe2009-disc, martin2011cycle, tanigawa2012, Ayliffe2012, Gressel2013b, szulagyi2014-acc, szulagyi2016-disc, Szulagyi2017}. 
Models indicate that the temperature in the circumplanetary disk is $\gtrsim 1000~K$ in steady state \citep{szulagyi2017-temp} and perhaps 
much higher during an outburst \citep{Zhu2015b}.  At these sizes and temperatures, 
ro-vibrational emission of CO in the NIR is also detectable \citep[e.g.,][]{najita2003}. 

The ro-vibrational CO emission from the Herbig star with a pre-transition disk, HD~100546, may arise from such 
a scenario. The presence of an inner companion has been inferred from a component of 
the $v=1-0$ ro-vibrational CO emission that varies relative to the stable hot band 
(i.e., $v^\prime \geq 2$) emission (\citealt{Brittain2019a}). The profile of the $v=1-0$ ro-vibrational line varied relative to the lines with $v^{\prime}\geq2$ (i.e., the hot band lines). In 2003 the line profile of the $v=1-0$ lines matched the profile of the hot band lines. In 2006 the red side of the $v=1-0$ line brightened. In 2010, the $v=1-0$ line remained elevated relative to 2003, but the Doppler shift of the emission was -1~km~s$^{-1}$. In 2013, the blue side of the $v=1-0$ line was brighter than in 2003. By 2017, the line profile returned to the profile of the gas in 2003 \citep{Brittain2019a}. The Doppler shift and time between observations is consistent with a source of warm gas orbiting at 11.6 -- 12.3 au.  
In the case of HD~100546, the CO flux is consistent with 
emission from gas in a circumplanetary disk with a radius of $\sim 
\unit[0.3]{au}$ \citep{Brittain2013a}. \citet{Pyerin2021} model the 0.9mm ALMA images of this HD~100546 and find that an 8M$\rm_{Jup}$ planet best reproduces the ring structure that is observed. The Hill sphere of an object with this mass 12au from HD~100546 is $\sim$1au, so a circumplanetary disk that fills one-third of the Hill sphere is comparable to the size of the emitting region inferred from the CO emission. \citet{Oberg2022} applied thermochemical modeling of such a circumplanetary disk and found that the luminosity of the emission of the CO rovibrational emission inferred from their modeling is consistent with the observed luminosity.

 The source of the CO emission is now 
behind the near side of the disk. When it emerges in 2031, the source of emission can be studied with  \unit[30]{m} class telescopes.   In the meantime, ongoing 
monitoring of CO emission from similar transition disks around Herbig stars may provide 
additional candidate sources \citep{Banzatti2022}.  Expansion of this sample will  
open the door to more detailed studies of this important environment.

\bigskip

\noindent \textit{Summary}: There has only been one robust detection of a planet in a disk around a Herbig star at the time of this writing \citep{Currie2022}. Improvements in instrumentation and the commissioning of 30m class telescopes will likely enable further detections. A complimentary approach to detecting the presence of the planet is to observe gas lines from the circumplanetary disk. \citet{Rab2019} show that rotational lines of CO from circumstellar disks should be observable for wide-orbit systems. \citet{Brittain2019a} provide evidence of ro-vibrational CO emission arising from a circumplanetary disk that is consistent with expectations from models \citep{Oberg2022}. 

\section{\textsf{FUTURE PROSPECTS}}\label{sec:future}
Immense progress has been made since the last dedicated review of Herbig stars \citep{waters1998}, and there
are several exciting lines of inquiry available to astronomers that promise to advance our understanding of 
these important objects even more over the next 25 years. Here we summarize a few of the key areas of investigation that we believe will be particularly fruitful. 

Since George Herbig first proposed a class of intermediate mass pre-main sequence stars, identifying objects of this class has been a challenge. Over the years, catalogs of Herbig stars
have included controversial candidates. As the number of objects in these catalogs has been limited, the study of intermediate mass pre-main sequence stars has necessarily been limited. Furthermore, once pre-main sequence intermediate mass stars shed their disk, they are difficult to identify due to a lack of activity signatures (the so-called ``Naked Herbigs" that are analogs to weak lined T Tauri stars). Large scale surveys such as from telescopes such as Spitzer, WISE, and Gaia have improved this situation. For example, \citet{Mooley2013} used Spitzer and Wise data to identify young A stars in Taurus.  Using the Gaia database, \citet{Vioque2020} used Machine Learning to increase the number of Herbig-candidates by an order of magnitude. Upcoming large scale surveys enabled by instrumentation such as WEAVE and facilities such as Pan-STARRS and the Vera Rubin Observatory promise to enable further such advances in our identification of Herbig stars. 

The source of the X-ray emission observed from Herbig stars is still not well understood. Roughly 70\% of Herbig stars are binaries (Sec. \ref{sec:binaries}) and roughly 70\% of Herbig stars are detected in X-rays (Sec. \ref{sec:xray}).  However, there is little overlap among these samples. A dedicated study of multiplicity
among Herbig stars for which X-ray observations exist will clarify the 
extent to which lower mass companions can account for the X-ray properties of Herbigs. 

Related to this is the structure of the stellar magnetic field. There is compelling evidence that Herbig stars with masses 
$\lesssim$4~\Msun \ accrete magnetospherically, but the geometry of these fields has not been established. The more 
massive Herbig stars appear to be accreting by some other process, with boundary layer accretion being the leading contender. 
Modeling this for Herbig stars will clarify the situation and perhaps provide a more accurate means of converting accretion 
luminosity to accretion rate. There is some evidence that the accretion rate of 2-2.5\Msun  \ Herbig stars declines as 
$t_{age}^{-2}$ from $\sim$3-10Myr. However, this may underestimate the rate of decline as non-accreting A stars in this age bin 
are not included in this sample. Combining data from the evolutionary precursors to Herbig stars (intermediate mass 
T Tauri stars) and non-accreting A-stars in this age bin will shed light on how the accretion rate of intermediate 
mass stars evolves. 

As the UV excess that veils the Balmer discontinuity is difficult to measure for accretion rates less than 
10$^{-8}$\Msun~yr$^{-1}$, we must rely on proxies such as H\textsc{i} emission lines. However, the underlying 
physics that drives this correlation is not known. Optical and NIR interferometry at longer baselines is 
necessary to determine the origin of these lines - particularly for very low accreting sources. As it stands, 
the accretion rates inferred for Herbig stars imply much higher disk masses than typically observed by other 
indirect tracers such as (sub)mm CO lines (and their isotopologues), (sub)mm dust emission, and HD emission. 

Excellent progress has been made on understanding the distribution of dust
in disks and the resultant SED. The evolutionary pathways that lead to GI and GII disks and the 
metamorphosis of pre-transitional disks to transition disks to debris disks remains an open question. 
New instrumentation such as MATISSE on the VLT will provide unparalleled interferometric 
imaging in the thermal IR ($L^{\prime}-, M-, $ and $N-$) bands. This will clarify the radial structure of the inner disks around GI 
and GII Herbig stars. The James Webb Space Telescope, with its dramatic increase in sensitivity, will enable the study of 
solid state features for a much larger sample of Herbig stars allowing better characterization of the role of environment on 
the dust properties of these stars, and to detect weak emission from gas in the inner disk. Such studies will also enable studies of the role stellar multiplicity plays in 
the development of GI and GII disks. ALMA has revolutionized our understanding of mm grains in disks, and now the Square Kilometer Array (SKA) promises to do the same for cm-sized grains \citep{Ilee2020}. Looking even further ahead the Next Generation Very Long Array will provide unprecedented resolution and sensitivity at frequencies bridging SKA and ALMA.  

Our knowledge of the gas content of Herbig disks has increased remarkably over the past 25yrs. Thermo-chemical 
modeling of disks is able to reproduce many of the trends observed among various gas emission lines spanning the 
NIR to the mm. JWST, with its enhanced resolution and sensitivity, promises to open new windows into the gas 
content of disks albeit many known Herbig stars being too bright to be studied. Understanding how the gas content and excitation varies among stars with different effective 
temperatures, degrees of flaring, and dust properties will improve the characterization of the initial chemical 
conditions of forming planets; especially in the inner disk ($<\!10$~au) the synergy between continuum and emission line studies can provide unique insights. We note that ALMA observations have only scratched the surface of understanding the chemistry in disks (MWC\,480 and HD\,163296 the only two disks being studied in detail) and more efforts are required to push beyond the commonly used tracer CO. Perhaps the most direct tracer of disk mass is HD, but the conversion of line flux 
from this molecule into disk masses requires careful characterization of the disk temperature which in turn depends 
sensitively on the dust properties of the disk. The mass of disks around Herbig stars remains an ongoing puzzle. 

Finally, the detection of planets around Herbig stars remains a challenge. The advent of 30m class telescopes in the 
coming decade will enhance our sensitivity to the presence of forming planets and move the inner working angle closer to the star. 
The detection of a significant sample of forming planets will provide the means to probe the early evolution of forming planets 
and clarify the range of initial conditions (bounded by the cold start and hot start scenarios) are reflected in the 
population of planets and thus clarify the status of direct imaging surveys of gas giant planets around young main sequence
stars in the solar neighborhood.

\section{\textsf{A NEW DEFINITION OF HERBIG STARS}}
While the empirical classification criteria that defined the classes of Herbig and T Tauri stars have been, and still are, 
very useful for many studies of star- and planet formation, we already noted in Sect.~\ref{subsec:def} that a full view 
on the evolution of intermediate mass stars should ideally be based on the mass of the star (more difficult to derive directly 
from observations) and not directly on its temperature (evident from spectral type). This thought motivates the definition 
of a new, stellar mass-based definition of the group of Herbig stars, i.e. leaving out the spectral type limitation of the 
Herbig star or T Tauri star definitions. As a lower mass boundary we propose to use the criterion that the atmosphere has to 
be radiative at the ZAMS. We propose to define the Herbig stars as the class of intermediate mass young stars that are evolving 
towards the main sequence, with mass $\gtrsim$1.5 M$_{\odot}$, that are surrounded by a remnant accretion disk, as evidenced by 
the detection of circumstellar gas at optical and/or longer wavelengths, and an IR excess caused by circumstellar 
dust. Herbig stars can be subdivided into warmer Herbig Ae/Be stars and cooler IMTT stars. The upper mass boundary is more 
difficult establish. The stars with masses $\gtrsim$ 8-10\Msun\ likely reach the main sequence prior to the dissipation of the surrounding envelope and 
are all but impossible to detect on a pre-main sequence track. However there are well established pre-main sequence B stars that exceed this mass (e.g., MWC 297). 
This definition carves out a general area in the HR diagram 
whose boundaries are set by the birth line, the ZAMS line, and evolutionary tracks of stars with mass $\gtrsim$1.5\Msun. 

\begin{acknowledgements}

IK acknowledges funding from the European Union H2020-MSCA-ITN-2019 under Grant Agreement no. 860470 (CHAMELEON). GM acknowledges funding from the Spanish project "On the Rocks II" (PGC2018-101950-B-100). RDO acknowledges funding for the STARRY project which received funding from the European Union’s Horizon 2020 research and innovation programme under MSCA ITN-EID grant agreement No 676036.

\end{acknowledgements}

%
\section*{Conflict of interest}
The authors declare that they have no conflict of interest.

\bibliographystyle{mnras}
\bibliography{references.bib}   

\end{document}